\documentclass{aa}  
\usepackage{url}
\usepackage{natbib}
\usepackage[colorlinks, linkcolor=blue, citecolor=blue, urlcolor=blue]{hyperref} 
\bibpunct{(}{)}{;}{a}{}{,} 

\usepackage{etoolbox}
\makeatletter
\newcount\c@additionalboxlevel
\newcount\c@maxboxlevel
\patchcmd\@combinedblfloats{\box\@outputbox}{%
  \stepcounter{additionalboxlevel}%
  \box\@outputbox
}{}{\errmessage{\noexpand\@combinedblfloats could not be patched}}

\AtBeginShipout{%
  \ifnum\value{additionalboxlevel}>\value{maxboxlevel}%
    \typeout{Warning: maxboxlevel might be too small, increase to %
      \the\value{additionalboxlevel}%
    }%
  \fi 
  \@whilenum\value{additionalboxlevel}<\value{maxboxlevel}\do{%
    \typeout{* Additional boxing of page `\thepage'}%
    \setbox\AtBeginShipoutBox=\hbox{\copy\AtBeginShipoutBox}%
    \stepcounter{additionalboxlevel}%
  }%
  \setcounter{additionalboxlevel}{0}%
}
\makeatother

\usepackage{graphicx}
\usepackage{txfonts}
\usepackage{siunitx}
\usepackage{multirow}
\usepackage{amsmath}	
\usepackage{amssymb}	
\usepackage{ae,aecompl}
\usepackage{euscript}
\usepackage{setspace}
\usepackage{mathabx}
\usepackage{bigints}
\usepackage{booktabs,makecell}
\usepackage{caption}
\usepackage{subcaption}
\usepackage{multirow}
\usepackage[table]{xcolor}
\usepackage[rightcaption]{sidecap}
\sidecaptionvpos{figure}{t}

\begin{document}

   \title{Starburst and post-starburst high-redshift protogalaxies}
   \titlerunning{Impact of CR feedback in Starburst and post-Starburst Protogalaxies}
   \subtitle{The feedback impact of high energy cosmic rays}

   \author{Ellis R. Owen
          \inst{1, 2}
          \and
          Kinwah Wu\inst{1, 3}
          \and
          Xiangyu Jin\inst{4, 5}
          \and
          Pooja Surajbali\inst{6}
          \and
          Noriko Kataoka\inst{1, 7}
          }

   \institute{Mullard Space Science Laboratory, University College London, Dorking, Surrey, RH5 6NT, United Kingdom\\
             \email{ellis.owen.12@ucl.ac.uk}
         \and
             Institute of Astronomy, Department of Physics, National Tsing Hua University, Hsinchu, Taiwan (ROC)
         \and
             School of Physics, University of Sydney, NSW 2006, Australia
         \and
             School of Physics, Nanjing University, Nanjing 210023, China
         \and
             Department of Physics and McGill Space Institute, McGill University, 3600 University St., Montreal QC, H3A 2T8, Canada
         \and
             Max-Planck-Institut f\"{u}r Kernphysik, Saupfercheckweg 1, Heidelberg 69117, Germany
         \and
             Department of Physics, Faculty of Science, Kyoto Sangyo University, Kyoto, 603-8555, Japan
             }

   \date{Received ----; accepted ----}

 
  \abstract{Quenching of star-formation has been identified in many starburst and post-starburst galaxies, indicating burst-like star-formation histories (SFH) in the primordial Universe. 
  Galaxies undergoing violent episodes of star-formation are expected to be rich in high energy cosmic rays (CRs). We
  have investigated the role of these CRs in such environments, particularly how they could contribute to this burst-like SFH via quenching and feedback. 
   These high energy particles interact with the baryon and radiation fields of their host via hadronic processes to produce secondary leptons. The secondary particles then also interact with ambient radiation fields to generate X-rays through inverse-Compton scattering. In addition, they can thermalise directly with the semi-ionised medium via Coulomb processes. 
   Heating at a rate of $\sim 10^{-25} \; \text{erg}~\text{cm}^{-3}~\text{s}^{-1}$ can be attained by Coulomb processes in a star-forming galaxy with one core-collapse SN event per decade, and this is sufficient to cause quenching of star-formation. At high-redshift, a substantial amount of CR secondary electron energy can be diverted into inverse-Compton X-ray emission. 
   This yields an X-ray luminosity of above $10^{41}~\text{erg}~\text{s}^{-1}$ by redshift $z=7$ which drives a further heating effect, operating over larger scales. This would be able to halt inflowing cold gas filaments, strangulating subsequent star-formation. 
   We selected a sample of 16 starburst and post-starburst galaxies at $7\lesssim z \lesssim 9$ 
  and determine the star-formation rates they could have sustained. 
We applied a model with CR injection, propagation and heating to calculate energy deposition rates in these 16 sources. Our calculations show that CR feedback cannot be neglected as it has the strength to suppress star-formation in these systems. We also show that their currently observed quiescence is consistent with the suffocation of cold inflows, probably by a combination of X-ray and CR heating.
}


   \maketitle
%

\section{Introduction}
\label{sec:introduction}

Post-starburst galaxies are systems in which star-formation appears to have been rapidly quenched after a violent starburst episode~\citep{Dressler1983ApJ, Couch1987MNRAS}. 
Such systems are seen both at low redshift~\citep{Dressler1983ApJ, Couch1987MNRAS, French2015ApJ, Rowlands2015MNRAS, Alatalo2016ApJ, French2018ApJ} and high redshift~\citep{Watson2015Natur, Hashimoto2018Nat, Owen2018sub} and the abundant reservoirs of interstellar molecular gas often found in these objects would ordinarily enable vigorous star-formation to continue -- however, it instead appears to be quenched~\citep{French2015ApJ, Rowlands2015MNRAS, Alatalo2016ApJ}. Researchers have therefore started to explore alternative mechanisms that would explain the observed quiescence.
 Proposed mechanisms include:
 starvation~\citep{Gunn1972ApJ, Cowie1977Natur, Nulsen1982MNRAS, Moore1999MNRAS, Boselli2006PASP}, 
 strangulation~\citep{Larson1980ApJ, Boselli2006PASP, Peng2015Natur}, 
feedback from the starburst and associated stellar end-products~\citep{Goto2006MNRAS, Kaviraj2007MNRAS, French2015ApJ}, 
 and the removal of dense molecular gas in outflows~\citep{Narayanan2008ApJS} 
 (potential evidence of which would be low-ionisation emission lines in the central regions of post-starburst outflows, for example ~\citealt{Yang2006ApJ, Yan2006ApJ, Tremonti2007ApJ}).
 
The presence of the large molecular reservoir, dust and a substantial interstellar medium (ISM) in many of these systems disfavours the action of outflows~\citep{Roseboom2009ApJ, Rowlands2015MNRAS, Alatalo2016ApJ, French2018ApJ, Smercina2018ApJ}. However, 
a deficiency of dense molecular gas pockets recently observed in low-redshift post-starburst 
systems~\citep{French2018ApJ}
implies the operation of processes
able to provide additional pressure support to the molecular gas reservoir against fragmentation and collapse.
There is debate over the exact process which may be at work here but, if it is feedback from the starburst episode itself, a 
promising candidate would be the action of high energy (HE) cosmic rays (CRs): 
these are able to directly provide additional pressure support~\citep[e.g.][]{Salem2014MNRAS, Zweibel2016APS}, as well as drive interstellar heating effects~\citep[e.g][]{Begelman1995ASPC, Zweibel2013PhPl, Zweibel2016APS, Owen2018MNRAS} so as to also increase thermal pressure support. 
Indeed, such CR heating would be particularly focussed in denser regions of the host galaxy's ISM. 
This is where increased thermal pressure support would be particularly required to hamper gravitational collapse and halt the formation of the dense pockets required for star-formation.

Not only do CRs provide the required feedback channels, they are also known to be abundant in starburst galaxies -- evidence of which includes the bright $\gamma$-ray glow of,
for example, Arp 220~\citep[e.g.][]{Griffin2016ApJ, Peng2016ApJ, Yoast-Hull2017MNRAS}, NGC 253 and M82~\citep{Acero2009Sci, VERITAS2009Natur, Abdo2010ApJ, Rephaeli2015NucPhysBProc} and M31~\citep{Abdo2010A&A-b} among others~\citep[e.g.][]{Abdo2010A&A-a, Abdo2010A&A, Hayashida2013ApJ, Tang2014ApJ, Rojas-Bravo2016MNRAS} which results from CR interactions.
The source of these CRs is thought to be stellar end products, for example supernova (SN) remnants~\citep{Hillas1984ARAA, Ackermann2013Sci, Kotera2011ARAA, Morlino2017hsn}, which can accelerate CRs to relativistic energies by, for example, Fermi processes \citep{Fermi1949PhRv}.
The high star-formation rates of starburst protogalaxies leads to frequent core-collapse SN explosions from their rapidly ageing population of massive, low-metallicity Type O and B stars.  
 As such, starburst environments are effective CR factories~\citep{ Karlsson2008AIPC, Lacki2011ApJ, Lacki2012AIPC, Wang2014AIPC, Farber2018ApJ}, and it 
 would therefore not be surprising for CRs to have an important role in feedback and quenching, both in high-redshift protogalaxies as well as their more local counterparts. 
 
 This study particularly considers the role of CRs in high-redshift protogalactic environments, when these systems were forming their first populations of stars.
We organise this paper as follows: 
In \S\ref{sec:star_forming_protogalaxies} we 
consider the indications of bursty star-formation histories and the presence of quenching in starburst and post-starburst systems observed at high redshift.
We then introduce a
 parametric model of a characteristic protogalaxy environment in \S\ref{sec:protogalaxies}, in which we consider the injection and interactions of CRs.
In \S\ref{sec:cr_transport}, we model CR propagation in protogalaxies, accounting for their interactions, secondaries, cooling and the ambient magnetic field. In \S\ref{sec:heating}, we then consider how CR heating arises and can be driven by two processes -- either by direct Coulomb thermalisation, or via inverse-Compton radiative emission -- and we specify a computational scheme to calculate the CR heating power in each case.
We apply our heating model to a sample of observed high-redshift starburst and post-starburst systems in \S\ref{sec:application}, and discuss the implications of the results for galactic evolution and star-formation histories. 
We summarise the work in \S\ref{sec:conclusions}.


\section{Quenching of star-formation in high-redshift protogalaxies}
\label{sec:star_forming_protogalaxies}

\begin{table*}
	\centering
	\resizebox{\textwidth}{!}{\begin{tabular}{l||cccc|ccc|c}
		\multicolumn{1}{c}{}  & \multicolumn{4}{c}{Literature Values}  &  \multicolumn{3}{c}{Estimated Quantities} & \\
		 \multicolumn{1}{c}{}  & \multicolumn{4}{c}{$\overbrace{\rule{7.6cm}{0pt}}$}  &  \multicolumn{3}{c}{$\overbrace{\rule{5.6cm}{0pt}}$} & \\
		Galaxy ID & $z^{a}$ & $\mathcal{R}_{\rm SF}$/$M_{\odot}~\text{yr}^{-1}$ & $M_{*}$/$10^9 \; M_{\odot}$ & $\tau_{*}$/Myr$^{b}$ & $t_{\rm SB}$/Myr$^{c}$ & $\mathcal{R}^{*}_{\rm SB}$/$M_{\odot}~\text{yr}^{-1}$$^{d}$ & $\tau_{\rm dyn}$/Myr$^{e}$ & Refs$^{f}$ \\
		&&&&&&&& \\[-0.5em]
		\hline
		 &&&&&&&& \\[-0.75em]
		A1689-zD1 & $7.60$ & $2.7^{+0.3}_{-0.3}$ & $1.7^{+0.7}_{-0.5}$ & $81^{+67}_{-34}$ & 350 & 4.9 & 13 & (1, 2, 3, 4) \\ 
		&&&&&&&& \\[-0.75em]
		UDF-983-964 & $\textit{7.3}^{\it +0.4}_{\it -0.3}$ & $7.4^{+2.4}_{-1.4}$ & $2.2^{+2.0}_{-1.5}$ & $170^{+70}_{-120}$ & 290 & 7.6 & 26 & (5, 6)\\ 
		&&&&&&&& \\[-0.75em]
		GNS-zD2 & $\textit{7.1}^{\it +1.5}_{\it -0.6}$ & $5.9^{+6.4}_{-1.6}$ & $2.5^{+1.8}_{-2.2}$ & $250^{+70}_{-120}$ & 240 & 10 & 15 & (5)\\ 
		&&&&&&&& \\[-0.75em]
		\multirow{ 2}{*}{MACS1149-JD1$^{g}$} & \multirow{ 2}{*}{9.11} & \multirow{ 2}{*}{$4.2^{+0.8}_{-1.1}$} & \multirow{ 2}{*}{$1.1^{+0.5}_{-0.2}$} & $3.0^{+2.0}_{-1.0}$ & \multirow{ 2}{*}{100}  & \multirow{ 2}{*}{11} & \multirow{ 2}{*}{23} & \multirow{ 2}{*}{(7, 8)} \\
		 & & & & $290^{+190}_{-120}$ & & & & \\ 
		&&&&&&&& \\[-0.75em]
		CDFS-3225-4627 & $\textit{7.1}^{\it +1.5}_{\it -0.5}$ & $7.4^{+5.2}_{-1.8}$ & $3.5^{+2.0}_{-2.3}$ & $280^{+70}_{-120}$ & 210 & 17 & 16 & (5, 9)\\ 
		&&&&&&&& \\[-0.75em]
		UDF-3244-4727$^{h}$ & $\textit{7.9}^{\it +0.8}_{\it -0.6}$ & $5.4^{+2.0}_{-1.2}$ & $2.8^{+0.5}_{-2.1}$ & $320^{+70}_{-120}$ & 80 & 35 & 24 & (5, 10)\\ 
		&&&&&&&& \\[-0.75em]
		HDFN-3654-1216 & $\textit{6.3}^{\it +0.2}_{\it -0.2}$ & $9.8^{+1.7}_{-0.9}$ & $6.9^{+0.3}_{-3.8}$ & $430^{+70}_{-120}$ & 190 & 36 & 38 & (5)\\ 
		&&&&&&&& \\[-0.75em]
		GNS-zD3 & $\textit{7.3}^{\it +0.9}_{\it -0.4}$ & $7.1^{+2.0}_{-1.1}$ & $4.2^{+0.4}_{-1.5}$ & $350^{+70}_{-120}$ & 110 & 38 & 22 & (5)\\ 
		&&&&&&&& \\[-0.75em]
		UDF-640-1417 & $\textit{6.9}^{\it +0.1}_{\it -0.1}$ & $10.5^{+0.7}_{-0.7}$ & $6.6^{+0.3}_{-0.9}$ & $380^{+70}_{-120}$ & 140 & 47 & 26 & (5, 6) \\ 
		&&&&&&&& \\[-0.75em]
		GNS-zD4 & $\textit{7.2}^{\it +0.4}_{\it -0.2}$ & $11.2^{+1.4}_{-1.2}$ & $6.8^{+0.3}_{-0.7}$ & $360^{+70}_{-120}$ & 120 & 57 & 19 & (5)\\ 
		&&&&&&&& \\[-0.75em]
		GNS-zD1 & $\textit{7.2}^{\it +0.2}_{\it -0.2}$ & $12.6^{+1.2}_{-1.1}$ & $7.6^{+0.4}_{-0.5}$ & $360^{+70}_{-120}$ & 120 & 63 & 12 & (5)\\ 
		&&&&&&&& \\[-0.75em]
		GNS-zD5 & $\textit{7.3}^{\it +0.2}_{\it -0.2}$ & $20.9^{+1.5}_{-1.4}$ & $12.3^{+0.3}_{-2.1}$ & $350^{+70}_{-120}$ & 110 & 110 & 14 & (5)\\ 
	\end{tabular}}
	\caption{\small High-redshift systems with a comparatively suppressed rate of star-formation, but show evidence of a developed stellar population indicative of an earlier starburst episode. 
	The listed objects are among the most distant observed for which suitable information about the stellar population and star-formation rate is known. Notes: \\
	$^{a}$\!\;Redshifts: photometric redshifts are indicated in italics, otherwise redshift values are calculated spectroscopically. In the spectroscopic cases, the uncertainty in redshift was less than the precision to which these values are quoted, and so is not shown. Uncertainties are shown for the less precise photometric redshifts.\\
	$^{b}$\!\;Stellar population ages: we quote the best-fit stellar ages only as constraints depend strongly on the assumed star-formation history.\\
	$^{c}$\!\;Starburst timescale: this is estimated assuming a redshift of galaxy formation of $z_{\rm f}=15.4$, the highest value suggested in the literature (see ref. 7, and also~\citealt{Thomas2017A&A}, which suggests a similar but slightly lower redshift of formation of $z_{\rm f}=14.8$), with the exception of A1689-zD1, for which there is evidence that a later epoch of formation is more appropriate, of $z=9$ (see ref. 4). In general, these values are intended to provide an over-estimate for the starburst time-scale to give a conservative star-formation rate, and so should be taken as an upper limit.\\
	$^{d}$\!\;Star-formation rate during starburst phase: these are estimated from the stellar mass and the starburst timescale. As such, they should be considered a lower limit.\\
	$^{e}$\!\;Dynamical timescales: these are estimates and should be treated as an upper limit. In general the stellar mass was used as a minimum mass to estimate the maximum plausible value of $\tau_{\rm dyn}$, as estimates of the galaxy mass were not stated or could not be estimated from the literature. The exceptions are {MACS1149-JD1} and {A1689-zD1} for which the full-width of half maximum of the O\!\;III (ref. 7) and C\!\;III lines (ref. 4) respectively were used to estimate the dynamical mass. Radial estimates were determined either by the half-light radius of the observed system, or the quoted galaxy radius where available, although we note that values for {GNS-zD1}, {GNS-zD1}, {GNS-zD3}, {GNS-zD4}, {GNS-zD5}, {CDFS-3225-4627} and {HDFN-3654-1216} were calculated based on half-light radii estimated by eye from the photometric images in ref (5), and so should be treated with caution. \\
	$^{f}$\!\;References: (1) \citet{Bradley2008ApJ}, (2) \citet{Watson2015Natur}, (3) \citet{Knudsen2017MNRAS}, (4) \citet{Mainali2018MNRAS}, (5) \citet{Gonzalez2010ApJ}, (6) \citet{Bouwens2004ApJ}, (7) \citet{Hashimoto2018Nat}, (8) \citet{Owen2018sub},  (9) \citet{Bouwens2006Natur}, (10) \citet{Oesch2009ApJ}.\\
	$^{g}$\!\;Stellar population modelling suggests the presence of two distinct stellar populations in MACS1149-JD1 --see ref (7). Here, we quote the best-fit stellar population ages for the young component (top line) and the old component associated with the earlier starburst phase (second line). \\
	$^{h}$\!\;Also referred to as {HUDF-708} in ref (10).}
	\label{tab:high_z_galaxies_table_post_sb}
\end{table*}

Recent studies have shown substantial star-forming activity in galaxies at high redshifts~\citep[e.g][]{Watson2015Natur, Hashimoto2018Nat}. 
These galaxies can be separated into two groups: those currently undergoing a starburst, and those where the starburst appears to have been quenched. 
An example of such quiesence is {MACS1149-JD1}, in which there is evidence of a second, older population of stars 
alongside those which would have been born more recently during the observed ongoing star-formation phase.
Thus, the onset of star-formation had to arise in this system at a time long before the epoch at which it was observed ($z=9.11$), 
possibly as early as $z\approx 15$, only 250 Myr after the Big Bang~\citep{Hashimoto2018Nat, Owen2018sub}. 
Consideration of the observed star-formation rate, stellar mass and the spectroscopic best-fit ages of the young and old stellar populations in this system leads to an intriguing conclusion:
to attain its estimated stellar mass by $z=9.11$, {MACS1149-JD1} must have formed stars at a much higher rate on average than the measured star-formation rate.
Hence, the observed star-formation must be relatively suppressed compared to the earlier episode during which the bulk of the stellar population would have formed. 
Moreover, the star-formation history (SFH) must have been burst-like but subsequently quenched -- not unlike the SFHs inferred from closer post-starburst systems~\citep{French2015ApJ, Rowlands2015MNRAS, French2018ApJ}.

A number of similar high-redshift galaxies were found to exhibit burst-like SFHs, and we have identified 11 other examples -- see Table~\ref{tab:high_z_galaxies_table_post_sb}.
These systems tend to show a relatively high stellar mass combined with a relatively low inferred star-formation rate, being high-redshift galaxies in a post-starburst, quenched stage of their evolution.
If enforcing such an assumption on their SFH,
sufficient information is available in these 12 cases to estimate the timescale 
over which their rapid star-formation episode progressed, if also adopting an upper estimate for the redshift of formation of the first stars in these galaxies, $z_{\rm f}$.
Recent studies (e.g.~\citealt{Hashimoto2018Nat}) estimate this redshift to be around $z_{\rm f}=15^{+2.7}_{-2.0}$ (see also~\citealt{Thomas2017A&A} which suggests a similar but slightly lower redshift of formation of $z_{\rm f}=14.8$), which would correspond to a time 
when the age of the Universe was around 250-300 Myr~\citep{Wright2006PASP}. It is likely that this would be an extreme case for the formation of the earliest galaxies, so could reasonably be regarded as a lower age limit, with many galaxies forming their first stars at later times -- for example, evidence suggests a later $z_{\rm f} \approx 9$ for {A1689-zD1} is more appropriate~\citep{Mainali2018MNRAS}. 

It is also informative to consider certain ongoing starbursts at high-redshift where observations allow for the star-formation rate to be found, and the characteristic stellar population age to be estimated to give indications as to the duration of the active starburst. We have been able to identify four examples where this is already possible from the literature -- see Table~\ref{tab:high_z_galaxies_table_sb}. This provides us with information about pre-quenched systems as well as their quenched counterparts.

\subsection{Post-starburst protogalaxies and indications of quenching}
\label{sec:galaxies_post_starburst}

The 12 examples of post-starburst high-redshift candidates are listed in Table~\ref{tab:high_z_galaxies_table_post_sb} (from lowest to highest inferred burst-phase star-formation rates), with literature references indicated.
In all cases, redshift, star-formation rate, stellar mass, stellar population age and dynamical timescale were measured or could be inferred from the literature. 
Estimates for the upper limit of the starburst period length, the lower limit for the star-formation rate during this starburst period, and upper limit of the dynamical timescale, $\tau_{\rm dyn}$, for each system are also shown. $\tau_{\rm dyn}$ is estimated
either from the velocity dispersion of emission lines in the spectrum~\citep{Binney2008_book, Forster-Schreiber2009ApJ, Epinat2009A&A, Gnerucci2011A&A} if available\footnote{
The dynamical mass may be estimated from observations using the velocity dispersion of the line profile. By the virial theorem,
$M_{\rm dyn} = f \text{(FWHM)}^2 \;\! r_{\rm gal}/{\rm G}$
~\citep[e.g.][]{Gnerucci2011A&A}, where $f \approx 2.25$~\citep{Binney2008_book, Forster-Schreiber2009ApJ, Epinat2009A&A}, $r_{\rm gal}$ is the approximate radius of the galaxy (inferred from the half-light radius) and ${\rm G}$ is the Newtonian gravitational constant. The FWHM is the full width of half-maximum of the line profile, being equal to the velocity dispersion of the system. This accounts for the difference in relative velocity due to the circular orbiting motion of the emitting gases around the galaxy. The dynamical timescale is $\tau_{\rm dyn} = (\pi/2) \sqrt{r^3/{\rm G} M_{\rm dyn}}$~\citep[e.g.][]{Binney2008_book}.
}, 
or by taking the stellar mass as a lower bound on the galaxy's dynamical mass (presumably there would be components in the galaxy other than stars, for example interstellar gas, which means the true dynamical mass would be larger) to allow a lower bound on $\tau_{\rm dyn}$ to be found~\citep[e.g.][]{Binney2008_book}.
We note that, physically, $\tau_{\rm dyn}$ represents the minimum time required for the effects of a feedback process to manifest themselves (whether by strangulation, heating or other mechanisms) and impact on the progression of subsequent star-formation.

It is possible to crudely estimate the timescale over which the starburst phase would have been active using
\begin{equation}
t_{\rm SB} \approx \left[t(z_{\rm obs})-\tau_{*}\right]-t(z_{\rm f}),
\label{eq:sb_timescale}
\end{equation}
which is the difference between the approximate time since the starburst ended ($t_{\rm end}\approx \tau_{*}$, the characteristic stellar population age), and the estimated lower limit of the formation time of the galaxy, at redshift $z_{\rm f}$. This gives a rough upper limit to the plausible timescale over which the starburst proceeded, and may be justified when considering that $t_{\rm SB}$ is considerably smaller than $t(z_{\rm obs})-t(z_{\rm f})$ such that the starburst is bound to be triggered near the $t(z_{\rm f})$ end of the timeline. Indeed, we later find that this approximation is not the main source of uncertainty in our model, and so is sufficient for our purposes.
This allows a lower bound for the starburst phase star-formation rate $\mathcal{R}_{\rm SB}^{*}$ to be estimated as
\begin{equation}
\mathcal{R}^{*}_{\rm SB} \approx \frac{(M_{*}/M_{\odot})}{(t_{\rm SB}/\text{yr})}~M_{\odot}~\text{yr}^{-1} \ ,
\end{equation}
which holds if the vast majority of the galaxy's stellar mass forms within the starburst period and does not significantly evolve after the end of the starburst episode (while some fraction of the more massive stars in these systems will have gone through their evolutionary lifecycle by the time these galaxies are observed, the number of stars with masses sufficiently large to yield a SN within a few hundred Myr would only comprise a small fraction of their total stellar population and would not have a large impact on the overall galaxy stellar mass). The quantities $\mathcal{R}^{*}_{\rm SB}$ and $t_{\rm SB}$ are calculated in Table~\ref{tab:high_z_galaxies_table_post_sb} for each of the 12 post-starburst galaxies, which begins to give some insight into the physical processes operating within them:
For instance, it can be seen that quenching does not occur immediately after the onset of star-formation. Instead, it takes several dynamical timescales to reduce star-formation to a relatively quiescent state, suggesting feedback is a gradual process rather than instantaneous. 

One outlier in this sample would appear to be {A1689-zD1}. 
This has a particularly long estimated starburst period, with a correspondingly low starburst star-formation rate.
Moreover, it has also been found to harbour an abundance of interstellar dust -- at a level of around 1\% of its stellar mass~\citep{Watson2015Natur, Knudsen2017MNRAS}. Dust in galaxies is thought to be built up over time as stellar populations evolve, being injected by, for example, AGB stars~\citep{Ferrara2016MNRAS}, but a large fraction would be destroyed by the frequent SN events in a starburst~\citep{Bianchi2007MNRAS, Nozawa2007ApJ, Nath2008ApJ, Silvia2010ApJ, Yamasawa2011ApJ}. In this case, it would seem that the SN rate is insufficient to destroy forming dust on a competitive timescale to its formation, given that the dust destruction time would be of the order 0.1 - 1 Gyr for a star-formation rate of $\mathcal{R}_{\rm SF} \approx 5.0~\text{M}_{\odot}~\text{yr}^{-1}$~\citep{Temim2015ApJ, Aoyama2017MNRAS}.

\subsection{Protogalaxies with ongoing starburst activity}
\label{sec:ongoing_starbursts}

\begin{table*}
	\centering
	\begin{tabular}{l||cccc|cc|c}
		\multicolumn{1}{c}{}  & \multicolumn{4}{c}{Literature Values}  &  \multicolumn{2}{c}{Estimated Quantities} & \\
		 \multicolumn{1}{c}{}  & \multicolumn{4}{c}{$\overbrace{\rule{6.8cm}{0pt}}$}  &  \multicolumn{2}{c}{$\overbrace{\rule{3.2cm}{0pt}}$} & \\
		Galaxy ID & $z^a$ & $\mathcal{R}_{\rm SF}$/$M_{\odot}~\text{yr}^{-1}$ & $M_{*}$/$10^9 \; M_{\odot}$ & $\tau_{*}$/Myr & $t_{\rm SB}$/Myr & $\tau_{\rm dyn}$/Myr$^b$ & Refs$^c$ \\
		&&&&&& \\[-0.5em]
		\hline
		&&&&&& \\[-0.75em]
		GN-z11 & $11.1$ & $24^{+10}_{-10}$ & $1.0^{+1.5}_{-0.6}$ & $40^{+60}_{-24}$ & $42^{+140}_{-30}$ & 11 & (1)\\ 
		&&&&&& \\[-0.75em]
		EGS-zs8-1$^{d}$ & $7.73$  & $79^{+47}_{-29}$ & $7.9^{+4.7}_{-2.9}$ & $100^{+220}_{-68}$ & $100^{+150}_{-60}$ & 23 & (2, 3)\\ 
		&&&&&& \\[-0.75em]
		GN-108036 & $7.21$ & $100^{+5.0}_{-2.0}$ & $0.58^{+0.14}_{-0.14}$ & $5.8^{+1.5}_{-1.4}$ & $5.8^{+1.6}_{-5.8}$ & 61 & (4) \\ 
		&&&&&& \\[-0.75em]
		SXDF-NB1006-2 & $7.21$ & $350^{+170}_{-280}$ & $0.35^{+1.8}_{-0.14}$ & $1.0^{+9.0}_{-0.0}$ & $1.0^{+33}_{-1.0}$ & 13 & (5) \\ 
		&&&&&& \\[-0.75em]
	\end{tabular}
	\caption{\small Sample of high-redshift starburst galaxies identified in the literature which appear to be undergoing a starburst event during the epoch at which they are observed. $z$ is the redshift of observation, $\mathcal{R}_{\rm SF}$ is the inferred star-formation rate, $M_{*}$ is the estimated total stellar mass, $\tau_{*}$ is the estimated characteristic mean age of the stellar population, $t_{\rm SB}$ is the timescale over which star-formation would appear to have been active during the current burst and $\tau_{\rm dyn}$ is an estimated upper limit for the dynamical timescale of the system. These galaxies are among the most distant known in the Universe, for which suitable information is available regarding the star-formation rate, stellar mass and stellar population. Notes: \\
	$^{a}$\!\;Redshifts: these redshifts are determined spectroscopically. In all cases, the uncertainty in the observed redshifts of these objects is substantially less than the precision to which these values are quoted. \\
	$^{b}$\!\;Dynamical timescales: these are estimates and should be treated as an upper limit. In general the stellar mass was used as a minimum mass to estimate the maximum plausible value of $\tau_{\rm dyn}$ as estimates of the galaxy mass was not stated or could not be derived from the literature. This is with the exception of {SDXF-NB1006-2}, where the dynamical mass estimate quoted in ref. (5) was used. Radial estimates were determined either by the half-light radius of the observed system, or the quoted galaxy radius where available. \\
	$^{c}$\!\;References: (1) \citet{Oesch2016ApJ}, (2) \citet{Oesch2015ApJ}, (3) \citet{Grazian2012A&A}, (4) \citet{Ono2012ApJ}, (5) \citet{Inoue2016Sci}. \\
	$^{d}$\!\;Also referred to as {EGSY-0348800153} and {EGS 8053}.}
	\label{tab:high_z_galaxies_table_sb}
\end{table*}
The four examples of high-redshift systems with an ongoing starburst episode are listed in Table~\ref{tab:high_z_galaxies_table_sb} in order of their star-formation rates, with literature references indicated. Stellar ages, star-formation rates, the timescales over which each starburst has been ongoing as well as associated dynamical timescales are shown. These are found by spectral fitting methods~\citep[as outlined in e.g.][]{Robertson2010Natur, Stark2013ApJ, Oesch2014ApJ, Mawatari2016PASJ}.
The length of the starburst phase of these systems can be estimated as:
\begin{equation}
\label{eq:sb_timescale}
t_{\rm SB} \approx \frac{(M_{*}/{\rm M}_{\odot})}{(\mathcal{R}_{\rm SF}/{\rm M}_{\odot}~\text{yr}^{-1})}~\text{yr} \ ,
\end{equation}
where $M_{*}$ is the stellar mass of the system, and $\mathcal{R}_{\rm SF}$ is the star-formation rate. ${\rm M}_{\odot}$ is the mass of the Sun.

From Table~\ref{tab:high_z_galaxies_table_sb}, the estimated stellar population ages in the two most actively star-bursting systems ({GN-108036} and {SXDF-NB1006-2}) are substantially lower than their dynamical timescales. This means that there has not been sufficient time for the impacts of any feedback or quenching process to have taken hold, and so is consistent with their ongoing high star-formation rates.
By contrast, the minimum dynamical timescale for the other two star-forming systems ({GN-z11} and {EGS-zs8-1}) is shorter than the estimated timescale over which their observed starburst appears to have been ongoing, $\tau_{\rm dyn} < t_{\rm SB}$. Thus, in these cases it is possible that feedback effects may be starting to influence their ongoing star-forming activity, perhaps accounting for their comparatively lower star-formation rates. 
We note that ISM cooling is relatively fast, arising on timescales of around 
\begin{equation}
\tau_{\rm cool} \approx 0.1 \left(\frac{n_{\rm e}}{10\;\! \text{cm}^{-3}}\right)^{-1} \left(\frac{T_{\rm e}}{10^5\;\!\text{K}}\right)^{1/2}\;\text{Myr} \ .
\label{eq:cooling_timescale}
\end{equation}
for $n_{\rm e}$ as the electron (or ionised gas) number density and $T_{\rm e}$ as the electron (gas) temperature. This means that, for star-formation to remain quenched, the underlying heating process must continue to operate for hundreds of Myr to reconcile the observations.
It is clear that stellar radiation would be substantially reduced after the end of the starburst episode, and so is not a plausible candidate to maintain ongoing heating. 
While heating due to CRs would be more powerful~\citep[see][]{Owen2018MNRAS}, the CR escape timescale (via diffusion) is too short to allow direct CR heating to maintain prolonged quenching,
\begin{equation}
\tau_{\rm diff} \approx \frac{\ell^2}{4 D}
\end{equation}
where $\ell$ is the characteristic length-scale of the protogalaxy (around 1 kpc), and $D$ is the characteristic diffusion coefficient (around $3\times10^{28}~\text{cm}^2~\text{s}^{-1}$ -- see section~\ref{sec:cr_transport} and, for example ~\citealt{Berezinskii1990book, Aharonian2012SSR, Gaggero2012thesis}), which gives a CR escape timescale of just a few Myr. This would be even shorter in the presence of a fast galactic outflow~\citep{Owen2018sub}. Given that other usual candidates for driving ongoing quenching are not present in these starburst systems (e.g. AGN), a second feedback mechanism must be operating to account for the long-lasting effect. \cite{Owen2018sub} proposed a strangulation mechanism~\citep{Larson1980ApJ, Boselli2006PASP, Peng2015Natur}, where cold filamentary inflows of gas able to reignite star-formation are stifled and prevented from returning to the galaxy for hundreds of Myr, which would be consistent with the quenching timescales seen in some of the post-starburst systems considered here. 
In such a picture,
we may reconcile the gas cooling with the molecular gas reservoirs by noting that the reservoir density would be substantially lower than the inner ISM where star-formation would be concentrated. This would allow for much longer cooling timescales and prolonged pressure support of the reservoir, but requires feedback activity operating on two scales: 
locally, with direct heating able to quench star-formation in the ISM, 
and more broadly with some larger scale heating or ionisation process required to strangulate star-formation for a substantial period of time. 


\section{Cosmic rays in protogalaxies}
\label{sec:protogalaxies}

As actively star-forming environments, high redshift starburst protogalaxies are expected to be rich in CRs. 
The high star formation rates of such galaxies~\citep[see e.g.][]{DiFazio1979A&A, Solomon2005ARAA, Pudritz2012book, Ouchi2013ApJ, Knudsen2016MNRAS} yield frequent SN events which produce the environments such as SN remnants known to be CR accelerators~\citep[see e.g.][]{Dar2008PhR, Kotera2011ARAA}, and these can efficiently accelerate CRs up to PeV energies~\citep{Bell1978MNRAS, Hillas1984ARAA, Kotera2011ARAA, Schure2013MNRAS, Bell2013MNRAS} via, for example diffusive shock acceleration processes such as Fermi acceleration~\citep{Fermi1949PhRv, Berezhko1999ApJ, Allard2007APh}. We specify an appropriate parametric model for a protogalaxy to assess the impacts of these CRs. The model is comprised of a radiation and density field with which the CRs interact, together with a magnetic field through which the CRs propagate~\citep[see also][which adopts a similar parametrisation]{Owen2018MNRAS}. 
  
  \subsection{Modelling protogalactic environments}
\label{sec:protogalaxy_environment}

\subsubsection{Radiation field}
\label{sec:radiation_field}

A protogalactic radiation field is comprised of photons from stellar radiation as well as the cosmological microwave background (CMB).
The CMB is modelled as a spatially uniform black body of a characteristic temperature determined by the redshift, for which the photon number density is given by
\begin{equation}
n_{\rm ph}(z) = \frac{8\pi}{\lambda_{\rm C}^3} \Theta^3(z) \Gamma(3) \zeta(3) \ ,
\end{equation}
  where $\lambda_{\rm C}$ is the Compton wavelength of an electron (${\rm h}/{m_{\rm e}} {\rm c}$), $\Gamma(...)$ is the gamma function and $\zeta(...)$ is the Riemann zeta function. Further, $\Theta(z) = {{\rm k_{\rm B}} T(z)}/{m_{\rm e} \rm{c}^2}$, 
  and $T(z) = T_0 (1+z)$ is the CMB temperature
  with $T_0=2.73~{\rm K}$ as its present value \citep{Planck2015A&A}.
  At a typical redshift of $z=7$, the energy density in CMB photons is around $U_{\rm CMB} \approx 1,100 ~\text{eV cm}^{-3}$.
  The stellar radiation field may be modelled as the distributed emission from an extended ensemble of sources, weighted according to the protogalactic density profile. In a high-redshift rapidly star-forming galaxy, we may assume a stellar population dominated by low-metallicity, massive Type O and B stars with short lifetimes. A system of $N = 10^6$ such stars can be shown to give a total protogalactic luminosity $N\;\!L_{*} \approx 2.8\times 10^{44}~\text{erg s}^{-1}$, and this would correspond to a galaxy with $\mathcal{R}_{\rm SN} = 0.1~\text{yr}^{-1}$, or $\mathcal{R}_{\rm SF} \approx 16~\text{M}_{\odot}~\text{yr}^{-1}$~\citep[e.g.][]{Owen2018MNRAS}.
  The stellar lifetime may be considered as independent of $\mathcal{R}_{\rm SN}$, so it would follow that this stellar luminosity should scale with $\mathcal{R}_{\rm SN}$ as stars would build-up in proportion to their birth rate. 
  The energy density of the stellar radiation field then follows as $U_{*} \approx 150 \;\! (\mathcal{R}_{\rm SF}/16~\text{M}_{\odot}~\text{yr}^{-1}) \;\! (r_{\rm gal}/1~\text{kpc})^{-2} ~\text{eV cm}^{-3}$ where $r_{\rm gal}$ is the characteristic radius of the galaxy, meaning that the stellar radiation is dominated by the CMB until the star-formation rate is well in excess of $\mathcal{R}_{\rm SF} \approx 120~\text{M}_{\odot}~\text{yr}^{-1}$ for a 1 kpc sized galaxy.
  
\subsubsection{Density and magnetic field}
\label{sec:density_magnetic_field}
  
We model the density field as an over-density on a uniform background intra-cluster medium (ICM) using the profile 
\begin{equation}
   \label{eq:density_profile}
   n_{\rm b}(r) = \frac{n_{\rm b, 0}}{\left(1+\xi_{\rm c}\right)\left(1+\xi_{\rm h}\right)}  + n_{\rm p}
\end{equation}
  \citep{Tremaine1994ApJ, Dehnen1993MNRAS}, 
  with $\xi_{\rm c} = (r/r_{\rm c})^2$ and $\xi_{\rm h} = (r/r_{\rm h})^2$ as the dimensionless form of the parameters $r_{\rm c}=1\;\! \rm{kpc}$ 
  and $r_{\rm h}= 2\;\! \rm{kpc}$ as the protogalaxy core and halo radius respectively. We adopt values of $n_{\rm b, 0} = 10$ cm$^{-3}$ for the central value of the interstellar density profile and $n_{\rm p} = 10^{-3}$ cm$^{-3}$ for the background density. We find the exact choice of these parameters, if reasonable, bears little influence on the conclusions drawn from the results presented in this study.
  
Observational studies favour the rapid development of magnetic fields in protogalaxies,
 reaching strengths comparable to the Milky Way within a few Myr of their formation~\citep{Bernet2008Nat, Beck2012MNRAS, Hammond2012arXiv, Rieder2016MNRAS, 2018MNRASSur}. 
 The development of a magnetic field in young galaxies is often attributed to a turbulent dynamo mechanism driven by the frequent SN explosions during a starburst phase~\citep[see e.g.][]{Rees1987QJRAS, Balsara2004ApJ, Beck2012MNRAS, Latif2013MNRAS, Schober2013A&A}, which allows for the emergence of an ordered $\mu$G magnetic field in the required timescales. 
 The saturation level of such a mechanism, for example as that introduced in the model of ~\citet{Schober2013A&A}, may be approximated by invoking equipartition with the kinetic energy of the turbulent gas, $B_{\rm L,sat} = \mathcal{S}(r) f$ where 
\begin{equation}
\label{eq:sfunc}
\mathcal{S}(r) = \left[{4 \pi \; \rho}\right]^{1/2} v_{\rm f}
\end{equation}
   with $\rho$ as the local density and $v_{\rm f}$ as the fluctuation velocity ($v_{\rm f}  \approx R_{\rm gal} ({2}\pi \rho G/3)^{1/2}$ for the protogalaxy, if adopting a pressure with gravity equilibrium approximation -- see~\citealt{Schober2013A&A}). $f$ represents a deviation from exact equipartition to account for the efficiency of energy transfer from the turbulent kinetic energy to magnetic energy, and simulation work estimates this to be around 10\%~\citep[see, e.g.][]{Federrath2011PRL, Schober2013A&A}.

\subsection{Cosmic ray injection}
\label{sec:cr_injection}

The power of injected primary CR protons is directly related to the total power of the SNe injecting them. Hence, following~\cite{Owen2018MNRAS}, a function describing the injection of hadronic CRs at a position $r$ may be described as a product of two separable components,
\begin{equation}
\label{eq:source_term}
   \mathcal{I}_{\rm p}(E, r) =\mathcal{L}_{0} \left(\frac{E}{E_0}\right)^{-\Gamma}\;\! S_{\rm N}(r) \ ,   
\end{equation} 
  where the volumetric CR injection rate is $S_{\rm N}(r)$, and the power-law energy spectrum component, typical of that arising from stochastic acceleration processes, has the normalisation
  \begin{equation}
\label{eq:cr_norm}
   \mathcal{L}_0 = \frac{L_{\rm CR, eff}(1-\Gamma)E_0^{-\Gamma}}{E_{\rm max}^{1-\Gamma} - E_0^{1-\Gamma}}  \ , 
\end{equation}
where we adopt a power-law index for the protons of $\Gamma = 2.1$, similar to regions where CRs are expected to be freshly injected~\citep[see, e.g.][]{Allard2007APh, Kotera2010JCAP, Kotera2011ARAA, Owen2018sub}.
We take a reference energy $E_0$ of 1 GeV and a maximum energy of interest of 1 PeV. $L_{\rm CR, eff}$ is the total power in the CR protons, which can be expressed in terms of SN event rate ${\cal R}_{\rm SN}$ :
\begin{equation} 
\label{eq:cr_scale}
   L_{\rm CR, eff} = \varepsilon \xi E_{\rm SN} {\cal R}_{\rm SN} 
     =   \alpha \left[\frac{\varepsilon \xi E_{\rm SN} \mathcal{R}_{\rm SF}}{M_{\rm SN}}  \right]   \ .  
\end{equation}
We use the parameters
     $E_{\rm SN}$ as the total energy generated per SN event (around $10^{53}$ erg for core-collapse Type II P SNe), 
     $\varepsilon$ as the fraction of SN power converted into to CR power (0.1 is chosen as a conservative value\footnote{See~\citet{Fields2001A&A, Strong2010ApJ, Lemoine2012A&A, Caprioli2012JCAP, Dermer2013A&A, Morlino2012A&A, Wang2018MNRAS} for studies regarding this parameter and a range of possible, reasonable values.}) 
     and $\xi$ as the fraction of SN energy available after accounting for losses to neutrinos (we use a value of 1\% for the energy retained\footnote{   
This value depends on the SN type as well as the way it interacts with its specific environment 
\citep[see e.g.][]{Iwamoto2006AIPC}. 
  In the case of core-collapse SNe, most of the energy is carried away by neutrinos with perhaps a fraction as low as 0.001, but more likely around 0.01, being retained by the system to, for example, accelerate CRs and drive shocks
    \citep[see reviews, e.g.][]{Smartt2009ARAA, Janka2012ARNPS}.}),
     $\alpha\approx 0.05$ as the fraction of stars able to yield a core-collapse SN event, which is set by the nature of the initial mass function (IMF) of the stars developing in a galaxy
  and the upper cut-off mass $M_{\rm SN}  = 50\;\!{\rm M}_{\odot}$~\citep[see, e.g.][]{Fryer1999ApJ, Heger2003ApJ} of a star to ultimately produce a SN event (use of the upper mass limit here ensures a conservative estimate for the CR luminosity). 
    
\subsection{Cosmic ray interactions in protogalaxies}
\label{sec:cr_interactions_protogalaxy}
 
  \begin{figure*}
 \centering
	\includegraphics[width=0.85\textwidth]{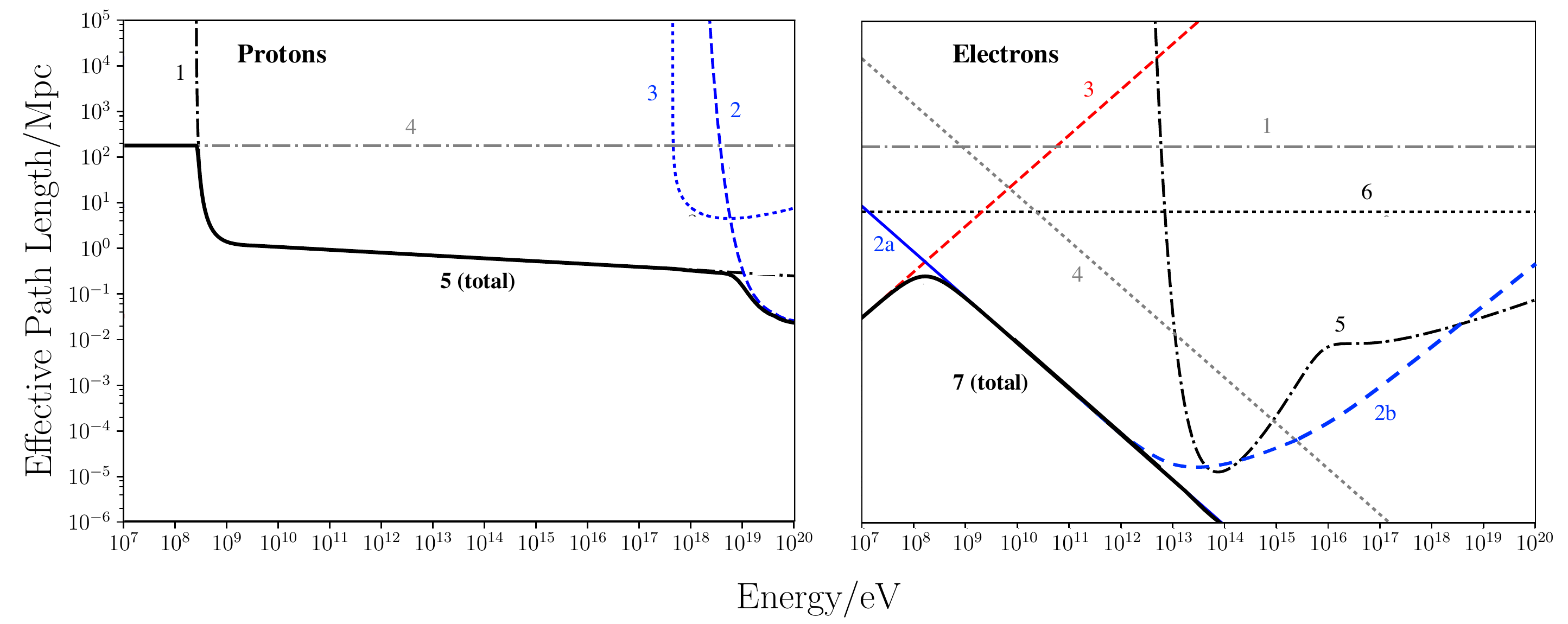}
    \caption{\small {\textbf{Left:}} CR proton losses in terms of effective interaction path lengths. This results from a protogalaxy model with target density of $n_{\rm p} = 10~\rm{cm}^{-3}$, a CMB radiation field specified at a redshift of $z=7$. The thick black line indicates the total effect that would be experienced by a CR proton at the indicated energy accounting for all energy loss contributions. The lines, as labelled, are (1) ${\rm pp}$ Pion-production (${\rm pp},{\rm p}\pi^{\pm}\pi^{0}$); (2) CMB Photo-pion (${\rm p}\gamma,\pi^{\pm}\pi^{0}$); (3) CMB Photo-pair (${\rm p}\gamma, e^{\pm}$); (4) Adiabatic losses due to cosmological expansion; (5) Total. {\textbf{Right:}} Secondary CR electron loss lengths with the same model. The lines, as labelled, are (1) Adiabatic losses due to cosmological expansion; (2) inverse-Compton (radiative) cooling due to the CMB with (2a) as the estimated path length in the Thomson limit, and (2b) being the estimated path length if accounting for the transition to Klein-Nishina behaviour at higher energies (we find this does not significantly impact our later results, so the Thomson limit is used in our main calculations); (3) Coulomb losses (responsible for directly heating the ISM plasma); (4) Synchrotron (radiative) cooling in a $5~\mu$G magnetic field, being an appropriate strength for this model; (5) Triplet Pair-Production ($\text{e}\gamma,\text{e}^{\pm}$); (6) Free-free (Bremsstrahlung) losses; (7) Total.}
    \label{fig:total_losses}
\end{figure*}

Above a threshold energy of around a GeV~\citep[e.g.][]{Kafexhiu2014PRD}, HE CR
particles can undergo hadronic interactions with the interstellar gases and also with the radiation fields permeating the galaxy~\citep[see][]{Dermer2009book, Owen2018MNRAS}. 
Hadronic interactions allow high energy CRs to deposit energy much more rapidly than their lower-energy counterparts and this enables them to drive a heating effect, although the eventual
thermalisation of this energy is governed by the secondary particles of these interactions and the processes they undergo. These secondaries are comprised of further hadrons, charged leptons, photons and neutrinos~\citep{Pollack1963PhRv, Gould1965AnAp, Stecker1968ApJ, Berezinsky1988A&A, Mucke1999PASA}. 
Neutrinos and high energy photons interact only minimally with their environment and so these effectively free-stream away from the vicinity of the interaction, taking their energy with them. The leptons injected from the pion decays typically each inherit around 3-5\% of the energy of the original CR primary and can thermalise into their medium to drive a heating effect, or cool through other mechanisms~\citep[e.g.][]{Blumenthal1970PRD, Rybicki1979book, Dermer2009book, Schleicher2013A&A} -- notably, radiative emission:
at high-redshift, the increased number density of CMB photons allow CR secondary electrons to undergo inverse-Compton scattering~\citep{Schober2015MNRAS}. This may also arise in intensely star-forming environments where the energy density in the stellar radiation field is sufficiently high. This process up-scatters relatively low-energy CMB photons to keV energies, creating an X-ray glow~\citep{Lacki2013MNRAS, Schober2015MNRAS}. This X-ray emission can then heat the environment by scattering off ambient free electrons.

The injection of secondary CR electrons is discussed in more detail in Appendix~\ref{sec:interaction_mechanisms} (specifically sections ~\ref{sec:cosmic_ray_proton_interactions} and~\ref{sec:cosmic_ray_electron_interactions}) where these interactions have been presented in general terms. Of interest to this study is the role these processes have on the protogalactic environment specified in sections~\ref{sec:radiation_field} and~\ref{sec:density_magnetic_field}. 
In the following, we consider the propagation path lengths in the centre of the protogalaxy model where CR losses would be most severe (as the density and magnetic fields are the strongest) to assess the relative importance of each process in cooling and absorbing the CR primary protons and secondary electrons.
We define the energy loss path length of freely-streaming particles undergoing interactions (either cooling or absorption) as
\begin{equation}
r_{\rm int} \equiv {\rm c} \; \beta \; \tau_{\rm int} \approx \frac{\gamma_{\rm e}}{b(\gamma_{\rm e})} \; {\rm c} \ ,
\label{eq:free_streaming_path_length}
\end{equation}
where $\beta$ is the velocity of the particle normalised to the speed of light (at high energies $\beta \approx 1$) and $\tau_{\rm int} \approx [ \dot{N}_{\rm int} ]^{-1}$ is the timescale of an interaction (the inverse of the rate at which interaction events occur). $\gamma_{\rm e}$ is the electron Lorentz factor (proportional to its energy) and $b(\gamma_{\rm e})$ is the electron cooling rate at the energy $E_{\rm e} = \gamma_{\rm e} m_{\rm e} {\rm c}^2$. This can be used to determine the relative importance of different processes in a given environment. The interactions which are experienced most strongly by a particle occur more rapidly and therefore have a shorter associated path length. While a beam of particles will experience all relevant absorption or cooling mechanisms, their relative path lengths give the weighted contribution of each process -- in many cases, one will dominate.

The effective average path lengths experienced by CR protons and electrons due to each of the cooling and absorption processes discussed in Appendix~\ref{sec:cosmic_ray_proton_interactions} and~\ref{sec:cosmic_ray_electron_interactions} are shown in Fig.~\ref{fig:total_losses}, with proton losses plotted in the left panel and electron losses in the right panel. These are calculated for an infinite uniform environment of conditions equivalent to those found at the centre of the protogalaxy model, at a redshift of $z=7$ and with an ambient magnetic field strength of $5~\mu$G which itself is not assumed to perturb or influence the propagation of the CR particles (an idealistic scenario such that the synchrotron cooling rate can be reasonably compared to other processes). This shows that, for the CR primary protons, losses are mainly dominated by the ${\rm pp}$ interaction above its threshold energy. Only at energies above $10^{19}$ eV, CMB photo-pion (dashed blue line) losses begin to become important, however the CR flux associated with such high energies would typically be negligible. At the lowest energies, below the ${\rm pp}$ interaction threshold, the free-streaming CRs are essentially only cooled by the adiabatic expansion of the Universe (if freely streaming). This indicates that the vast majority of secondary electrons are injected into the protogalactic medium by the ${\rm pp}$ pion-production interaction, and so the absorption parameter $\alpha*$ used in equation~\ref{eq:alpha_star} is dominated by the $n_{\rm b}\hat{\sigma}_{\rm p\pi}$ term. For the secondary CR electrons, losses are dominated by Coulomb losses and inverse-Compton losses with the CMB (additionally stellar photons may contribute to inverse-Compton losses in extremely active star-forming systems), with other mechanisms having a comparatively negligible effect. 


  \begin{figure*}
  \centering
	\includegraphics[width=0.85\textwidth]{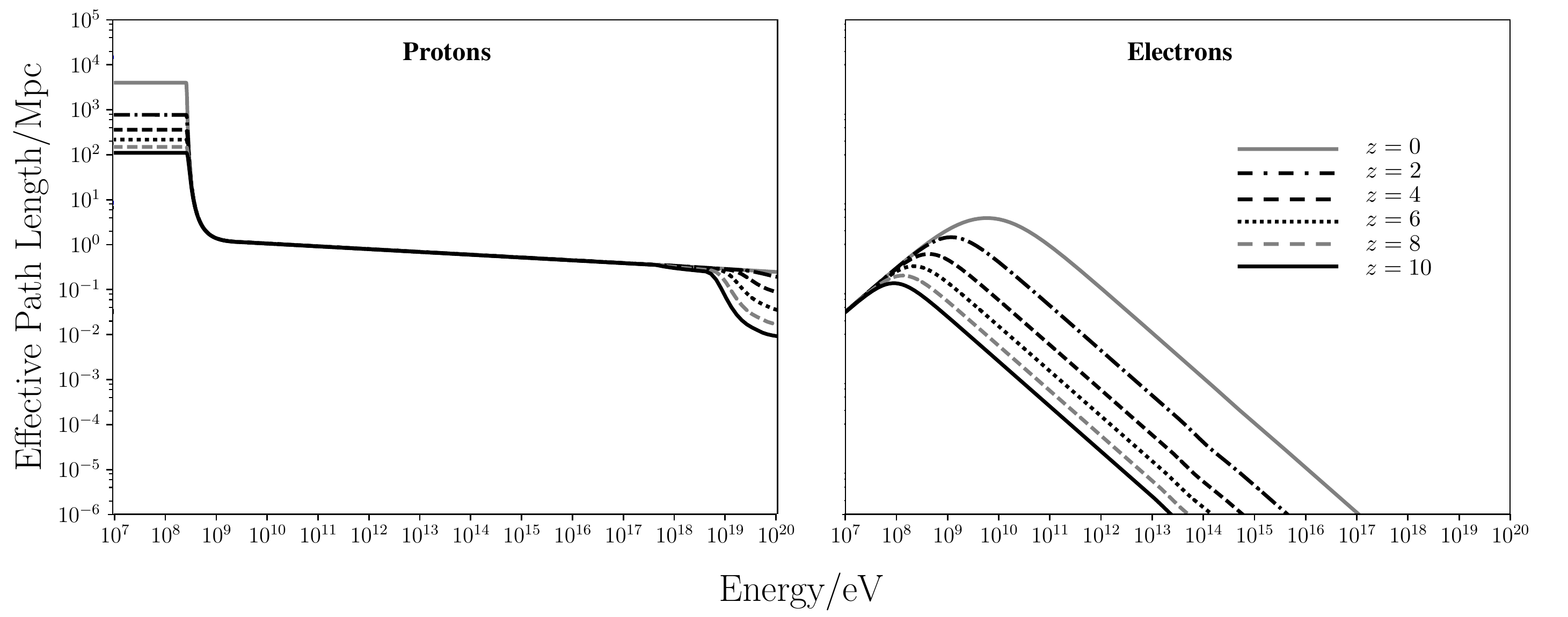}
    \caption{\small Interaction loss lengths accounting for all relevant processes (i.e. the solid line of Fig.~\ref{fig:total_losses}), but calculated at redshifts of $z=0$ to $10$ as shown by the legend. {\textbf{Left:}} Path lengths for protons;  {\textbf{Right:}} Path lengths for electrons.}
    \label{fig:redshift_losses}
\end{figure*}
The cooling and absorption mechanisms are also redshift dependent. Fig.~\ref{fig:redshift_losses} shows the total CR losses in the protogalactic environment at redshifts of $z=0, 2, 4, 6, 8$ and $10$, indicated by the solid grey, dot-dashed black, dashed black, dotted black, dashed grey and solid black lines respectively. The protons (left panel) are largely unaffected by the redshift evolution. This is because their losses are dominated by the ${\rm pp}$-pion production channel which is governed by the local density field of the galaxy (this does not evolve with redshift) and is therefore effectively decoupled from cosmological expansion. The electrons (right panel) are more affected by the expanding Universe in that their losses become much more influenced by inverse-Compton cooling at high redshift. This dominates over Coulomb cooling at lower energies as redshift increases, with the turn-over (the peak in the cooling curve) occurring above 10 GeV at $z=0$, falling to 0.1 GeV by $z=10$. This is consistent with the increasing photon number density of the CMB at higher redshifts.


\section{Propagation of cosmic rays}
\label{sec:cr_transport}

The path lengths calculated according to equation~\ref{eq:free_streaming_path_length} assume that the CR particles freely stream from their point of emission, propagating in a straight line. 
However, in general the propagation of charged CRs in ambient galactic (and protogalactic) magnetic fields is more complicated. A high energy (relativistic) particle of charge $q$ in a uniform magnetic field describes a curved path, of size characterised by the gyro-radius (or Larmor radius):
\begin{equation}%
r_{\rm L} = \frac{3.3 \times 10^{12}}{|q|} \left(\frac{E}{10^{9}\;\!{\rm eV}}\right) \left( \frac{{\rm \mu G}}{B}\right) ~\rm{cm} \ .%
\end{equation}%
\newline
This can be used to set the scale of a phenomenological description of CR propagation in terms of their diffusion through a turbulent interstellar magnetic field. For this, we may invoke the transport equation:
\begin{align}
\label{eq:transport_equation}
\frac{\partial}{\partial t} \; n(E, r) &= \nabla \cdot \left[D(E, r)\nabla n(E, r) \right] \nonumber\\
&+ \frac{\partial}{\partial E} \left[b(E, r) n(E, r)\right] + Q(E, r) - S(E, r) \ ,
\end{align}
where processes such as advection of CRs in bulk flows, winds and galactic outflows have been neglected.\footnote{Previous studies~\citep[see, e.g.][]{Owen2018sub} indicate that advection of CRs in outflows does not substantially reduce their abundance within a protogalaxy ISM (by only around 10\%). Internal interstellar winds and flows will have a more local impact, in that they push CRs from one part of the galaxy to another -- but this arises on a sub-galactic scale, with the galactic-scale picture being unaffected.}
Here, $Q(E, r)$ is the particle source term, $S(E, r)$ is the absorption term, $b(E, r)$ is the particle cooling rate (being the sum of all relevant cooling processes) and $D(E, r)$ is introduced as the diffusion coefficient which takes a parametric form of
\begin{equation}
D(E, r) = D_0 \left[ \frac{r_L(E,\langle |B| \rangle)|_{r}}{r_{L,0}}\right]^{\delta} \ ,
\end{equation}
where $\langle |B| \rangle\vert_{r} = |B(r)|$ is the characteristic mean strength of the magnetic field at some position $r$, and $D$ is normalised to $D_0 = 3.0\times 10^{28}$ cm$^2$ s$^{-1}$, a value comparable to empirical measurements of the Milky Way ISM diffusion coefficient~\cite[see, e.g.][]{Berezinskii1990book, Aharonian2012SSR, Gaggero2012thesis}\footnote{While the ISM of high redshift protogalaxies cannot be directly observed, we argue that it is not unreasonable to expect that processes driving turbulence within them are not greatly different to those seen in the local Universe. As such, we argue that the diffusion coefficient would likely be similar to that in the Milky Way and, in the absence of more detailed studies or observations, adopting an alternative prescription would not imply more correct physics.} for a 1 GeV CR proton in a 5$\mu$G magnetic field (with corresponding Larmor radius $r_{L,0}$).  $\delta$ is introduced to encode the interstellar magnetic turbulence. We adopt a value of 1/2 here~\citep[e.g.][]{Berezinskii1990book, Strong2007ARNPS}. This is appropriate for a Kraichnan-type turbulence spectrum following a power law of the form $P(k) \; {\rm d} k \approx k^{-2+\delta}$, which is thought to be a reasonable description for the turbulence in an ISM~\citep[see][]{Yan2004ApJ, Strong2007ARNPS}. For incompressible Kolmogorov turbulence $\delta = 1/3$, but previous studies suggest that this would not allow for the required scattering and diffusion of CRs to be consistent with Milky Way observations~\citep[see][]{Goldreich1995ApJ, Lazarian2000ApJ, Stanimirovic2001ApJ, Brandenburg2013SSRv}.
We solve the transport equation to model steady-state distributions of CR protons and electrons in a protogalaxy.

\subsection{Proton transport}
\label{sec:proton_transport}

Fig.~\ref{fig:total_losses} shows that CR protons are predominantly absorbed rather than cooled. We may therefore simplify the transport equation~\ref{eq:transport_equation} to
\begin{align}
\label{eq:proton_transport_equation}
\frac{\partial n_{\rm p}}{\partial t} = \nabla \cdot \left[D(E_{\rm p}, r)\nabla n_{\rm p} \right] + Q_{\rm p}(E_{\rm p}, r) - S_{\rm p}(E_{\rm p}, r) \ ,
\end{align}
where we have used the contraction $n_{\rm p} = n_{\rm p}(E, r)$. This is the (differential) number density of protons per energy interval between $E_{\rm p}$ and $E_{\rm p} + {\rm d} E_{\rm p}$, 
\begin{equation}
n_{\rm p}(E_{\rm p}, r) = \frac{{\rm d} N_{\rm p}}{{\rm d} V \; {\rm d} E_{\rm p}} \biggr\vert_{E_{\rm p}, r}
\end{equation}
for $N_{\rm p}$ as the number of protons and ${\rm d} V$ as a differential volume element. 
The source term $Q_{\rm p}$ is the local injection rate of protons per unit volume by SN events: equation~\ref{eq:source_term} gives the energy injection rate of CR protons at a specific energy $E_{\rm p}$; since we require the (differential) injected rate of CRs 
per energy interval between $E_{\rm p}$ and $E_{\rm p} + {\rm d} E_{\rm p}$, we use
\begin{equation}
Q_{\rm p}(E_{\rm p}, r) = \frac{\partial \mathcal{I}_{\rm p}}{\partial E_{\rm p}} \biggr\vert_{E_{\rm p}, r} \ .
\end{equation}
The sink term is dominated by ${\rm pp}$ pion production over the energies of interest, with the CR absorption rate given as $ {\rm c} ~n_{\rm p}(E_{\rm p}, r) \; \hat{\sigma}_{\rm p\pi}(E_{\rm p})~n_{\rm b}(r) $; see equation~\ref{eq:cr_absorption}. 
Given that our model protogalaxy is axisymmetric, equation~\ref{eq:proton_transport_equation} may be written in spherical coordinates as:
\begin{align}
\label{eq:proton_transport_equation_sph}
\frac{\partial n_{\rm p}}{\partial t} = \frac{1}{r^2} \frac{\partial}{\partial r} \left[r^2 \; D(E_{\rm p}, r) \frac{\partial n_{\rm p}}{\partial r} \right] + \mathcal{I}_{\rm p}(E_{\rm p}, r) - {\rm c} ~n_{\rm p} \; \hat{\sigma}_{\rm p\pi} \; n_{\rm b} \ .
\end{align}


It is useful to rewrite equation~\ref{eq:proton_transport_equation_sph} as an initial value problem for a single proton injection event from a source region of size $\mathcal{V}_{\rm S}$:
\begin{align}
\label{eq:proton_refactor_1}
\frac{\partial n_{\rm p}}{\partial t} = \frac{1}{r'^2} \frac{\partial}{\partial r'} \left[r'^2 \; D(E_{\rm p}, r') \frac{\partial n_{\rm p}}{\partial r'} \right] - {\rm c} ~n_{\rm p} \; \hat{\sigma}_{\rm p\pi} \; n_{\rm b} \ .
\end{align}
where $r'$ is introduced as the radial distance between the injection point ${\textbf r}_i$ and some general location ${\textbf r}$, given by $|{\textbf r}-{\textbf r}_i|$.
This inhomogeneous partial differential equation (PDE) can be converted to a homogenous form by use of the substitution,
\begin{equation}
\label{eq:sub_expr}
\xi = n_{\rm p} \exp \left\{ \int_0^t  {\rm c} \; {\rm d}t'  \; \hat{\sigma}_{\rm p\pi} \; n_{\rm b} \right\} \ ,
\end{equation}
where we invoke the approximation that the ISM density, $n_{\rm b}$ is locally uniform. While this is not strictly true, it is a close enough approximation for our purposes: we will model the CR propagation only in the inner regions of the protogalaxy ISM where radial density variations are small. Beyond this, matters such as how the protogalactic magnetic field connects to that of the circumgalactic and intergalactic medium have an important effect on CR propagation, and must be considered. These are complicated matters worthy of a dedicated investigation, and robust modelling of this interfacing region lies outside of the scope of the present work.
Here, the element ${\rm c}\;{\rm d}t = {\rm d}s'$ represents a displacement~\textit{along the path actually traversed by a CR proton}. 
This is the quantity that determines the amount of attenuating material through which a proton traverses.
 The propagation of CRs is macroscopically modelled diffusively, but the true speed of these relativistic particles remains close to ${\rm c}$. 
 Their frequent scatterings in the magnetic field causes them to exhibit a random walk such that their movement away from their injection point is substantially less than ${\rm c}$ when considered in terms of their large-scale displacement. 
 Thus, we have introduced the coordinate $s'$ as the proton path coordinate (i.e. along the path followed by a CR as it scatters). 
 The relation between $s'$ and the spatial dimension $r'$ is thus ${\rm d} s' = \psi(r') \; {\rm d}r'$, where $\psi(r')$ quantifies the magnetic scattering (calculated locally at $r'$). 
 Following~\cite{Owen2018MNRAS}, $\psi(r')$ can be determined from the ratio of the free-streaming (i.e. the free attenuation length due to CR interactions $\ell_{\rm p\pi}$) and diffusive path lengths of the CRs,
\begin{equation}
\psi(E_{\rm p},  r') = \frac{\ell_{\rm p\pi}}{\ell_{\rm diff}} = \sqrt{\frac{{\rm c}}{4 \; D \; n_{\rm b} \; \hat{\sigma}_{\rm p\pi}}} \ ,
\end{equation}
where we have used the short-hand notation $D = D(E_{\rm p}, r')$, and $\hat{\sigma}_{\rm p\pi} = \hat{\sigma}_{\rm p\pi}(E_{\rm p})$.
It then follows that  ${\rm c} \; {\rm d}t'  \; \hat{\sigma}_{\rm p\pi} \; n_{\rm b} = \psi(E_{\rm p}, r') \; {\rm d}r' \; \alpha^{*}(E_{\rm p}, r')$, where we have substituted for the absorption coefficient (in the case of ${\rm pp}$ pion production dominating the CR absorption), as defined in eq.~\ref{eq:alpha_star}. 
From this, we may define the CR attenuation factor in terms of~\ref{eq:sub_expr} as:
\begin{equation}
\mathcal{A}(E_{\rm p}, {\textbf r}; {\textbf r}_i) = \exp \left\{ - \int_{{\textbf r}_i}^{{\textbf r}} \psi(E_{\rm p}, {\textbf r}') \; \alpha^*(E_{\rm p}, {\textbf r}') \; {\rm d}{\textbf r}' \right\} \ ,
\end{equation}
which quantifies the level of attenuation experienced by a beam of CR protons between a source at location ${\textbf r}_{i}$, and some general location ${\textbf r}$. 

The substitution~\ref{eq:sub_expr} now allows equation~\ref{eq:proton_refactor_1} to be written as the homogeneous PDE:
\begin{align}
\label{eq:proton_homogenous}
\frac{\partial \xi}{\partial t} = \frac{1}{r'^2} \frac{\partial}{\partial r'} \left[r'^2 \; D(E_{\rm p}, r') \frac{\partial \xi}{\partial r'} \right] \ ,
\end{align}
where an injection episode at ${\textbf r} = {\textbf r}_i$ is taken as an initial condition and provides a normalisation constant $\xi_0$. The solution is well known and can be found by method of Green's functions to give:
\begin{equation}
\xi = \frac{\xi_0}{\left[4 \pi D(E,r')\;\! t' \right]^{m/2}}\exp{\left\{-\frac{r'^2}{4 D(E,r')\;\! t'}\right\}} \ ,
\end{equation}
\citep{Owen2018MNRAS} where $t'$ is the time elapsed since the injection took place. $m$ is introduced as a geometrical parameter which, in the case of a three-dimensional spherical system takes the value $m=3$. The normalisation $\xi_0$ depends on the strength of the injection (being the product of the volumetric injection rate $Q_{\rm p}(E_{\rm p})$ and a characteristic assigned source size, $\mathcal{V}_{\rm S}$) and the time $\Delta t$ over which it was active, $\xi_0 = Q_{\rm p}(E_{\rm p}, {\textbf r}_i) \;\! \mathcal{V}_{\rm S}\;\! \Delta t$. 
Transposing this back to a solution of $n_{\rm p}$ yields:
\begin{align}
\label{eq:final_diffusion_protons}
n_{\rm p} =& \; \frac{Q_{\rm p}(E_{\rm p}, {\textbf r}_i) \;\! \mathcal{V}_{\rm S} \;\! \Delta t \; \mathcal{A}(E_{\rm p}, {\textbf r}; {\textbf r}_i)}{\left[4 \pi D(E_{\rm p},r')\;\! t' \right]^{3/2}} \;\exp{\left\{-\frac{r'^2}{4 D(E_{\rm p},r')\;\! t'}\right\}} \ ,
\end{align}
which practically separates the diffusive evolution from the attenuation term, 
thus allowing the attenuation to be calculated separately from the diffusion of CRs, and combined afterwards.
 The solution for CRs injected in multiple bursts from a single source follows 
 as the convolution of individual, instantaneous injections. 
 However, for analytical tractability, we may consider the discrete injection episodes from each source as a continuous process such that
the convolution becomes a time-integral. The solution follows as:
\begin{align}
\label{eq:np_equation}
n_{\rm p} =& \; \frac{Q_{\rm p}(E_{\rm p}, {\textbf r}_i) \;\! \mathcal{V}_{\rm S} \; \mathcal{A}(E_{\rm p}, {\textbf r}; {\textbf r}_i)}{4 \pi^{3/2} \; r' D(E_{\rm p}, r')} \; \Gamma\left(\frac{1}{2}, x \right) \ ,
\end{align}
for each source at location ${\textbf r}_i$, 
where
\begin{equation}
\label{eq:x_estimate}
x = \frac{r'^2}{4\; D(E_{\rm p}, r') \; t} \ .
\end{equation}
The result evidently tends rapidly towards the steady-state solution which, as $t\rightarrow\infty$, becomes:
\begin{equation}
\label{eq:proton_solution}
n_{\rm p} = \; \frac{Q_{\rm p}(E_{\rm p}, {\textbf r}_i) \;\! \mathcal{V}_{\rm S} \; \mathcal{A}(E_{\rm p}, {\textbf r}; {\textbf r}_i)}{4\pi r' D(E_{\rm p}, r')} \ ,
\end{equation}
where the upper incomplete Gamma function has been evaluated as
\begin{equation}
\Gamma\left(\frac{1}{2}, x\right) = \sqrt{\pi} \left[1-\text{erf}(\sqrt{x})\right]
\end{equation}
and $x \rightarrow 0$ as $t\rightarrow\infty$, such that $\Gamma(\frac{1}{2}, x)\rightarrow {\pi}^{1/2}$.
CR protons rapidly settle into this steady state distribution, ${\rm d}n_{\rm p}/{\rm d}t = 0$ (compared to galactic timescales), and so remain in effective equilibrium while CR injection is ongoing. The full galactic steady-state solution can be found by convolving this continuous single-point solution over a weighted distribution of injecting sources. This lends itself naturally the Monte-Carlo (MC) numerical scheme outlined in section~\ref{sec:computational_scheme}.
  
 \subsection{Electron transport}
\label{sec:electron_transport}

Unlike CR protons, HE electrons predominantly interact with their environment by continuous energy transfer, which may be regarded as cooling. As a result, equation~\ref{eq:transport_equation} is instead reduced to:
\begin{align}
\label{eq:electron_transport_equation}
\frac{\partial n_{\rm e}}{\partial t} = \nabla \cdot \left[D(E_{\rm e}, r)\nabla n_{\rm e} \right] - \frac{\partial}{\partial E_{\rm e}} \left[b \; n_{\rm e}\right] + Q_{\rm e}(E_{\rm e}, r) \ ,
\end{align}
where the contraction $n_{\rm e} = n_{\rm e}(E, r)$ has been used for the CR electron number density per energy interval between $E_{\rm e}$ and $E_{\rm e} + {\rm d}E_{\rm e}$,
\begin{equation}
\label{eq:electron_spec_def}
n_{\rm e}(E_{\rm e}, r) = \frac{{\rm d} N_{\rm e}}{{\rm d} V \; {\rm d} E_{\rm e}} \biggr\vert_{E_{\rm e}, r} \ ,
\end{equation}
and $b = b(E_{\rm e}, r) = |{\rm d}E_{\rm e}/{\rm d}t|$ is the magnitude of the sum of the relevant cooling processes experienced by the CR electrons. From Fig.~\ref{fig:total_losses}, it can be seen this is dominated by inverse-Compton and Coulomb processes. The electron source term $Q_{\rm e}(E_{\rm e}, r)$ is the secondary CR electron injection rate per unit volume per energy interval due to CR primary interactions, as determined by equation~\ref{eq:cre_injection_spec1}. Like the CR protons, invoking an axisymmetric system allows equation~\ref{eq:electron_transport_equation} to be written in spherical coordinates as:
\begin{align}
\label{eq:electron_transport_equation_sph}
\frac{\partial n_{\rm e}}{\partial t} = \frac{1}{r^2} \frac{\partial}{\partial r} \left[r^2 D(E_{\rm e}, r) \frac{\partial n_{\rm e}}{\partial r} \right] - \frac{\partial}{\partial E_{\rm e}} \left[b \; n_{\rm e}\right] + Q_{\rm e}(E_{\rm e}, r) \ .
\end{align}


 \begin{SCfigure*}
    \caption{\small Electron spectra following their injection from a single discrete source (normalised to the injection level at the minimum energy indicated by the dashed red line, $n_{\rm e, 0}$), accounting for variations in the energy densities of the radiation fields governing inverse-Compton cooling as well as electron diffusion. In all panels, the minimum energy at which CR secondary electrons may be injected is indicated by the dashed red vertical line -- this would correspond to the secondaries produced when a 0.28 GeV CR proton primary undergoes a $\text{pp}$ interaction. The injection spectral index would be approximately $\Gamma=-2.1$, as indicated in blue in the upper panel. Indices of subsequent steady-state spectra are steeper than this due to the energy dependence of the electron cooling rate.\\
     \textbf{Upper panel:} Spectral variation with distance: dashed light grey line shows the steady-state spectrum expected at the injection point (we note that this is not the injection spectrum); remaining lines (solid black through to solid grey) denote the spectrum at 0.1-0.5 kpc in 0.1 kpc increments when adopting a CMB energy density at $z=7$ and no stellar radiation field. The steepening of the high-end of the spectrum with distance is due to the energy-dependence of the radiative cooling function and of electron diffusion (with higher energy particles diffusing and undergoing radiative cooling more quickly).\\
     \textbf{Middle panel:} As above, but spectra calculated at 0.2 kpc (black lines) and 0.5 kpc (blue lines) from their injection location, with CMB at $z=7$ and the spectral lines indicating evolution according to: no stellar radiation field (solid lines), stellar radiation field consistent with $\mathcal{R}_{\rm SF} = 3.5~\text{M}_{\odot}~\text{yr}^{-1}$ (dashed lines), $\mathcal{R}_{\rm SF} = 35~\text{M}_{\odot}~\text{yr}^{-1}$ (dot-dashed lines) and $\mathcal{R}_{\rm SF} = 350~\text{M}_{\odot}~\text{yr}^{-1}$ (dotted lines) where star-formation is concentrated in a 1 kpc volume. \\
     \textbf{Lower panel:} As above, but with all lines calculated at 0.2 kpc from their injection source location, lines in black corresponding to a CMB-only radiation field and lines in blue corresponding to a strong radiation field with $\mathcal{R}_{\rm SF} = 350~\text{M}_{\odot}~\text{yr}^{-1}$, shown at different redshifts over a similar range to our high-redshift galaxy sample: $z=6$ (solid line), $7$ (dashed line), $8$ (dot-dashed line) and $9$ (dotted line).}
    \includegraphics[width=0.46\textwidth]{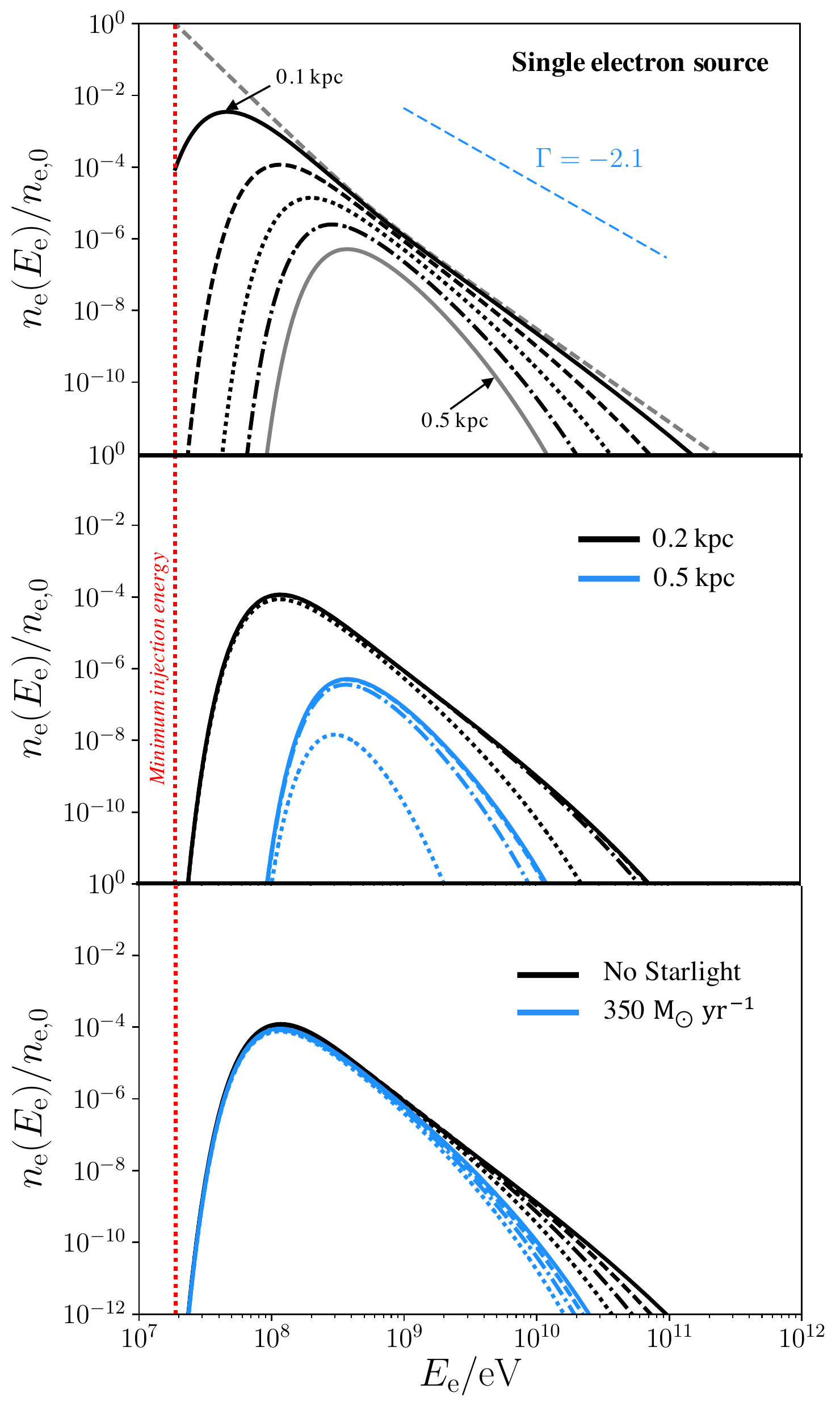}
    \label{fig:electron_spectra_plots}
\end{SCfigure*}
We require the CR electron distribution in the protogalaxy in its steady-state, when ${\rm d}n_{\rm e}/{\rm d}t = 0$, because this represents the persisting condition into which the system settles. After the proton distribution has stabilised to a steady-state, the injection of electrons will do so too as it is balanced by cooling and diffusion. This enables equation~\ref{eq:electron_transport_equation_sph} to be written as:
\begin{align}
\frac{\partial (b \; n_{\rm e})}{\partial E_{\rm e}}  = \frac{1}{r^2} \frac{\partial}{\partial r} \left[r^2 D(E_{\rm e}, r) \frac{\partial n_{\rm e}}{\partial r} \right] + Q_{\rm e}(E_{\rm e}, r) \ .
\end{align}
As with equation~\ref{eq:proton_transport_equation_sph}, we may restate this as a boundary value problem for a single electron injection event from a source region of size $\mathcal{V}_{\rm s}$ at some location $\textbf{r}_i$,
which yields the homogenous PDE:
\begin{align}
\label{eq:homogeneous_electron_pde}
\frac{\partial (b \; n_{\rm e})}{\partial E_{\rm e}}  = \frac{1}{r'^2} \frac{\partial}{\partial r'} \left[r'^2 D(E_{\rm e}, r') \frac{\partial n_{\rm e}}{\partial r'} \right] \ .
\end{align}
Here, $r'$ is again introduced as the distance between an injection point $\textbf{r}_{i}$ and an observation location $\textbf{r}$
with the injection term giving the boundary condition as
the steady-state number of electrons at the source,
given by the product of their total injection rate and cooling rate in an energy interval between $E_{\rm e}$ and ${\rm d}E_{\rm e}$,
that is $N_{\rm e, 0}(E_{\rm e}, \textbf{r}_i) = Q_{\rm e}(E_{\rm e}, \textbf{r}_i) \;\! \mathcal{V}_{\rm S}\;\! \tau_{\rm cool}(E_{\rm e})$.
This may be thought of as the number of electrons injected by the source point over a cooling timescale. 
Such a boundary condition invokes the assumption that the contribution from higher-energy electrons cooling into a given energy interval is negligible compared to fresh injections, which we argue is a reasonable approximation due to the 
underlying power-law nature of the electron injection spectrum.
 
Fig.~\ref{fig:total_losses} (right panel) shows that the electron cooling term, $b$, is dominated by inverse-Compton processes at higher energies, with Coulomb losses becoming important at lower energies.
According to equation~\ref{eq:synch_ic_cooling}, the inverse-Compton component depends on the energy density of the ambient radiation field.
This is governed by the spatial distribution of photons, including both those from the CMB and stellar radiation fields. 
Section~\ref{sec:radiation_field} indicates that, at the redshifts of interest, 
low to moderate star-formation rates result in the radiation energy density being 
easily dominated by the contribution from CMB photons,
which thus govern the inverse-Compton cooling of electrons. The CMB is isotropic and homogeneous and, as such, the inverse-Compton cooling function is likewise.
However, in systems with a relatively high star-formation rate, the stellar radiation photons can begin to have an important effect on the electron cooling:
for example, at a redshift of $z=7$, if the star-formation rate of a galaxy with a characteristic size of 1 kpc were to exceed $120~\text{M}_{\odot}~\text{yr}^{-1}$, the stellar radiation field would instead provide a dominant contribution to the photon energy density and thus principally govern the inverse-Compton cooling function.
Such a stellar radiation field would not be spatially independent, and it is therefore important to assess whether this contribution needs to be taken into account for this work:
When considering the sample of galaxies in Tables~\ref{tab:high_z_galaxies_table_post_sb} and~\ref{tab:high_z_galaxies_table_sb}, some systems have a remarkably high star-formation rate. 
However, their compactness is also relevant in calculating their stellar radiation energy density: for example, while a 1 kpc radius galaxy with a star-formation rate of $120~\text{M}_{\odot}~\text{yr}^{-1}$ would offer a comparable energy density in stellar photons to the CMB at $z=7$, another galaxy with the same star-formation rate and at the same redshift, but with a size of 10 kpc would have a negligible energy density compared to that of the CMB. 
When accounting for their size, we find that none of the star-forming galaxies in our sample exhibit an energy density in their stellar radiation fields which would exceed (or would even be comparable to) that of the CMB -- see Table~\ref{tab:high_z_galaxies_energy_density}. 

However, these estimates assume that our sample of galaxies exhibit a relatively even distribution of star-formation throughout their volume.
There are suggestions that, instead, high-redshift star-formation may be clumpy~\citep[e.g.][]{Elmegreen2009ApJ, ForsterSchreiber2011ApJ, Murata2014ApJ, Tadaki2014ApJ, Kobayashi2016ApJ}.
  It is therefore useful to consider an extreme case in which star-formation is clumped in the inner 1 kpc of a galaxy, in order to assess the impact this would have on the peak stellar radiation energy density compared to the energy density of the CMB. In the most active case of our sample, 
 {SXDF-NB1006-2} with $\mathcal{R}_{\rm SF} \approx 350~\text{M}_{\odot}~\text{yr}^{-1}$, 
 if all the star-formation were concentrated in the central 1 kpc of this system, its stellar radiation energy density would reach around $U_{*} \approx 3,300~\text{eV}~\text{cm}^{-3}$. 
 This is substantially greater than the contribution from the CMB in this case ($U_{\rm CMB}\approx 1,200~\text{eV}~\text{cm}^{-3}$) and so it cannot immediately be argued that the stellar radiation is unimportant.
 We find that the overall impact is still not particularly substantial: in Fig.~\ref{fig:electron_spectra_plots}, we show various (normalised) electron spectra when injected from a single source and subjected to inverse-Compton (and Coulomb) cooling process\footnote{Other relevant cooling processes are also applied in calculating these spectra, but their effect is not of great importance.}. Of particular note is the lower panel: The spectra shown in black correspond to the case where there is no significant contribution from stars to the ambient radiation field (i.e. it is dominated by the CMB energy density). In blue, the spectra are for the case when accounting for the presence of stellar photons, with their energy density consistent with a 1 kpc radius galaxy violently forming stars at a rate of $\mathcal{R}_{\rm SF} = 350~\text{M}_{\odot}~\text{yr}^{-1}$. The different lines in each case relate to different redshifts (and thus CMB energy densities), as noted in the caption. 
 It is clear that, while there is a notable difference at higher energies between the two cases, the underlying power-law nature of the electron spectrum means that the energy density of the electrons in this part of the distribution is relatively small compared to lower energies nearer to the spectral peak.
 The contribution from the part of the spectrum above 10$^9$ eV would account for less than 0.1\% of the total electron energy density.
 The middle plot confirms this assertion: here the black lines correspond to the spectra at 0.2 kpc from the discrete electron source, while those in blue show the result further away, at 0.5 kpc. The different lines in these cases relate to different star-formation rates associated with the stellar radiation field (none for the solid line, then respectively $3.5, 35$ and $350~\text{M}_{\odot}~\text{yr}^{-1}$ for the dashed, dot-dashed and dotted lines). 
 This shows that CR electrons must traverse through a substantial fraction of their host galaxy before their spectra differ considerably between the case where a strong stellar radiation field is present and the case where inverse-Compton cooling is entirely dominated by the CMB. 
 In our later calculations
 (see section~\ref{sec:computational_scheme}), we will consider a distribution of injecting sources -- by the time electrons have propagated far enough for any spectral difference between the two cases to be apparent, their contribution to the overall galactic electron spectrum would have been `drowned out' by another source emitting electrons nearby, which would be well within a 0.2 kpc radius.
 We therefore argue that it is reasonable for us to ignore the contributions from stellar photons in determining the radiative cooling rates of our CR secondary electrons, and duly resort to a CMB-only model which is spatially independent.

\begin{table}
	\centering
	\begin{tabular}{l|cc}
		&& \\[-0.5em]
		Galaxy ID & $U_{*}$/eV cm$^{-3}$ & $U_{\rm CMB}$/eV cm$^{-3}$  \\
		&&\\[-0.5em]
		\hline
		&&\\[-0.5em]
		HDFN-3654-1216 	& 50 & 800 \\ 
		UDF-640-1417 		& 110 & 1,000 \\ 
		GNS-zD2 			& 92 & 1,200 \\ 
		CDFS-3225-4627 	& 120 & 1,200 \\
		GNS-zD4 			& 200 & 1,200 \\  
		GNS-zD1 			& 370 & 1,200 \\ 
		GN-108036 			&  380 & 1,200 \\ 
		SXDF-NB1006-2 		& 530 & 1,200 \\ 
		UDF-983-964 		& 36 & 1,300 \\ 
		GNS-zD3 			& 150 & 1,300 \\  
		GNS-zD5 			& 370 & 1,300 \\ 
		A1689-zD1 			& 11 & 1,500 \\ 
		EGS-zs8-1 			& 190 & 1,600 \\ 
		UDF-3244-4727 		& 160 & 1,700 \\  
		MACS1149-JD1 		& 6.6 & 2,800 \\  
		GN-z11 				& 630 & 5,800 \\ 
		\end{tabular}
	\caption{\small Peak energy densities of the radiation fields for all 16 of our sample of star-forming galaxies during their starburst phase, compared to the energy density of the CMB radiation field at their observed redshift. This shows that, in all cases, radiative cooling of the CR electrons would be dominated by the CMB.}
	\label{tab:high_z_galaxies_energy_density}
\end{table}

Coulomb CR electron cooling depends on the local number density and ionisation state of the interstellar gas. Although the ionisation state is assumed to be fixed, the density profile does exhibit some spatial dependence in our model (see equation~\ref{eq:density_profile}). 
The gas density varies significantly on kpc scales, but can be assumed to be comparatively uniform more locally. The length-scale over which electron Coulomb cooling operates is:
\begin{equation}
\label{eq:coulomb_length}
\ell_{\rm C} \approx \sqrt{4\;\!  D(E_{\rm e}, r) \;\! \tau_{\rm C}} \ ,
\end{equation}
\citep{Owen2018MNRAS} where $\tau_{\rm C}$ is the electron Coulomb cooling timescale. For a 40 MeV secondary CR electron diffusing through an interstellar medium of density $n_{\rm b} = 10~\text{cm}^{-3}$, this gives a length-scale of $\ell_{\rm C} \approx 0.2$ kpc, being substantially less than the size of the protogalaxy. 
As a first approximation, we therefore model the Coulomb cooling to be spatially independent locally. As this is also the case with the radiative cooling function discussed earlier, we take $b$ to be spatially independent in general for each individual injection location in the following treatment (we note that this is only a local approximation, and our model does account for variation in the cooling function at different source injection locations).
Thus, using the substitution $\zeta = b \; n_{\rm e}$, equation~\ref{eq:homogeneous_electron_pde} may be expressed as:
\begin{equation}
\label{eq:homogeneous_electrons_with_sub}
\frac{\partial \zeta}{\partial E_{\rm e}} = \frac{1}{r'^2} \frac{\partial} {\partial r'}\left[ \frac{D(E_{\rm e}, r') ~r'^2}{b} \frac{\partial \zeta}{\partial r'}\right] \ ,
 \end{equation}
 where we have invoked the spatial independence of $b$. 
  As this PDE has a similar structure to equation~\ref{eq:proton_homogenous}, it may be solved by the same method. This gives:
  \begin{equation}
\zeta = \frac{\zeta_0}{\left[4 \pi D(E_{\rm e},r')\;\! E_{\rm e} \; b^{-1} \right]^{3/2}}\exp{\left\{-\frac{r'^2}{4 D(E_{\rm e},r')\;\! E_{\rm e} \; b^{-1}}\right\}} \ .
\end{equation}
 Here, the normalisation constant $\zeta_0 = b \; N_{\rm e, 0}$ (from the boundary condition), and we note that $\tau_{\rm cool}(E_{\rm e}) = E_{\rm e} \; b^{-1}$.
  Back-substituting for $\zeta = b \; n_{\rm e}$ yields the steady-state result for electrons continuously injected at a single point source:
  \begin{equation}
\label{eq:electron_solution}
n_{\rm e} = \frac{ Q_{\rm e}(E_{\rm e}, \textbf{r}_i) \;  \mathcal{V}_{\rm S}}{\left[4 \pi D(E_{\rm e},r')\right]^{3/2} \; \tau_{\rm cool}^{1/2}(E_{\rm e})} \;\! \exp{\left\{\frac{- r'^2}{4 D(E_{\rm e},r')\;\! \tau_{\rm cool}(E_{\rm e})}\right\}} \ .
 \end{equation}
  
\section{Cosmic ray heating}
\label{sec:heating}

Previous studies have considered a range of different mechanisms by which CR heating may operate.
These include damping of magnetohydrodynamical (MHD) waves~\citep[e.g.][]{Wentzel1971ApJ, Wiener2013ApJ, Pfrommer2013ApJ}, direct collisions and ionisations~\citep[see][]{Schlickeiser2002book, Schleicher2013A&A} as well as Coulomb interactions between charged CRs and a plasma~\citep[e.g.][]{Guo2008MNRAS, Ruszkowski2017ApJ}. 
Some of these mechanisms can yield heating rates\footnote{When adopting $\mathcal{R}_{\rm SN} = 1.0~\text{yr}^{-1}$.} of up to $10^{-27}~\text{erg}~\text{cm}^{-3}~\text{s}^{-1}$ in appropriate conditions~\citep{Wiener2013ApJ, Walker2016ApJ}, 
however the scales over which heating power may be imparted depends on the mechanism at work.
Some operate chiefly
 within~\citep{Field1969ApJ, Walker2016ApJ} their host galaxy (including within molecular clouds via ionisation, see~\citealt{Papadopoulos2010ApJ, Papadopoulos2011MNRAS, Juvela2011ApJ, Galli2015arXiv}), but 
 others are more effective in heating the medium around it~\citep[e.g.][]{Loewenstein1991ApJ, Guo2008MNRAS, Ruszkowski2017ApJ, Ruszkowski2018ApJ} or far beyond~\citep[e.g.][]{Nath1993MNRAS, Samui2005ICRC, Sazonov2015MNRAS, Leite2017MNRAS}. 
Hadronic interactions are one such proposed mechanism~\citep[e.g.][]{Ensslin2007A&A, Guo2008MNRAS, Ensslin2011A&A, Ruszkowski2017ApJ}, and could operate within the ISM of a protogalactic host. 
This mechanism is particularly important in star-forming galaxies due to the relative abundance of CR protons above the hadronic interaction threshold energy, perhaps accounting for as much at 99\% of their total energy density~\citep[e.g.][]{Benhanbiles-Mezhoud2013ApJ}. 

In general, hadronic heating of a medium by CRs arises after the (${\rm pp}$) interaction of the CR primary injects secondary electrons (see Appendix~\ref{sec:interaction_mechanisms}). 
These secondary electrons can then cool and thermalise into the ISM. 
As discussed previously in section~\ref{sec:electron_transport}, the cooling of these electrons can proceed through two channels: via Coulomb interactions with the ambient ISM plasma or, at higher energies, via inverse-Compton interactions with (predominantly) CMB photons. The Coulomb channel leads to direct thermalisation of the electron energy into the ISM. We refer to this hereafter as \textit{direct Coulomb heating}, or DC heating.
Conversely, the inverse-Compton channel does not itself lead to direct energy transfer to the ambient gases: instead (and particularly in high-redshift environments where the CMB is of higher energy density), CR electrons up-scatter CMB photons into (principally) the X-ray band, causing the host galaxy to glow~\citep{Schleicher2013A&A, Schober2015MNRAS}. 
It is thought that the X-ray luminosity in starburst galaxies at high-redshift could become 
very intense: calculations in \citealt{Schober2015MNRAS} suggest luminosities of above $10^{39}~\text{erg s}^{-1}$ in the 0.5-10 keV band\footnote{This is the energy range which the \textit{Chandra} X-ray observatory is sensitive to (see \url{http://chandra.harvard.edu}) and was chosen in the~\citet{Schober2015MNRAS} study to make observational predictions for the X-ray flux from starbursts to diagnose the presence of CRs and magnetic fields.} could be achieved by $z=7$ for systems with $\mathcal{R}_{\rm SN} \approx1.0~\text{yr}^{-1}$, and scaling in proportion to the SN event rate (or star-formation rate), however Appendix~\ref{sec:xray_lum} shows this value could be even higher.
These keV X-rays can then propagate and scatter in the ionised or semi-ionised interstellar and circumgalactic medium to drive a heating effect. We refer to this process as \textit{Indirect X-ray heating}, or IX heating hereafter.

\subsection{Direct coulomb heating}
\label{sec:direct_coulomb_heating}

The volumetric heating rate due to the DC mechanism at a location $r$ is given by:
\begin{equation}
    H_{\rm DC}(r) =  \int_{E_{\rm min}}^{E_{\rm max}} {\rm d}E_{\rm p} \; n_{\rm p}(E_{\rm p},r) \; {\rm c}\; f_{\rm C}(E_{\rm p}, r)\; \alpha^*(E_{\rm p}, r)  \ ,   
\label{eq:heating_calc} 
\end{equation} 
\citep{Owen2018MNRAS}
where $E_{\rm min} = 0.28~\text{GeV}$ is the minimum energy required for a hadronic interaction~\citep[e.g.][]{Kafexhiu2014PRD}, and $E_{\rm max} \approx 1~\text{PeV}$ is the approximate maximum energy a CR could be accelerated to in environments expected inside the host system, for example SN remnants~\citep[see][]{Kotera2011ARAA, Schure2013MNRAS, Bell2013MNRAS}.
$n_{\rm p}(E_{\rm p},r)$ is the local number density of primary CR protons (as calculated in section~\ref{sec:proton_transport}),
 $\alpha^*(E_{\rm p}, r)$ is the absorption coefficient (given by equation~\ref{eq:alpha_star}) 
 and $f_{\rm C}(E_{\rm p})$ is the Coulomb CR heating efficiency factor, 
 which encodes the fraction of energy transferred from the interacting CR primary proton into this particular heating channel. It is defined as:
\begin{equation}
\label{eq:efficiency_coulomb}
f_{\rm C}(E_{\rm p}, r) = \bar{K} \;\! \mathcal{M}\vert_{E_{\rm p}} \;\! \left( \frac{\tau_{\rm C}^{-1}}{\tau_{\rm C}^{-1} + \tau_{\rm rad}^{-1} + \tau_{\rm brem}^{-1}} \right)\Bigg\vert_{E_{\rm e}} \ ,
\end{equation}
which is the rate of energy transfer from the secondary electrons to the ISM by Coulomb exchanges as a fraction of the total electron cooling rate by all processes, specified at an electron energy $E_{\rm e}$. This is related to the CR primary energy by
$ E_{\rm e} = {E_{\rm p} \; \bar{K}}/{\mathcal{M}\vert_{E_{\rm p}}}$,
which suggests that the energy of a typical CR secondary electron is around 40 MeV for a 1 GeV primary proton.
The term $\bar{K} \;\! \mathcal{M}\vert_{E_{\rm p}}$  is introduced to account for electron production multiplicity (denoted $\mathcal{M}$) and energy transfer efficiency to the secondary electrons ($\bar{K}$) -- see Appendix section~\ref{sec:particle_injection} for details.
This approach implies that local and instantaneous thermalisation arises. 
This is justified given that we have previously shown the thermalisation length-scale for 40 MeV electrons (from a 1 GeV primary) is $\ell_{\rm C} \approx 0.2~\text{kpc}$, being substantially smaller than the 1 kpc characteristic size of the host galaxy. 
Instantaneous thermalisation can also be argued over galactic timescales: a 0.4 Myr thermalisation timescale is negligible compared to the dynamical timescale of the host system -- typically 10s of Myr.

\subsection{Indirect X-ray heating}
\label{sec:indirect_xray_heating}

The total inverse-Compton cooling rate of a single electron with energy $E_{\rm e} = \gamma_{\rm e} \;\! m_{\rm e} {\rm c}^2$ 
in a radiation field of energy density $U_{\rm ph}$ is given by
\begin{equation}
\label{eq:ic_powerloss}
\frac{{\rm d}E_{\rm e}}{{\rm d}t} = - \frac{4}{3}\sigma_{\rm T} {\rm c} \left(\frac{E_{\rm e}}{m_{\rm e}{\rm c}^2}\right)^2 U_{\rm ph} \ ,
\end{equation}
\citep{Rybicki1979book}
where $\sigma_{\rm T} = 6.65\times 10^{-25}~\text{cm}^2$ is introduced as the Thomson cross-section. 
This is the rate at which energy is transferred from a CR electron to photons in the radiation field, and so dictates the inverse-Compton power due to the single electron,
\begin{equation}
P_{\rm IC} = - \frac{{\rm d}E_{\rm e}}{{\rm d}t} \ .
\end{equation}
The characteristic energy  $E_{\rm ph}$ to which low-energy photons in the radiation field can be up-scattered by the high energy electrons (of characteristic energy $E_{\rm e}^*$) is
\begin{equation}
E_{\rm ph} \approx 2.82 \; \frac{\Theta(z)}{m_{\rm e}\;\!{\rm c}^2} \; E_{\rm e}^{*2} \ ,
\end{equation}
where $E_{\rm e}^*$ may be estimated from the peak of the local electron spectrum (e.g. those shown in Fig.~\ref{fig:electron_spectra_plots}). $\Theta(z)$ retains its earlier definition of ${{\rm k_{\rm B}} T(z)}/{m_{\rm e} \rm{c}^2}$. 
This would suggest a characteristic emitted up-scattered photon energy of around 10 to 100 keV, in the X-ray band.
For this energy range, we note that the inverse-Compton Klein-Nishina cross section does not exhibit a strong energy dependence, and is well approximated by the Thomson scattering cross-section to within 10\%, $|\sigma_{\rm KN} - \sigma_{\rm T}|/\sigma_{\rm KN} < 0.1$. 
This allows us to make a substantial simplification in our calculations hereafter, in that the specific frequency of the up-scattered photons is not of particular importance when considering their subsequent scatterings and interactions. As such, in the following, we will only need to consider the total X-ray inverse-Compton emission energy -- not the full spectrum of this radiation.

To calculate the local intensity of inverse-Compton X-rays produced by the high energy CR electrons, we may adopt a similar approach to that which was used in solving the transport equation. 
In general, the intensity of the X-ray radiation field due to some discrete source at a position $\textbf{r}_i$ and spectral luminosity $L_{\rm IX}(E)$
may be written as:
\begin{equation}
{I}_{\rm IX}(E, r; \textbf{r}_i) = \frac{L_{\rm IX}(E)\vert_{\textbf{r}_i}}{4\pi r'^2 {\rm c}} \;\! \exp\left\{-\int_0^{r'} {\rm d}\ell \;\! \sigma_{\rm KN}(E)\;\! x_i \;\! n_{\rm b}(\ell) \right\} \ ,
\end{equation}
where $r'$ retains its earlier definition as the distance between the source position $\textbf{r}_i$ and some general location $\textbf{r}$,
and $x_i$ is the local gas ionisation fraction such that the number of charged particles in the ambient medium is given by $x_i \;\! n_{\rm b}$. 
The exponential term encodes the attenuation effect experienced by the X-ray radiation as it propagates from its source\footnote{We note that, even in central galaxy conditions of $n_{\rm p}=10~\text{cm}^{-3}$, the mean free path of a keV photon is around 100~kpc, resulting in the attenuation term being very close to unity with its impact being negligible on our analysis.}.
Invoking our approximation whereby the spectral distribution of the X-rays is not important, this reduces to:
\begin{equation}
\mathcal{I}_{\rm IX}(r; \textbf{r}_i) = \frac{\mathcal{L}_{\rm IX}\vert_{\textbf{r}_i}}{4\pi r'^2 {\rm c}} \;\! \exp\left\{-\int_0^{r'} {\rm d}\ell \;\! \sigma_{\rm T}\;\! x_i \;\! n_{\rm b}(\ell) \right\} \ ,
\label{eq:xray_intensity_result}
\end{equation}
where $\mathcal{L}_{\rm IX}$ is now the total inverse-Compton X-ray luminosity of the source, related to spectral luminosity by 
\begin{equation}
\mathcal{L}_{\rm IX} = \int_0^{\infty} {\rm d}E \;\! L_{\rm IX}(E) \ .
\end{equation}
It is calculated as:
\begin{equation}
\mathcal{L}_{\rm IX}\vert_{\textbf{r}_i} = \frac{4}{3}\sigma_{\rm T} {\rm c} 
\;\! U_{\rm ph} \;\! \mathcal{V}_{\rm S} \int_{E_{\rm min}}^{E_{\rm max}} \;\!{\rm d}E_{\rm e} \left(\frac{E_{\rm e}}{m_{\rm e}{\rm c}^2}\right)^2 \;\! n_{\rm e}(E_{\rm e}; \textbf{r}_i)  \ ,
\label{eq:xray_intensity_profile_point}
\end{equation}
where $\mathcal{V}_{\rm S}$ retains its earlier definition as a characteristic source size, which contains $N_{\rm e} = n_{\rm e}\;\!\mathcal{V}_{\rm S}$ electrons.
The resulting total X-ray intensity $\mathcal{I}_{\rm IX}(r)$ through a distribution of such sources at a set of positions $\textbf{r}_{i}$ can be found by convolution of the distribution of points with the result given by equation~\ref{eq:xray_intensity_result}.
From this, the X-ray heating rate would then simply follow as:
\begin{equation}
    H_{\rm IX} (r) = \mathcal{I}_{\rm IX}(r) \;\! \;\! \alpha_{\rm IX}(r)   \ .
\label{eq:xray_heating_calc} 
\end{equation}
  
\subsection{Computational scheme}
\label{sec:computational_scheme}

Equations~\ref{eq:proton_solution} and~\ref{eq:electron_solution} give the steady-state diffusion solution for CR protons and electrons emitted by a single source of some characteristic volume $\mathcal{V}_{\rm S}$. Further, equation~\ref{eq:xray_intensity_result} gives the X-ray intensity due to a single point-like source.
In our calculations, we must extend these single point solutions to model a spatially extended distribution of CR and radiative emission. 
For CR protons, this distribution would presumably follow the underlying density profile of the protogalaxy, since CRs originate in stellar end products: stars form by gravitational collapse of their underlying matter distribution so it follows that their end products should correlate with this too.
We thus sample the $N_{\rm S}$ points throughout a spherical volume by a Monte-Carlo (MC) method, weighted by the density profile. 
In this case, we specify a maximum radius for the distribution of 2 kpc, beyond which no further points are placed. Each point is assigned a characteristic volume, $\mathcal{V}_{\rm S} = 4 \pi r_{\rm gal}^3/(3 N_{\rm S})$, where $r_{\rm gal}$ is the characteristic size of the protogalactic system (typically 1 kpc) and $N_{\rm S}$ is the number of individual sources sampled.
The resulting distribution of points is then convolved with the steady-state proton distribution for a single source to yield the full galactic CR proton steady-state profile.

\begin{figure}
  \vspace*{0.1cm}
	\includegraphics[width=\columnwidth]{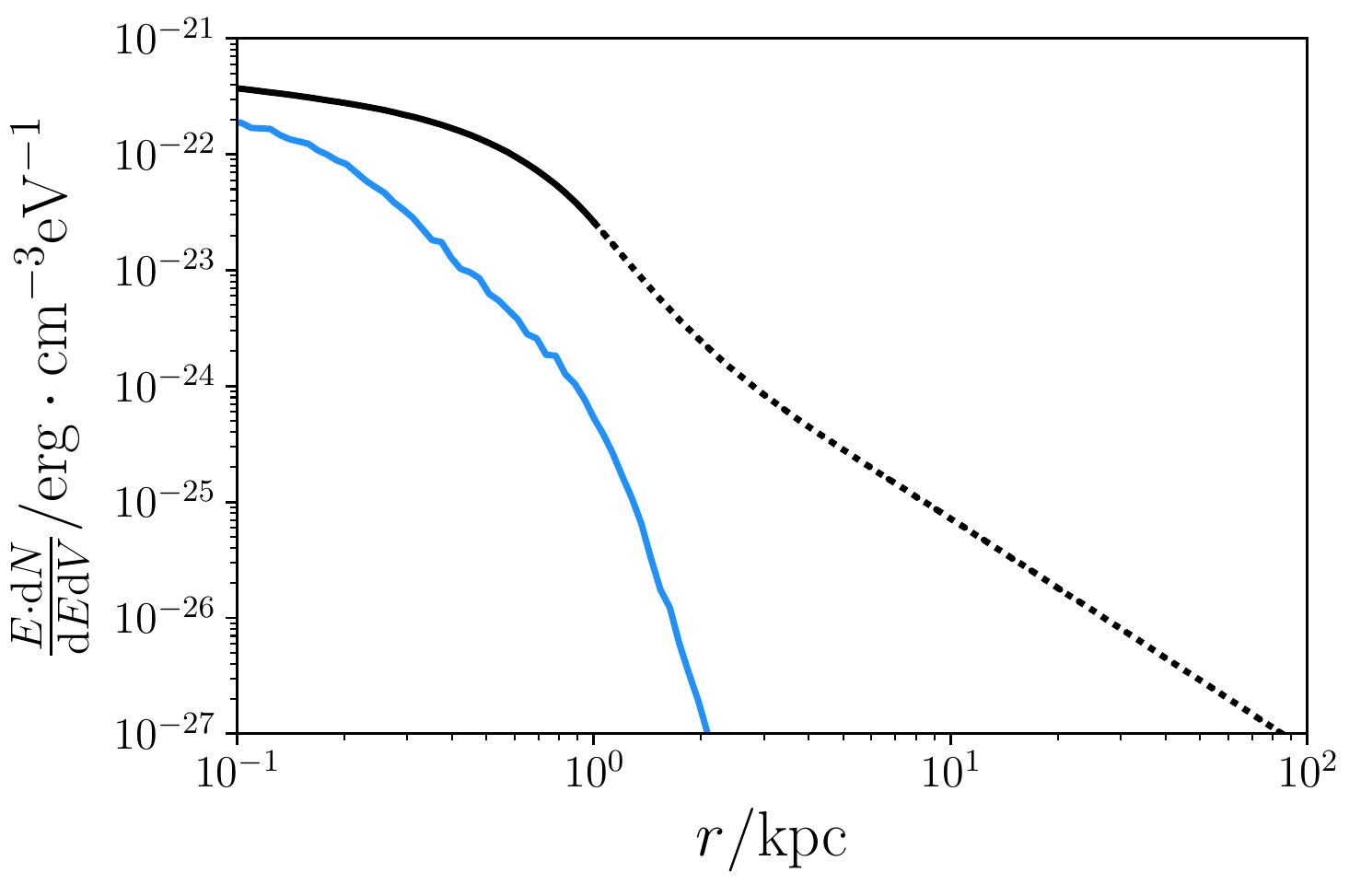}
    \caption{\small Steady-state protogalactic CR proton (primary) and electron (secondary) profile integrated over all energies
       for a SN rate $\mathcal{R}_{\rm SN}  = 0.1\; {\rm yr}^{-1}$. 
      The blue line shows the profile for electrons, while the black line corresponds to protons. The solid part of the black line is the proton distribution within the interstellar medium of the protogalaxy, where the magnetic field model is more robust. Outside the protogalaxy, the magnetic field is less well understood and so our result (the dashed part of the black line) should be regarded as an upper limit\protect\footnotemark.}
    \label{fig:electron_diffusion}
\end{figure}

The CR electron profile is calculated in a similar manner -- however, this must be a two-step process to properly account for the varying density profile of the galaxy in determining the level of electron injection at each position (in accordance with equations~\ref{eq:cre_injection_spec2} and~\ref{eq:alpha_star}). The algorithm proceeds as follows: first, sample a points distribution according to the underlying density profile (as was done for the CR protons). Next, calculate the steady-state proton distribution at each point independently and sample a further $N_{\rm S, e}$ points by the same spherical MC sampling method, this time weighted by the single-source steady-state proton solution. The resulting distribution is then convolved with the steady-state electron solution for a single source to give the resulting CR secondary electron profile for a single primary CR injection point. This result is then convolved with the original CR proton source point distribution to give the full galactic profile for electrons. We show the spectrally-integrated result as the CR electron energy density (blue line) together with that for the CR protons (black line) in Fig.~\ref{fig:electron_diffusion}, when adopting a SN event rate in the protogalaxy of $\mathcal{R}_{\rm SN} = 0.1~\text{yr}^{-1}$. Here (and hereafter) we have used 
$N_{\rm S} = 10,000$ and $N_{\rm e} = 1,000$ 
(the exact electron distribution is less instrumental to our final results), which we find gives an acceptable signal to noise ratio.
\footnotetext{The results shown in this plot are slightly different to those for CR protons presented in Fig. 5 of~\cite{Owen2018MNRAS}, but the two results are not directly comparable. This is because (1) here, we include the attenuation of CRs in our treatment, which was not accounted for in the earlier result (instead this was accounted for later in their simulation pipeline and applied to their final CR heating calculation); (2) the result here integrates over the full energy spectrum of the CR protons and electrons, whereas the result in the previous study shows the CR proton profile when setting the proton distribution to its mean energy of 2.5 GeV; (3) this result is computed using $\mathcal{R}_{\rm SN}  = 0.1\; {\rm yr}^{-1}$, whereas the previous study showed the case for $\mathcal{R}_{\rm SN}  = 1.0\; {\rm yr}^{-1}$.}

The scheme used to calculate the X-ray intensity profile proceeds as follows: a further $N_{\rm S, X}$ points are sampled by the MC method, but now are weighted according to the full galactic CR electron profile. The X-ray intensity profile for each point is determined using equation~\ref{eq:xray_intensity_result}, and this is then convolved with the weighted points distribution. The total galactic inverse-Compton X-ray intensity profile then results. We again adopt a value of $N_{\rm IX} = 10,000$. This, together with the CR electron profile, is used to determine the CR heating power via the two mechanisms according to equations~\ref{eq:heating_calc} and~\ref{eq:xray_heating_calc}.

\subsection{Heating power}
\label{sec:cr_heating_power_calc}

 \begin{figure}
 	\centering
	\includegraphics[width=\columnwidth]{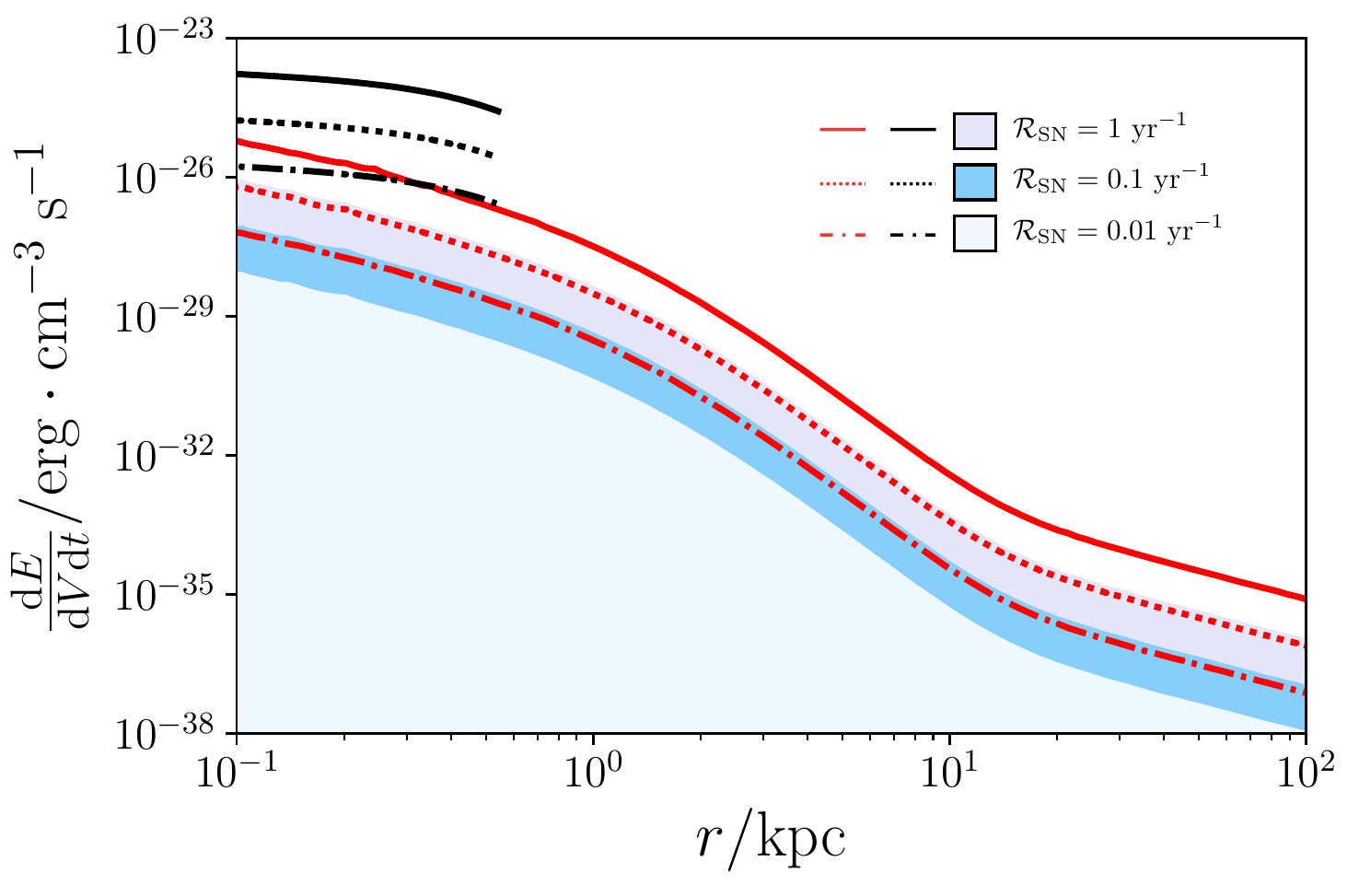}
    \caption{\small CR heating rates of the protogalaxy ISM by (1) direct Coulomb heating, in black, up to 0.6~kpc, and (2) indirect X-ray heating, in red. The solid lines are the results calculated for $\mathcal{R}_{\rm SN} = 1~\text{yr}^{-1}$, dotted lines for $\mathcal{R}_{\rm SN} = 0.1~\text{yr}^{-1}$ and dot-dashed for $\mathcal{R}_{\rm SN} = 0.01~\text{yr}^{-1}$. $\mathcal{R}_{\rm SN}$ determines the energy budget of the system, and so the CR heating power in both channels scales directly. The shaded regions show the stellar radiation `floor' for the IX process: any evolution that brings the IX line into this area will instead be dominated by the up-scattering of stellar-radiation by CR electrons, preventing the X-ray heating power from falling into this region. The lower region accounts for a stellar radiation field when $\mathcal{R}_{\rm SN} = 0.01~\text{yr}^{-1}$, the intermediate case above, for $\mathcal{R}_{\rm SN} = 0.1~\text{yr}^{-1}$, and the highest region arises for $\mathcal{R}_{\rm SN} = 1.0~\text{yr}^{-1}$. We note that the DC heating rate falls to negligible levels outside the protogalaxy (not shown) due to the magnetic containment of the CRs.}
    \label{fig:heating_powers}
\end{figure}
 \begin{figure}
	\includegraphics[width=\columnwidth]{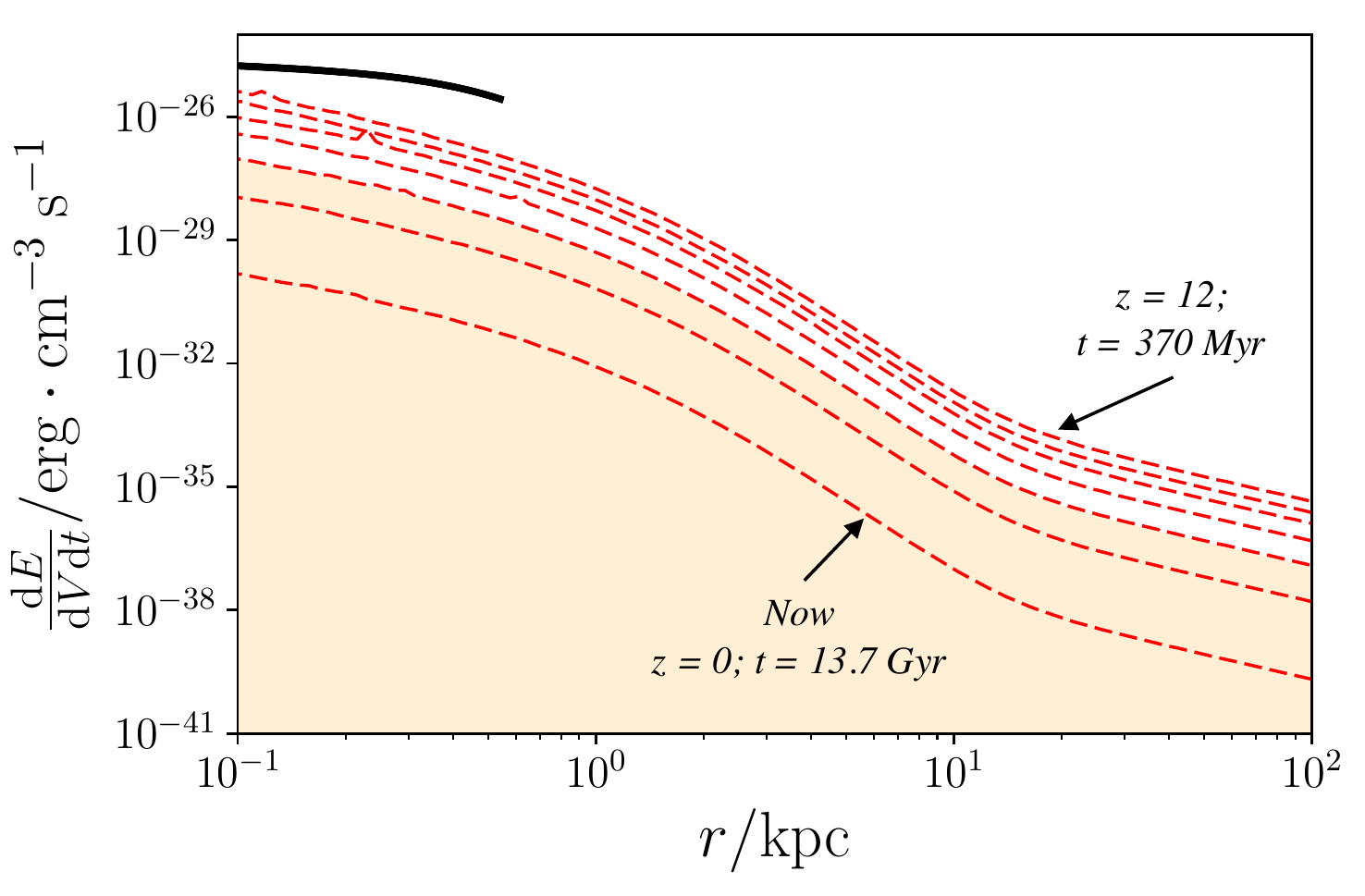}
    \caption{\small CR heating rates via the direct Coulomb (solid black lines) and indirect X-ray mechanisms (red dashed lines) for a protogalaxy with $n_{\rm b,\;\!0} = 10~\text{cm}^{-3}$ and $\mathcal{R}_{\rm SN} = 0.1~\text{yr}^{-1}$ for redshifts $z = 0, 2, 4, 6, 8, 10$ and $12$. The black DC lines (only one line is visible as they overlap) do not show any discernible evolution over this redshift range, but those for the IX effect show significant variation due to the evolving CMB. The central heating rate at $z=12$ by both mechanisms is comparable at a level of around $10^{-25}\;\!\text{erg}~\text{cm}^{-3}~\text{s}^{-1}$. The shaded orange region represents the stellar radiation `floor' for the IX process: any redshift evolution that brings the IX line into this area (i.e. below $z\approx 4$) will instead be dominated by the up-scattering of stellar-radiation, preventing the IX power from falling into this region. The DC heating power becomes negligible outside the protogalaxy (not shown).}
    \label{fig:z_evolution}
\end{figure}
In Fig.~\ref{fig:heating_powers}, we show the total power associated with the two CR heating processes: the direct Coulomb (DC) mechanism in black,
and the indirect inverse-Compton X-ray (IX) mechanism in red.
This uses a model protogalaxy at redshift $z=7$, 
characteristic size 1 kpc, 
central density $n_{\rm b, 0} = 10~\text{cm}^{-3}$,
and considers SN-rates of $\mathcal{R}_{\rm SN} = 0.01, 0.1$ and $1.0~\text{yr}^{-1}$ (broadly corresponding to star-formation rates of $\mathcal{R}_{\rm SF} \approx 1.6, 16$ and $160~\text{M}_{\odot}~\text{yr}^{-1}$). 
We show that a heating rate as high as $10^{-24}\;\!\text{erg}~\text{cm}^{-3}~\text{s}^{-1}$ could be sustained by the DC channel when $\mathcal{R}_{\rm SN} =1.0~\text{yr}^{-1}$ as CRs become trapped by the magnetic field\footnote{As the galactic magnetic field grows in strength, charged CRs can no longer freely stream from their origin. Instead they predominantly diffuse, with their effective diffusion velocity being substantially lower than their free-streaming velocity (see~\citealt{Owen2018MNRAS} for details).}, 
with a $z=7$ heating IX power of around $10^{-25}\;\!\text{erg}~\text{cm}^{-3}~\text{s}^{-1}$. This would correspond to a total X-ray luminosity of around $10^{41}~\text{erg}~\text{s}^{-1}$ -- see also Appendix~\ref{sec:xray_lum}.
With regard to the IX mechanism, we also indicate shaded regions in Fig.~\ref{fig:heating_powers} to illustrate the level that would be reached if only stellar photons were up-scattered. 
Thus, if the IX heating power due to up-scattering of CMB photons were to dominate (the plotted red lines), it must remain above this shaded area.

It is clear that up-scattering of CMB photons would be more important than that for stellar photons for any reasonable starburst model at $z=7$. 
At low redshifts the picture is different. Consider Fig.~\ref{fig:z_evolution}, which shows the redshift evolution of the two CR heating effects between $z=0$ and $z=12$.
The starlight provides an effective `floor' to the inverse-Compton emission (and resulting IX heating) at lower redshifts. 
For the particular model that we have adopted in Figs~\ref{fig:heating_powers} and~\ref{fig:z_evolution}, a cut-off redshift of approximately $z=4$ arises (we note that this is model-dependent).
Below this, the inverse-Compton scattering of stellar-photons dominates over CMB up-scattering, with the final IX heating channel attaining a power roughly $10^{3}$ times smaller than the corresponding DC channel, 
making the IX process relatively unimportant by the time such a cut-off redshift has been reached. 
In general, the IX heating would scale with $\mathcal{R}_{\rm SN}$ when in the starlight-dominated regime. 
Thus, we argue the IX mechanism could only operate at a competitive level with the DC mechanism at high-redshift (which itself is 
largely redshift independent -- as indicated by the coinciding black DC heating lines in Fig.~\ref{fig:z_evolution}).
Given the scales over which these processes deposit their energy, this is intriguing: 
the DC mechanism predominantly operates within the ISM of the host galaxy -- it is governed by the structure and extent of the magnetic field of the host, and so would drop to negligible levels as low as $10^{-40}~\text{erg}\;\!\text{cm}^{-3}\;\!\text{s}^{-1}$~\citep{Owen2018MNRAS} outside the protogalaxy\footnote{Since the interfacing region between the ISM and circumgalactic medium (CGM) would be entrained by complicated magnetic field structures -- the modelling of which remains beyond the scope of the current paper -- we only show the DC channel heating effect up to 0.6 kpc in our results, well within the ISM of the host.}. 
By contrast, the IX mechanism would operate on larger scales -- the emitted X-rays can propagate into the circumgalactic medium to drive heating effects around the host, rather than just within it.
The redshift evolution of the IX effect thus means that CR heating would be split comparably between the interior and exterior of a galaxy at high-redshift, but would become increasingly focussed onto the ISM over cosmic time. 
This means that the resulting CR feedback processes at work in and around starbursts could be fundamentally different in the early Universe compared to the present epoch (see also section~\ref{sec:application}).

\subsection{Comment on cosmic ray heating in interstellar and circumgalactic structures}
\label{sec:cr_heating_density_structures}
 
\begin{figure}
	\includegraphics[width=\columnwidth]{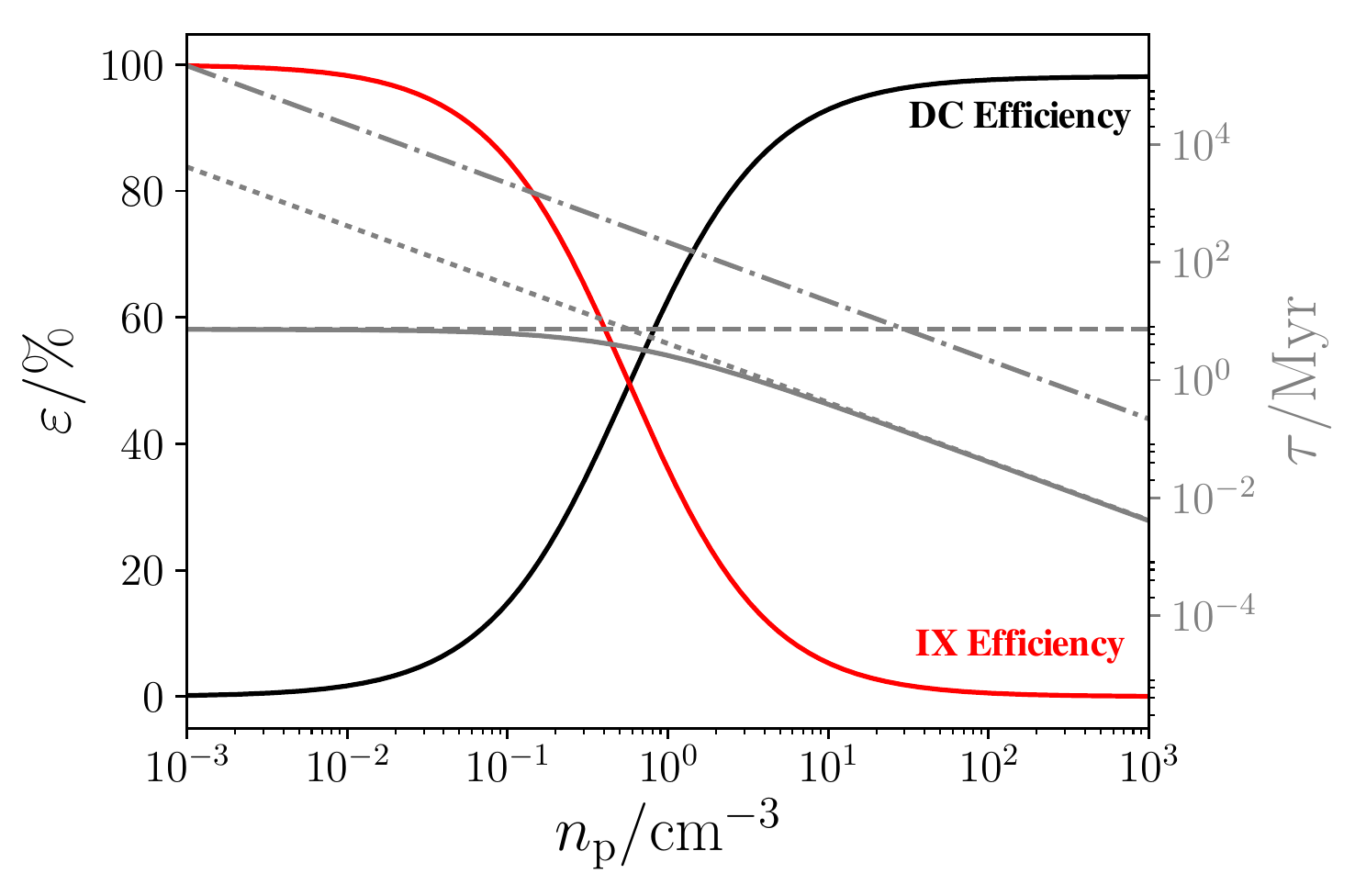}
    \caption{\small Fraction of 40 MeV CR secondary electron energy passed to the inverse-Compton X-ray heating channel (red solid line) compared to the direct Coulomb heating channel (black solid line) as a function of density. The left hand $y$-axis corresponds to the relative fraction, and this indicates a strong preference for the IX heating channel in low density media, or the DC channel in higher density media (if ionised). Timescales of respective processes are shown in grey to illustrate the underlying physics: the solid line gives the total energy loss timescale for CR electrons. This is calculated by combining the contributions from inverse-Compton (dashed horizontal line), Coulomb (dotted line) and free-free losses (dot-dashed line). The respective timescale $y$-axis is shown to the right of the plot.}
    \label{fig:channel_frac}
\end{figure}

 At high-redshifts, previous studies have suggested that the ISM of galaxies is clumpy:
a single galaxy could be split into just 5-10 large kpc scale clumps~\citep[][]{van_den_Bergh1996AJ, Cowie1996AJ, Elmegreen2004ApJa, Elmegreen2004ApJb, Elmegreen2005ApJ, Elmegreen2007ApJ, Bournaud2009ApJ, Krumholz2010MNRAS, Wardlow2017ApJ}, resulting from the high instability of the cool molecular-gas rich environment~\citep{Bournaud2009ApJ, Krumholz2010MNRAS} fuelled for example by inflows~\citep{Dekel2009ApJ}, with recent observations with ALMA pointing towards a range of different ISM conditions, from relatively smooth to highly clumpy~\citep[see][]{Gullberg2018ApJ}.
Moreover, the circumgalactic medium (CGM) may have a complex structure, comprised of low-density H\;\!II regions together with cold dense inflows~\citep[e.g.][]{Ribaudo2011ApJ, Sanchez2014A&ARv, Peng2015Natur, Owen2018sub} which themselves could be ionised and beaded with dense, neutral clumps~\citep{Fumagalli2011MNRAS}.
As such, more careful consideration of how CRs heat and propagate through different densities of medium is important to properly account for their interactions with their environment and the way in which CR feedback may operate in and around protogalaxies. 
Consider Fig.~\ref{fig:channel_frac}, which shows how the efficiency of the two CR heating channels depends on ambient density (assuming full ionisation of the medium) for a 40 MeV CR secondary electron (resulting from the hadronic interaction of a 1 GeV proton). This is estimated from the relative timescales of the key loss processes of the CR electrons responsible for driving the CR heating effects, as are indicated by the lines in grey (inverse-Compton, Coulomb losses and synchrotron cooling). 
Fig.~\ref{fig:channel_frac} points towards a two-phase heating mechanism:
CRs effectively heat dense pockets of gas (if exhibiting a reasonably high ionisation fraction) by Coulomb scattering, but are strongly impacted by the galactic magnetic fields and their substructures, effectively containing them to the ISM.
Conversely, the IX channel is very efficient in hot diffuse gas -- although it can still operate in higher density systems on large scales when the ionisation fraction is sufficiently high. 
This implies that the DC heating mechanism may be important for quenching star-formation, whereas the IX mechanism could play a role in strangulating the host system by heating inflowing filaments.\footnote{We note that regions of increased density would likely harbour stronger magnetic fields than the ISM average. This is seen in Milky Way observations~\citep[e.g.][]{Crutcher2010ApJ}, and would presumably also arise in gravitationally collapsed clumps and clouds in high-redshift galaxies. While increased magnetic energy densities would yield a larger fraction of CR secondary energy being lost to synchrotron processes, magnetic field strengths are unlikely to be sufficiently strong to compete with inverse-Compton losses until cloud densities exceed $n_{\rm b} \geq 10^6~\text{cm}^{-3}$~\citep[][]{Crutcher2010ApJ, Owen2018MNRAS}. 
Densities of this level only persist in a very small volume fraction of clouds on sub-pc scales~\citep{Draine2011book}, and would not be likely to bear a significant influence on the global astrophysics of the host galaxy considered in this study. Moreover, under such conditions, Coulomb scattering would easily dominate over radiative cooling such that the effect of stronger magnetic fields on CR electron losses would be inconsequential.
}

\subsection{Parametrisation}
\label{sec:parametrisation}

Our calculated CR heating values can be reasonably scaled for both channels, if conditions are comparable. 
This allows us to parameterise our simulation results, to make them more easily applicable to our sample of observed high-redshift galaxies (see section~\ref{sec:application}).
The direct Coulomb CR heating rate is principally governed by the input CR power (i.e. via the star-formation rate $\mathcal{R}_{\rm SF}$ of the system) and the interstellar density, which provides the target for the underlying hadronic interactions.
 As such, it may reasonably be scaled as
\begin{equation}
\label{eq:central_coulomb}
\mathcal{H}_{\rm C, 2} \approx \mathcal{H}_{\rm C, 1} \;\!\left( \frac{\mathcal{R}_{\rm SF, 2}}{\mathcal{R}_{\rm SF, 1}} \right)\;\!\left( \frac{n_2}{n_1}  \right) \ ,
\end{equation}
where reference values are $\mathcal{H}_{\rm C, 1} = 1.7\times10^{-25} \;\text{erg}\; \text{cm}^{-3}\;\!\text{s}^{-1}$,
$\mathcal{R}_{\rm SF, 1} = 16 \; \text{M}_{\odot}\;\!\text{yr}^{-1}$ and
$n_1 = 10\;\!\text{cm}^{-3}$. Subscript `2' denotes the scaled values. 
This requires that $\mathcal{H}_{\rm C, 2}$ and $\mathcal{H}_{\rm C, 1}$ are computed for systems where the ISM is of comparable density such that the CR heating efficiency (see Fig.~\ref{fig:channel_frac}) is largely unchanged -- this would usually be the case for our purposes, given that 
this CR heating mechanism operates within the host galaxy and is unaffected by the variations that might arise in the external circumgalactic environment.

Similarly, the internal IX heating effect can be estimated 
using a parametrised scaling -- however, this requires an additional dependence on redshift. 
The parametrisation in this case would be: 
\begin{equation}
\label{eq:central_xray}
\mathcal{H}_{\rm X, 2} \approx \mathcal{H}_{\rm X, 1}  \;\! \left(\frac{1+z_2}{1+z_{1}}\right)^4\;\!\left( \frac{\mathcal{R}_{\rm SF, 2}}{\mathcal{R}_{\rm SF, 1}} \right)\;\!\left( \frac{n_2}{n_{1}}  \right) \ ,
\end{equation}
where $\mathcal{H}_{\rm X, 1} = 6.4\times 10^{-27}  \;\text{erg}\; \text{cm}^{-3}\;\!\text{s}^{-1}$, $z_1=7$ and other quantities retain their earlier definitions. 
We note that, if the interstellar medium of a system were clumpy rather than relatively smooth, this scaling may not be strictly valid. 
For instance, if the ISM gas was comprised of clumps of around $10^{2}\;\!\text{cm}^{-3}$ interspersed with low-density H\;\!II regions of densities around $10^{-1}\;\!\text{cm}^{-3}$, 
it would be quite feasible for the inverse-Compton X-ray heating to dominate over the Coulomb CR heating at high-redshift in the majority of the volume of the galaxy due to the relative efficiencies of the two processes in different density media (see Fig.~\ref{fig:channel_frac}). 
While it is beyond the scope of the current paper, this effect could have important implications for star-formation and is being investigated in a dedicated study.

For external CR heating, it is only necessary to consider IX heating, as the DC process is confined to the host.
At distances where the circumgalactic medium density is not greatly influenced by the density profile of the host protogalaxy, 
the distance from the host is sufficiently large that
we may consider it to be a point source. 
At such distances, the above parametrisation~\ref{eq:central_xray} additionally has a distance dependence, 
however the circumgalactic medium has been assumed to be uniform and so is normalised to a much lower level (consistent with the profile seen for the external heating power in Fig.~\ref{fig:heating_powers}), as given by
\begin{equation}
\label{eq:outer_xray}
\mathcal{H}_{\rm X, 2} \approx \mathcal{H}_{\rm X, 1} \;\! \left(\frac{1+z_2}{1+z_1}\right)^4\;\!\left( \frac{\mathcal{R}_{\rm SF, 2}}{\mathcal{R}_{\rm SF, 1}} \right)\;\!\left( \frac{r_2}{r_1}  \right)^{-2} \;\!\left(\frac{n_{\rm e, 2}}{n_{\rm e, 1}}\right) \ ,
\end{equation}
where $\mathcal{H}_{\rm X, 1} = 7.9\times 10^{-37} \;\text{erg}\; \text{cm}^{-3}\;\!\text{s}^{-1}$. 
We introduce the distance dependence here, with $r_1 = 100\;\!\text{kpc}$ and $r_2$ as the distance of the required point from the centre of the protogalaxy. Furthermore, we use 
$n_{\rm e, 1} = x_i \;\! n_{\rm 1} = 10^{-2}\;\text{cm}^{-3}$, 
as the number density of electrons, with $x_i$ being the ionisation fraction. 
A check on these parameterisations is presented in Appendix~\ref{sec:param_xcheck}.


\section{Thermodynamics in high-redshift protogalaxies}
\label{sec:application}

\begin{figure}
 	\centering
	\includegraphics[width=0.8\columnwidth]{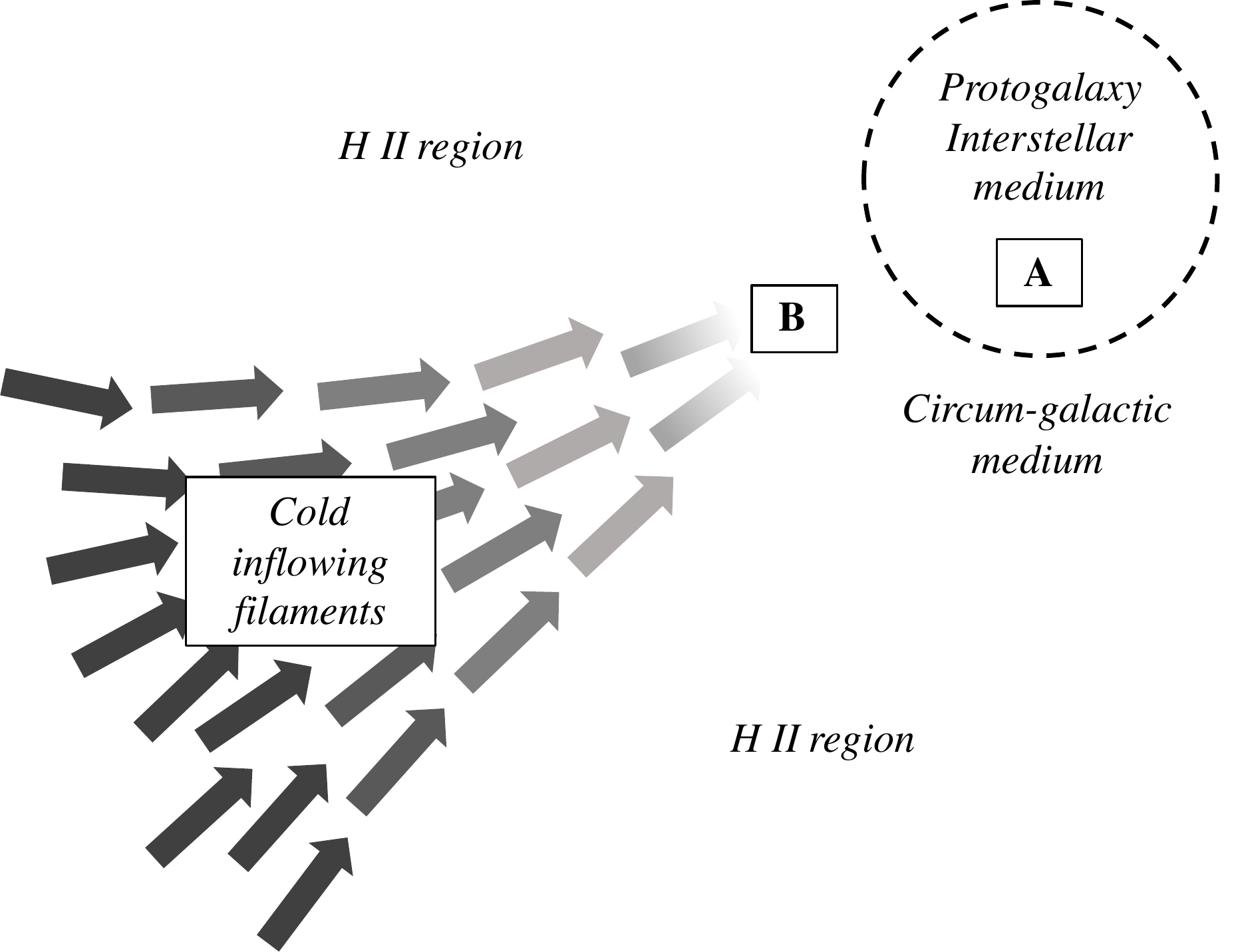}
    \caption{\small Schematic illustration of the CR heating processes operating in and around a starburst protogalaxy.
    At {\textbf A}, internal interstellar heating is dominated by the DC channel as CRs and their secondaries are not able to easily escape from the galactic magnetic field of their host system. This increases thermal gas pressure and leads to quenching~\citep{French2015ApJ}. At {\textbf B}, the IX channel heating mechanism is more important, as this does not rely on CR escape. The X-rays emitted from the galaxy begin to heat and evaporate any cold inflowing filaments, and can do so on a timescale shorter than the inflow time. Such filamentary inflows are required to drive star-formation by providing cold gas which can more easily undergo fragmentation and collapse than ambient interstellar gas~\citep{Ribaudo2011ApJ, Sanchez2014A&ARv}, so their destruction could strangulate the galaxy~\citep{Ribaudo2011ApJ, Sanchez2014A&ARv, Peng2015Natur}.}
    \label{fig:filaments_cartoon}
\end{figure}

The two CR heating channels impact the dynamics and evolution of their host systems and environments in different ways: 
the DC channel can raise the temperature of the ISM gases to a level where star-formation cannot continue because of the increased pressure support provided by the heated gas against gravitational collapse
~\citep[e.g.][]{French2015ApJ}
 -- this may be regarded as a \textit{quenching} mechanism;
the IX channel is more closely attributed to \textit{strangulation} -- cutting off the infall supply of cold gas from its source. If star-formation is driven by the supply of cold gas to the host galaxy through inflowing filaments
\citep[see][]{Ribaudo2011ApJ, Sanchez2014A&ARv}, 
external IX heating could be sufficient to heat them as they approach the galaxy. 
These two processes are illustrated in the context of a galaxy hosting an inflow and active, ongoing star-formation in the schematic in Fig.~\ref{fig:filaments_cartoon}.

\subsection{Feedback processes}
\label{sec:feedback_processes}

\subsubsection{Internal heating and quenching}
\label{sec:internal_heating_and_quenching}
 
The condition for quenching can be defined in terms of the virial temperature, being that required to prevent gravitational collapse. It is defined as:
\begin{equation}
T_{\rm vir} \approx 6.8\times10^4 \left(\frac{{M}}{10^{9}\;\!\text{M}_{\odot}}\right) \;\! \left(\frac{1\;\!\text{kpc}}{r_h}\right) \; \text{K}
\label{eq:virial_temperature}
\end{equation}
\citep{Binney2008_book}, where ${M}$ is the mass of the system and $r_h$ is the characteristic size of the system. For a high-redshift protogalaxy, this would strictly be the mass and size of proto-stellar clouds. However, since observations suggest galactic-wide reservoirs of cool gas being supported against gravitational collapse~\citep{French2015ApJ, Rowlands2015MNRAS, Alatalo2016ApJ}, we may apply the virial theorem to the galaxy as a whole to find the virial temperature for the full system. 
Thus, any system where the average ISM temperature exceeds this virial temperature would certainly be unable to collapse, and quenching would ensue.
With this cut-off temperature $T_{\rm vir}$, we can then assign a quenching timescale $\tau_{\rm Q}$ as the time required for this temperature to be attained by a gas subjected to a given heating rate.
However, the full CR heating power only develops after the magnetic field of the host galaxy has saturated. 
To estimate $\tau_{\rm Q}$, we must take this magnetic field evolution into account such that the quenching time would be the magnetic saturation time $\tau_{\rm mag}$ plus the time taken for entrenched CR heating to raise the ISM temperature above $T_{\rm vir}$. We may estimate the magnetic saturation time as:
\begin{equation}
\tau_{\rm mag} \approx 140\;\left(\frac{\mathcal{R}_{\rm SF}}{16\;\!\text{M}_{\odot}}\right)^{-1}\;\text{Myr}
\label{eq:magnetic_saturation}
\end{equation}
\citep{Schober2013A&A, Owen2018MNRAS}. The time taken for the virial temperature to be reached due to the DC heating channel then follows as
\begin{equation}
\tau_{\rm Q} \approx \tau_{\rm mag} + n_{\rm b} \;\! k_{\rm B} \;\! T_{\rm vir} \;\! \mathcal{H}_{\rm DC}^{-1} \ ,
\label{eq:quenching_timescale}
\end{equation}
where the second term gives the timescale for this temperature to be reached given a CR Coulomb heating power $\mathcal{H}_{\rm DC}$, 
and $n_{\rm b}$ is the local ISM gas density. 
We apply our model to the observed high-redshift galaxies introduced in section~\ref{sec:star_forming_protogalaxies} to find the DC heating rate for each system (column $\mathcal{H}_{\rm DC}$ in Table~\ref{tab:applied_model_table}). From this, we use the above analysis to estimate the associated quenching timescale in Myr, shown by column $\tau_{\rm Q}$ in Table~\ref{tab:applied_model_table}.

\subsubsection{External heating and strangulation}
\label{sec:external_heating_strangulation}

The IX channel operates on larger scales with its effects being principally felt in the circumgalactic medium and the structures therein, where the DC process cannot reach -- 
in particular, inflowing filaments~\cite{Dekel2009Natur, Stewart2013ApJ, Goerdt2015MNRAS, Goerdt2015MNRAS_b} which supply the cold gas required to feed star-formation in the host galaxy~\citep{Dijkstra2006ApJ, Sancisi2008A&ARv, Ribaudo2011ApJ, Sanchez2014A&ARv, Peng2015Natur} via, for example, cold mode accretion~\citep{Birnboim2003MNRAS, Keres2005MNRAS}. 
If sufficiently strong, heating and evaporation of these filaments can lead to to strangulation, if they are prevented from delivering a supply of cold gas to the ISM of the host  -- see Fig.~\ref{fig:filaments_cartoon}.
The power with which the IX heating channel operates depends both on the local density of the filament being irradiated by X-rays, and the ionisation fraction of that filament. 
The ionisation fraction is very uncertain, however we adopt a fiducial value of $x_i = 0.9$. This is in line with suggestions from simulation work (and may even be conservative) which suggest a largely ionised structure beaded with denser, cooler and less ionised clumps (see~\citealt{Fumagalli2011MNRAS, FaucherGiguere2011MNRAS}). 
We may estimate the characteristic density of a filament using mass continuity, if assuming that inflowing filaments are the principal driver of the star-formation in a protogalaxy at high-redshift
~\citep[e.g][]{Keres2005MNRAS, Fumagalli2011MNRAS, Sanchez2014A&ARv}.
The star-forming efficiency of the supplied gas from these inflows is around 30\% (i.e. $\epsilon = 0.3$) -- a value roughly consistent with simulations of high-redshift starbursts\footnote{See~\citealt{Sun2016MNRAS}, which finds star-formation efficiencies of tens of per cent for halo masses of above $10^{11}~\text{M}_{\odot}$, corresponding to stellar masses of around $10^{9}~\text{M}_{\odot}$, comparable to the systems used in the sample in this work -- see also~\citealt{Behroozi2015ApJ}, and~\citealt{Meier2002AJ, Turner2015Natur} which find similar efficiencies in the young starburst NGC 5253 that may also be indicative of the levels to be expected in the types of systems we are modelling.}.
We set the characteristic inflow velocity to be $v_{\rm Ly\alpha}\approx 400~\text{km}~\text{s}^{-1}$~\citep[based on the Lyman-$\alpha$ velocity offset observed in][]{Hashimoto2018Nat} 
and adopt a total covering fraction of the inflow(s) into the host system of 
10\% (this is towards the conservative end of the range plausible values, with literature estimates varying between 5-40\% -- see, e.g.~\citealt{Ribaudo2011ApJ, Fumagalli2011MNRAS, FaucherGiguere2011MNRAS}).
This yields an effective inflow radius of around 30\% of the radius of the host galaxy.
From this prescription, we may estimate the number density in the inflow filament as:
\begin{equation}
n_{\rm b} \approx \frac{\mathcal{R}_{\rm SF}}{\epsilon\;\!v_{\rm Ly\alpha}\;\!\sigma_{\rm f}\;\!m_{\rm p}} \ ,
\label{eq:filament_density}
\end{equation}
where $\sigma_{\rm f} = \pi r_{\rm f}^2$ is its approximate cross sectional area. 
This gives reasonable values within the expected range, that is $n_{\rm b} \approx 20~\text{cm}^{-3}$ for a system with $\mathcal{R}_{\rm SF} = 16~\text{M}_{\odot}\; \text{yr}^{-1}$ (values of the order $n_{\rm b}\approx 10^{-1}-10^{1}~\text{cm}^{-3}$ would arguably be sensible -- see, e.g.\citealt{Keres2009MNRAS, Ceverino2010MNRAS, Goerdt2012MNRAS, Goerdt2015MNRAS, Falgarone2017Natur}).
To analyse the impact of IX heating on the dynamics, we now require a suitable approximation for the virial temperature of a cold filament:
consider the temperature required to prevent a segment of filamentary cloud from collapsing. This will give the minimum temperature above which the cloud begins to disperse (giving a lower limit for the strangulation timescale for a given heating rate). Thus, we consider a spherical region of cloud with radius $r_{\rm f} = \sqrt{0.1}\;\! r_{\rm gal}$ 
(this follows from the 10\% covering fraction of the filament), 
and use this to estimate the mass of a filamentary region, $M_{\rm fil} \approx 4\pi r_{\rm f}^3 n_{\rm b} m_{\rm p}/3$. 
For the example considered above, with $\mathcal{R}_{\rm SF} = 16~\text{M}_{\odot}\; \text{yr}^{-1}$, this would indicate a filamentary region mass of around $6\times10^{7}~\text{M}_{\odot}$. 
By equation~\ref{eq:virial_temperature}, the corresponding virial temperature is $T_{\rm vir}\approx 1.2 \times 10^4\;\text{K}$. 
This is roughly consistent with the expectation for the temperature range of these structures (i.e between  $10^4-10^5~\text{K}$ -- see~\citealt{Dekel2006MNRAS, Dekel2009Natur}).
The strangulation timescale is calculated in a similar manner to the quenching timescale in the previous section, i.e the sum of the magnetic containment time (required for the X-ray glow to develop) and the heating timescale: $\tau_{\rm S} = \tau_{\rm mag} + \tau_{\rm IX}$.
The IX heating timescale $\tau_{\rm IX} = x_i \; n_{\rm b} \;  k_{\rm B} \;  T_{\rm vir} \;\! / \;\! {\overline{\mathcal{H}}_{\rm IX}}$, 
where $n_{\rm b}$ is the filament number density such that and $x_i\;\!n_{\rm b}$ gives the number density of electrons, as required for X-ray scattering driven heating.
We have introduced $\overline{\mathcal{H}}_{\rm IX}$ as the effective IX heating rate throughout the filament. 
The true IX heating power along the length of the filament would be subject to the inverse square law along its length, meaning its value would vary substantially. 
However, adopting an effective heating power for the inflow allows us to approximate the timescales associated with the heating process. 
This is estimated as follows:
$\overline{\mathcal{H}}_{\rm IX}$ varies as $r^{-2}$ away from the galaxy. 
Thus, consider a point $a$ at which the total (integrated) heating rate up to $r=a$ is equal to that above $r=a$, being an effective midpoint:
\begin{equation}
\int_{r_{\rm in}}^{a}r^{-2}\;{\rm d}r = \int_{a}^{R}r^{-2}\;{\rm d}r \ ,
\end{equation} 
where $r_{\rm in}$ is the innermost radius of the filament, and $R$ is its outermost extent. Taking $R = 50~\text{kpc}$ 
beyond which inflows are not clearly evident in modelling with simulations
~\citep[e.g.][]{Dekel2009Natur, Stewart2013ApJ, Goerdt2015MNRAS}, and $r_{\rm in} = 1~\text{kpc}$ gives $a = 1.96~\text{kpc}$. 
For an inflowing filament, the gas would experience equal amounts of heating on either side of $a$. 
Here, the proximity of $a$ to $r_{\rm in}$ suggests a rapid evaporation of the inflow filament as it nears the galaxy, with much more gentle heating arising further out. 
Using the earlier parameterisations (see section~\ref{sec:parametrisation}), the characteristic IX heating power at $a$ is estimated 
by applying an inverse-square law scaling to equation~\ref{eq:central_xray}, 
yielding\footnote{Given the continuous density profile along the filament and into the ISM, a scaling of the internal IX heating parametrisation was considered more meaningful than a re-scaling of the external IX heating power, which would be substantially lower due to the foreground density profile of the galaxy no longer being important at distances above 20 kpc.}
$\overline{\mathcal{H}}_{\rm IX} = \mathcal{H}_{\rm IX}\vert_{a} \approx 3.0\times 10^{-27}~\text{erg}\;\text{cm}^{-3}\;\text{s}^{-1}$ for a system with $\mathcal{R}_{\rm SF} = 16~\text{M}_{\odot}\;\text{yr}^{-1}$ and $n_{\rm b} = 20~\text{cm}^{-3}$ at $z=7$. 
This gives a strangulation timescale of around 200 Myr (when also accounting for the magnetic containment time) which, for an inflow traversing a distance of $50~\text{kpc}$ at a velocity of $400\;\text{km}\;\text{s}^{-1}$, compares to an infall timescale of over 100 Myr. 
The filament's dynamical timescale is around 
$\tau_{\rm dyn} \approx 20~\text{Myr}$, 
so this paints a picture of the cold inflowing filament being destroyed by the X-ray glow of the host galaxy, providing a reasonable expectation that the system could be strangulated by the IX heating process. 
We apply these calculations to determine the associated timescales for the observed high-redshift systems in Table~\ref{tab:applied_model_table}. 

\begin{table*}
	\centering
	\resizebox{0.9\textwidth}{!}{\begin{tabular}{l||cccc|cc|cc|cc}
		&&&&&&&&&& \\[-0.5em]
		\multirow{ 2}{*}{Galaxy ID} & \multirow{ 2}{*}{$z$} & $\mathcal{R}_{\rm SB}^{*}$ & $t_{\rm SB}$ & $t_{\rm qui}$ & \multirow{ 2}{*}{$\log\left[\dfrac{ \mathcal{H}_{\rm DC}}{\text{ /erg cm}^{-3}\;\!\text{ s}^{-1}}\right]^c$} &  $\tau_{\rm Q}$ & \multirow{ 2}{*}{$\log\left[\dfrac{ \overline{\mathcal{H}}_{\rm IX}}{\text{ /erg cm}^{-3}\;\!\text{ s}^{-1}}\right]^e$} & $\tau_{\rm S}$ & $\tau_{\rm cool}$ & $\tau_{\rm dyn}$ \\
		&&$/\text{M}_{\odot}~\text{yr}^{-1}$& /Myr$^a$ & /Myr$^b$ & & /Myr$^d$ & & /Myr$^f$ & /Myr$^g$ & /Myr \\
		&&&&&&&&&&\\ \hline
		 \multicolumn{11}{c}{\sc Quenched Post-Starbursts, High Redshift}\\ \hline
		 &&&&&&&&&& \\[-0.75em]
		A1689-zD1 & $7.60$ & 4.9 & 350 & 81 &						$-25.28$ & 510 & $-27.07$ & 					540 & 76 & 13  \\ 
		&&&&&&&&&& \\[-0.75em]
		UDF-983-964 & $\textit{7.3}^{\it +0.4}_{\it -0.3}$ & 7.6 & 290 & 170 &	$-25.09$ & 350 & $-27.43$ & 				390 & 100 & 26 \\ 
		&&&&&&&&&& \\[-0.75em]
		GNS-zD2 & $\textit{7.1}^{\it +1.5}_{\it -0.6}$ & 10 & 240 &	250 &		$-24.97$  & 280 & $-27.09$ & 				330 & 120 & 15 \\ 
		&&&&&&&&&& \\[-0.75em]
		MACS1149-JD1 & 9.11 & 11 & 100 & 290 &							$-24.93$  & 210 & $-27.68$ & 				250 & 43 & 23 \\ 
		&&&&&&&&&& \\[-0.75em]
		CDFS-3225-4627 & $\textit{7.1}^{\it +1.5}_{\it -0.5}$ & 17 & 210 & 280 &		$-24.74$  & 180 &$-26.63$ & 		240 & 140 & 16 \\ 
		&&&&&&&&&& \\[-0.75em]
		UDF-3244-4727 & $\textit{7.9}^{\it +0.8}_{\it -0.6}$ & 35 & 	80 &	320 &	$-24.43$  & 80 & $-26.02$ & 				140 & 110 & 24 \\ 
		&&&&&&&&&& \\[-0.75em]
		HDFN-3654-1216 & $\textit{6.3}^{\it +0.2}_{\it -0.2}$ & 36 & 190 & 430 &		$-24.42$  & 83 & $-26.86$ & 			230 & 130 & 38 \\  
		&&&&&&&&&& \\[-0.75em]
		GNS-zD3 & $\textit{7.3}^{\it +0.9}_{\it -0.4}$ & 38 & 110 & 350 &			$-24.39$  & 79 & $-26.11$ & 				160 & 140 & 22 \\ 
		&&&&&&&&&& \\[-0.75em]
		UDF-640-1417 & $\textit{6.9}^{\it +0.1}_{\it -0.1}$ & 47 & 140 &	380 &	$-24.30$  & 67 & $-26.25$ & 					170 & 150 & 26 \\  
		&&&&&&&&&& \\[-0.75em]
		GNS-zD4 & $\textit{7.2}^{\it +0.4}_{\it -0.2}$ & 57 & 120 &	360 &		$-24.22$ & 57 & $-26.00$ & 					140 & 150 & 19 \\  
		&&&&&&&&&& \\[-0.75em]
		GNS-zD1 & $\textit{7.2}^{\it +0.2}_{\it -0.2}$  & 63 & 120 &	 360 &		$-24.17$  & 59 & $-25.67$ & 					140 & 190 & 12 \\ 
		&&&&&&&&&& \\[-0.75em]
		GNS-zD5 & $\textit{7.3}^{\it +0.2}_{\it -0.2}$ & 110 & 110 & 350 &			$-23.93$ & 40 & $-25.25$ & 				120 & 230 & 14 \\  
		&&&&&&&&&&\\ \hline
		 \multicolumn{11}{c}{\sc Starbursts, High Redshift}\\ \hline
		 &&&&&&&&&& \\[-0.75em]
		GN-z11 & $11.1$ &  $24^{+10}_{-10}$ & 42*&	-- &				$-24.59$ & 110 & $-25.07$ & 					120 & 110 & 11 \\  
		&&&&&&&&&& \\[-0.75em]
		EGS-zs8-1 & $7.73$ & $79^{+47}_{-29}$ & 100* &		-- &				$-24.08$ & 34 & $-26.35$ & 				110 & 110 & 23 \\  
		&&&&&&&&&& \\[-0.75em]
		GN-108036 & $7.21$ & $100^{+5.0}_{-2.0}$ & 5.8* &	-- &				$-23.97$ & 23 & $-25.30$ & 					120 & 51 & 61 \\  
		&&&&&&&&&& \\[-0.75em]
		SXDF-NB1006-2 & $7.21$ & $350^{+170}_{-280}$ & 1.0* &		-- &		$-23.43$ & 6.4 & $-26.02$ & 				110 & 14 & 13 \\   
	\end{tabular}}
	\caption{\small Derived information for high-redshift quenched post-starburst and starburst systems, as introduced in Tables~\ref{tab:high_z_galaxies_table_post_sb} and~\ref{tab:high_z_galaxies_table_sb}. Columns for redshift $z$, star-formation rate during burst $\mathcal{R}^{*}_{\rm SB}$, length of starburst period $t_{\rm SB}$ and dynamical timescale $\tau_{\rm dyn}$ as shown previously and included here to aid comparisons between systems and timescales. This further introduces estimates for $t_{\rm qui}$, the time for which a post-starburst galaxy has remained relatively quiescent after its initial violent star-formation episode, heating rate via direct Coulomb CR heating in the interstellar medium of the host, $\mathcal{H}_{\rm DC}$, heating rate via inverse-Compton X-ray heating of the inflowing filaments at an estimated average level, $\overline{\mathcal{H}}_{\rm IX}$, estimated timescales for star-formation quenching $\tau_{\rm Q}$ and strangulation $\tau_{\rm S}$, and cooling timescale for the inflowing gas, $\tau_{\rm cool}$. Notes: \\
	$^{a}$\!\; Starburst time, $t_{\rm SB}$: estimated time over which the starburst phase has endured -- asterisk * indicates starburst is still ongoing.
	$^{b}$\!\;Quiescence timescale, $t_{\rm qui}$: estimated time over which the galaxy has maintained a low rate of star-formation, as determined from the age of the stellar population and starburst timescale. For our purposes, it is sufficient to estimate this as the stellar population ages, such that $t_{\rm qui} \approx \tau_{*}$ from Tables~\ref{tab:high_z_galaxies_table_post_sb} and~\ref{tab:high_z_galaxies_table_sb}.  \\
	$^{c}$\!\;Coulomb heating rate, $\mathcal{H}_{\rm DC}$:  Decimal log of Coulomb heating rate in the interstellar medium responsible for cosmic ray quenching activity. \\
	$^{d}$\!\;Quenching timescale, $\tau_{\rm Q}$:   Quenching timescale, estimated from the virial temperature of the host galaxy. \\
	$^{e}$\!\;Characteristic filament inverse-Compton heating rate, $\overline{\mathcal{H}}_{\rm IX}$:  Decimal log of the characteristic  inverse-Compton X-ray heating rate in the filament structures, as responsible for strangulating star-formation.\\
	$^{f}$\!\;Strangulation timescale $\tau_{\rm S}$:  Strangulation timescale, estimated from the virial temperature of the inflowing filaments (not of the total system). This should be regarded as an upper limit, and is highly model-dependent. \\
	$^{g}$\!\;Cooling timescale, $\tau_{\rm cool}$:  Cooling timescale of the circumgalactic gases, being the minimum time required for star-formation to re-ignite after quenching or strangulation.}
	\label{tab:applied_model_table}
\end{table*}

\subsection{Quiescence, cooling and inflow}
\label{sec:quiescence_and_cooling}

For star-formation to resume in a galaxy after quenching and strangulation has occurred,
either the heated interstellar gases must cool sufficiently, 
or inflowing cold filaments must be allowed to resume their supply of gas into the system. 
The timescale for the cooling of the ambient gases can be estimated using equation~\ref{eq:cooling_timescale}.
Quenching by CR heating can be sustained for some time after the cessation of burst activity in the host. 
Typically, CRs will be absorbed or will escape from the galaxy by diffusion. 
Those absorbed will contribute to the heating effect, otherwise the CRs can escape within a timescale of $\tau_{\rm diff}\approx \ell^2/4 D \approx 3~\text{Myr}$. 
Their heating cannot be sustained at a level required for quenching beyond a few Myr. 
If the CRs heat interstellar gas to above the virial temperature of the host, with an ISM density of around $10~\text{cm}^{-3}$, from ~\ref{eq:cooling_timescale} we estimate that the gas should cool and star-formation would resume in less than a Myr. 
Given that the quiescent periods for all the post-starburst systems in Table~\ref{tab:applied_model_table} are substantially more than a Myr (generally of order 100 Myr), the cooling of interstellar gas would not enable the reinstatement of star-forming activity. 
Indeed, the extreme processes at play during the starburst episode may have driven away much of the gas reservoir into the galactic halo by for example the action of starburst-driven outflows.

The discrepancy in the timescales indicates that fresh cold gas is required to rejuvenate the galaxy, and this can only be achieved when circumgalactic conditions allow for the reinstatement of cold filamentary inflows which would likely have been dispersed and evaporated by the processes driving strangulation during the starburst episode. 
This requires the cooling of the heated circumgalactic gas, which can occur after its irradiation by inverse-Compton X-rays from the host galaxy has declined -- again, this is driven by CRs in the galaxy, and will reduce a few Myr after the end of the starburst phase, when CRs have had time to diffuse away. 
Cooling of the circumgalactic gas can arise on a timescale of 100 Myr, if adopting a characteristic circumgalactic medium density of $10^{-2}~\text{cm}^{-3}$. 
Such circumgalactic flows may advect recycled circumgalactic gas with them, as well as ambient magnetic fields and cosmic rays. 
This could have important implications for future star-forming activities, with such components providing enhanced pressure support against gravitational collapse, or higher metallicities in recycled gas promoting the opposite. 
Fresh primordial gas could return to feeding the galaxy within a comparable timescale, if returning from a distance of 50 kpc with an inflow velocity of around $400~\text{km}~\text{s}^{-1}$ as before.

\subsection{Discussion}
\label{sec:timescales_discussion}

 \begin{figure*}
 \centering
	\includegraphics[width=0.8\textwidth]{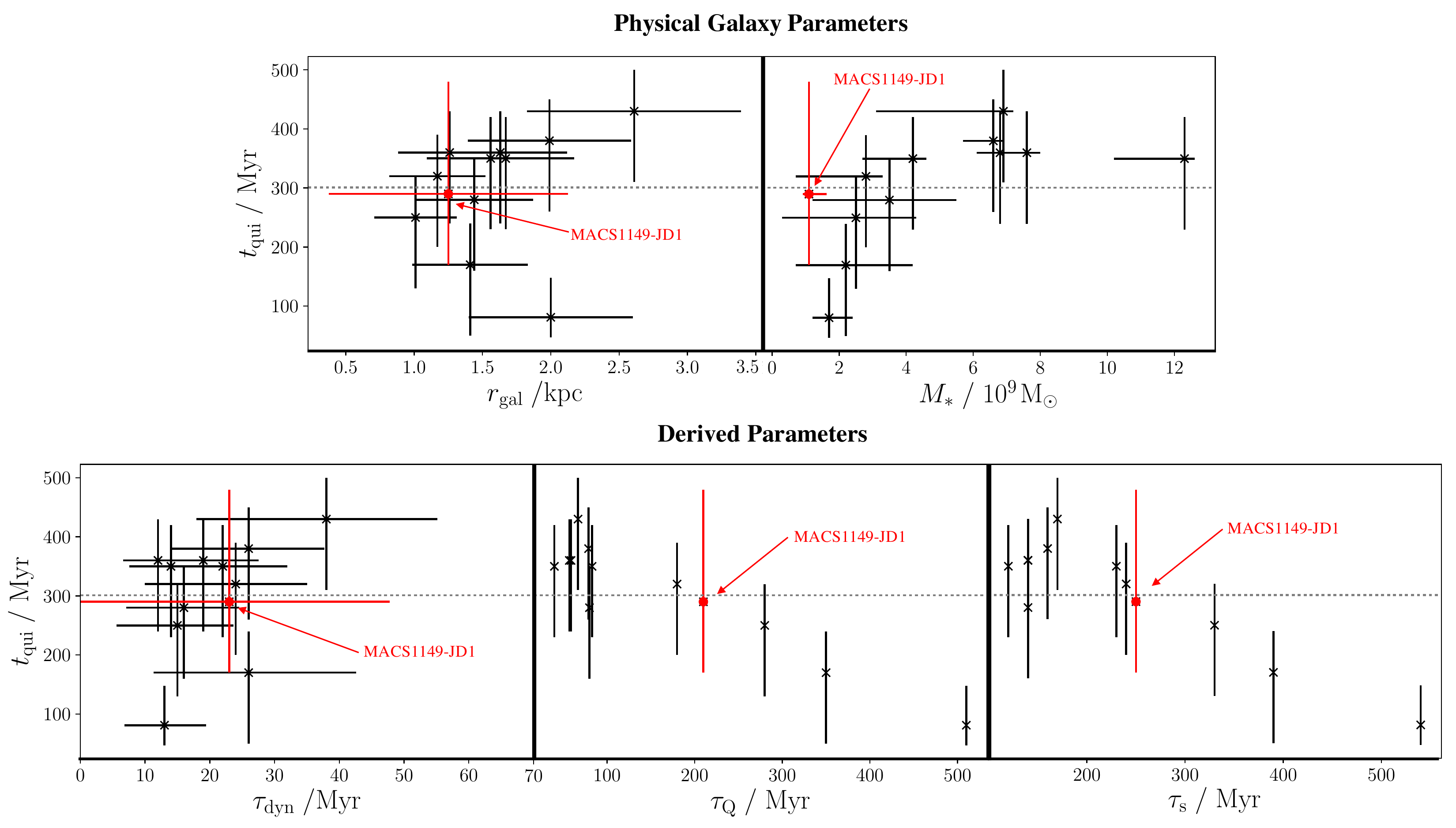}
    \caption{\small Quiescence timescales against generic physical (upper row) and derived (lower row) galactic parameters. The data of {MACS1149-JD1} are highlighted in red (this galaxy is at much higher redshift than the others in the sample, but does not otherwise appear to be unusual). The value of $t_{\rm qui}$ averaged over all data points is indicated by the horizontal dashed grey line. 
    {\bf Upper Left}: Plot against estimated galaxy radius. This indicates that the quiescent time scale is also independent of the size of the system, with a weighted Pearson correlation coefficient of 0.15. Error bars in $r_{\rm gal}$ are as quoted in the literature (A1689-zD1 at 50\% and MACS1149-JD1 at 71\%), otherwise estimated as 30\%.
 {\bf Upper Right}: Plot against stellar mass, which shows a positive correlation (coefficient of 0.58).
 {\bf Lower Left}: Plot against galaxy dynamical timescales, which $t_{\rm qui}$ appears to be independent of, with a correlation of 0.06.
 {\bf Lower Centre}: Plot against quenching timescale, $\tau_{\rm Q}$, which shows a negative correlation (coefficient of -0.81). This is consistent with $t_{\rm qui}$ being a marker for the strength of feedback -- the longer taken to quench star-formation, the weaker the CR driven feedback and the shorter the quiescence timescale. Uncertainties in $\tau_{\rm Q}$ are unclear and so are not plotted.
 {\bf Lower Right}: As per the centre panel, but against the strangulation timescale, $\tau_{\rm S}$. Again the negative correlation (of coefficient -0.78) is consistent with $t_{\rm qui}$ being influenced by feedback with shorter values for longer strangulation times. As with $\tau_{\rm Q}$, uncertainties in $\tau_{\rm S}$ are not clear so are omitted.}
    \label{fig:tqui_all_data}
\end{figure*}

The timescales in Table~\ref{tab:applied_model_table} suggest that CR feedback operates in all these systems via both the direct quenching and strangulation mechanisms, with some competition between them.
This is evident from the similarity between the starburst timescales, $t_{\rm SB}$ and the quenching and strangulation timescales, $\tau_{\rm Q}$ and $\tau_{\rm S}$.
We consider three scenarios below:

\textbf{Low $\mathcal{R}_{\rm SB}^{*}$, $t_{\rm SB} \leq \left\{ \tau_{\rm Q}, \tau_{\rm S} \right\}$:} At low star-formation rates (e.g. those seen in {A1689-zD1} and {UDF-983-964}), both $\tau_{\rm S}$ and $\tau_{\rm Q}$ are dominated by the magnetic saturation timescale, $\tau_{\rm mag}$. 
This would arise over 440 Myr for {A1689-zD1}, and 320 Myr for {UDF-983-964} (see equation~\ref{eq:magnetic_saturation}), with quenching only possible at times close to or after this.
The low rates of star-formation might be attributed to a lack of cold gas supply to the galaxies (e.g. by inflows) and suggest it would thus be relatively easy to heat the local reservoir of gas, leading to quenching. 
If suitably cool gas only persisted in pockets throughout these systems, or if gas were of relatively high temperature (near the virial temperature) so as to cool only enough to yield these relatively low star-formation rates, 
only a moderate amount of CR heating would then be required for quenching. 
This could plausibly be achieved even before magnetic saturation had completed (with the associated full CR heating power) and would account for the starburst period being shorter  even than the estimated quenching timescale.

{\bf Moderate $\mathcal{R}_{\rm SB}^{*}$, $t_{\rm SB} \simeq \tau_{\rm Q}$:} At moderate star-formation rates, $\tau_{\rm mag}$ is less important to $t_{\rm SB}$ (according to equation~\ref{eq:magnetic_saturation}).
In these systems, the quenching timescale $\tau_{\rm Q}$ and starburst time $t_{\rm SB}$ are reasonably consistent, suggesting they are more dependent on their cold gas reservoirs to form stars, with the quenching mechanism described in this paper operating. Before star-formation can be stopped, much of the cold gas reservoir must be heated by CRs after magnetic saturation. Strangulation is less important here but operates over longer timescales to maintain quenching during the quiescent period.

{\bf High $\mathcal{R}_{\rm SB}^{*}$, $t_{\rm SB} \simeq \tau_{\rm S}$:} For systems with high star-formation rates, the collapse of interstellar gas reservoirs may be insufficient to maintain star-forming activity at the estimated levels for a prolonged starburst period (with the quenching timescales being very short).
Systems like {GNS-zD1} and {GNS-zD5} would likely instead rely on the injection of cold gas by filamentary inflows to sustain their extreme starburst.
But the continued supply of cold gas allows star-formation to persist under strong CR heating of the ISM, until strangulation can cut off this external gas supply too. 
This would account for the similarity between $t_{\rm SB}$ and $\tau_{\rm S}$ in these cases.

After the starburst phase, quiescence would persist until the processes driving star-formation resume.
The reasonable consistency between the cooling timescale $\tau_{\rm cool}$ (the timescale over which circumgalactic gas cools and collapses) and the quiescence timescale\footnote{
This lower limit for $t_{\rm qui}$ is comparable to the stellar population age $\tau^{*}$ as the age of the stars would correspond to their age by the end of the starburst plus the quiescent time. A more appropriate value for this quantity could be estimated by modelling the star-formation rate \textit{within} the starburst episode, but this is beyond the scope of this paper and  deserving of a dedicated study.
}
$t_{\rm qui} \approx \tau^{*}$ 
indicates that star-formation resumes only when cold filamentary inflows are reinstated (after the X-ray heating driven by CRs has subsided).

The quiescent timescale,
$t_{\rm qui}$ can be regarded as a measure of the degree of star-formation feedback in the system, rather than simply the time elapsed since the end of the starburst episode (see Appendix~\ref{sec:sf_data_app}). 
This may be the case if post-starburst galaxies only tend to exceed detection thresholds when they resume star-formation, so the high-redshift post-starburst galaxies we would observe are preferentially those which have emerged from their quiescent phase.
Indeed, this would be supported by the two stellar populations found in MACS1149-JD1, where the youngest population is attributed to the observed lower star-formation rate, and is only a few Myr old -- see~\citealt{Hashimoto2018Nat} for details. 
Consider the relation of $t_{\rm qui}$ on system parameters: in Fig.~\ref{fig:tqui_all_data} 
we show how it relates to physical galaxy parameters (size $r_{\rm gal}$ and stellar mass $M_{*}$) as well as the dynamical timescale of each system $\tau_{\rm dyn}$, the quenching timescale $\tau_{\rm Q}$ and the strangulation timescale $\tau_{\rm S}$. Here (and subsequently), we have highlighted the high-redshift system MACS1149-JD1, but we note that it appears to be unremarkable compared to its lower-redshift companions in our analysis.

\begin{figure}
 \centering
	\includegraphics[width=0.9\columnwidth]{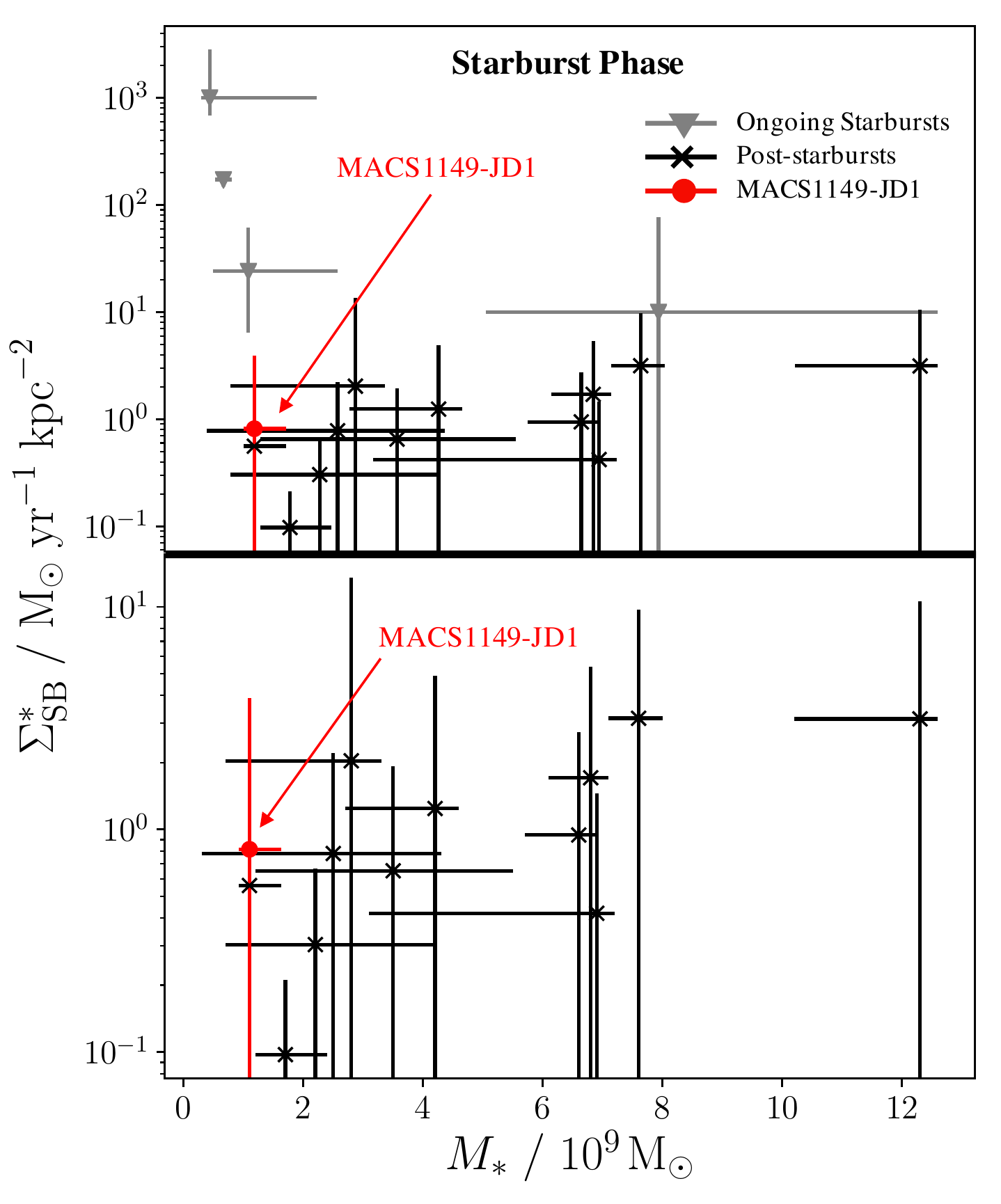}
    \caption{\small Star-formation rate surface density plotted against galaxy stellar mass, indicating these quantities are moderately correlated (weighted Pearson coefficient of 0.71 for the post-starburst systems). {\bf Upper Panel}: $\Sigma_{\rm SB}^{*}$ for the four galaxies observed during their starburst phase (see Table~\ref{tab:high_z_galaxies_table_sb}) are plotted in grey, with the post-starbursts are shown in black (see Table~\ref{tab:high_z_galaxies_table_post_sb}). We plot the MACS1149-JD1 in red (this system is observed at a much higher redshift than the others in the sample, but its behaviour appears to be relatively consistent). {\bf Lower Panel}: As above, with the removal of the starbursts to more clearly show the distribution of the post-starburst galaxies.}
    \label{fig:sfr_data_burst_compare}
\end{figure}

\begin{figure}
 \centering
	\includegraphics[width=0.9\columnwidth]{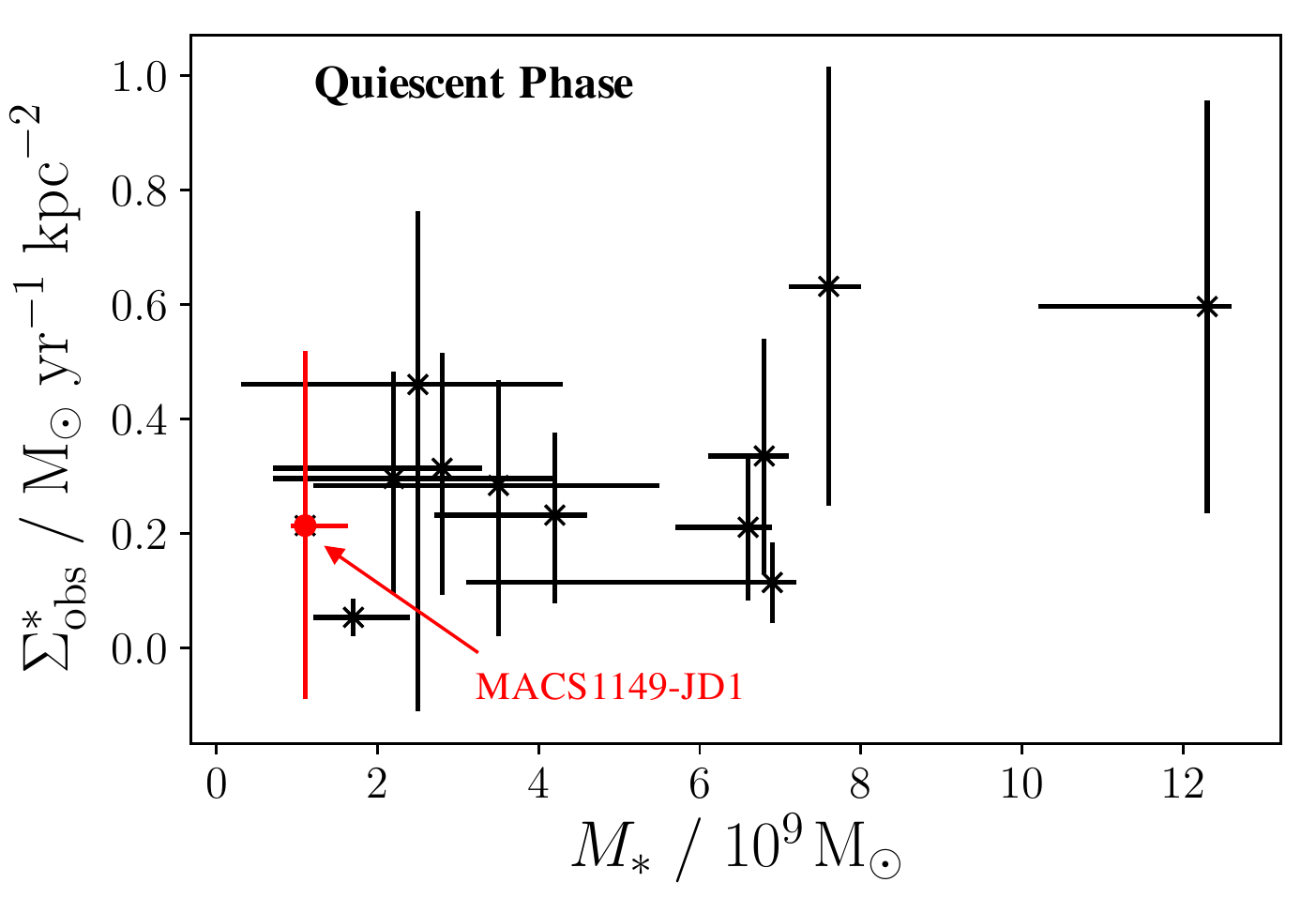}
    \caption{\small Observed star-formation rate surface densities for the post-starburst sample plotted against stellar mass. This shows a moderate correlation (weighted Pearson coefficient of 0.55). Again, the high redshift system MACS1149-JD1 is highlighted in red, but does not appear to be inconsistent with the rest of the sample.}
    \label{fig:sfr_mass_data}
\end{figure}

These panels show a positive correlation between stellar mass and $t_{\rm qui}$ (Pearson coefficient of 0.58), as well as negative correlations with $\tau_{\rm Q}$ (-0.81) and $\tau_{\rm S}$ (-0.78). 
The latter two negative correlations are relatively strong and show that, for longer quenching or strangulation timescales (i.e. when the feedback effect would be weakest), $t_{\rm qui}$ is shorter -- this is consistent with $t_{\rm qui}$ being an effective tracer for the level of feedback experienced by a system.
The positive correlation between $t_{\rm qui}$ and stellar mass $M_{*}$ is also consistent with feedback, given that the systems with higher stellar mass appear to harbour greater star-formation rates during their starburst episode when normalised by the galaxy size:
Fig.~\ref{fig:sfr_data_burst_compare} shows such a dependence of $\Sigma^{*}_{\rm SB}$ on stellar mass, while the quiescent star-formation rate appears to simply scale with that during the burst -- see Fig.~\ref{fig:sfr_mass_data}. 
There is no notable correlation between $t_{\rm qui}$ and galaxy size $r_{\rm gal}$ or dynamical timescale $\tau_{\rm dyn}$ but, in a feedback model, no such correlation would be expected between these parameters.
This shows there is no strong dependency of $t_{\rm qui}$ on extensive system quantities, only intensive ones relating to the level of feedback, being $\tau_{\rm Q}$ and $\tau_{\rm S}$.
Such results are consistent with the quiescence being strongly influenced by feedback from star-formation.

The additional four systems observed with ongoing starburst activity are also shown in grey in the upper panel of Fig.~\ref{fig:sfr_data_burst_compare}. These weakly indicate some inverse trend with stellar mass, as would result from the depletion of gas supplies as strangulation takes hold -- however it is difficult to draw strong conclusions with this small data set and large uncertainties.

\section{Summary}
\label{sec:conclusions}

In this study, we have shown the power that can be attained by CR heating in high-redshift starburst galaxies,
and how the underlying energy-transfer process in CR heating is split according to two distinctive channels: 
one in which the secondary electrons arising from the hadronic interactions of CR protons undergo Coulomb scattering
to thermalise into their medium over a length-scale of around 0.2 kpc, 
and the other in which the radiative emission from these secondary electrons (namely inverse-Compton scattering, predominantly with CMB photons) drives a strong X-ray glow which itself can cause a heating effect of the host galaxy and its circumgalactic environment. 
The second of these channels is particularly important in high-redshift environments, 
when the energy density of the CMB was substantially higher than it is in the present Universe.

We have demonstrated how these two CR heating channels can result in two distinct forms of CR-induced feedback on star-formation.
The direct Coulomb thermalisation process can readily heat the ISM of the host galaxy, up to levels around $10^{-25}~\text{erg}~\text{cm}^{-3}~\text{s}^{-1}$ in a system exhibiting SN events at a rate of $\mathcal{R}_{\rm SN} = 0.1~\text{yr}^{-1}$ -- this could exceed more conventional feedback mechanisms, for example stellar photons and diffuse X-rays from the hot ISM gases~\citep[see][]{Owen2018MNRAS}.  
We find that, in a sample of high-redshift starburst and post-starburst galaxies, 
such heating would be able to increase the ISM temperature to above its virial temperature (and hence decisively quench star-formation in the host) within a few tens of Myr (when star-formation rates during the starburst phase of the system are above a few tens $\text{M}_{\odot}~\text{yr}^{-1}$).
However, this Coulomb thermalisation process is strongly confined by the magnetic fields of the host galaxy, and is relatively ineffective in the wider circumgalactic environment.
By contrast, the heating due to inverse-Compton X-rays (referred to as indirect X-ray heating) can operate both within and outside the host system.
At $z=12$, corresponding to the earliest times at which star-formation has been inferred~\citep[see][]{Hashimoto2018Nat}, the CR heating via this channel can attain comparable levels to the direct Coulomb thermalisation process within the ISM. After this (i.e. at lower redshifts), it becomes less important within the galaxy, but remains important outside, where X-ray heating can halt the cold inflowing filaments required to feed star-formation in the most active systems.
This can lead to strangulation~\citep[e.g.][]{Peng2015Natur} -- by effectively cutting off the supply of cold gas in inflows, this indirect CR X-ray heating can bring star-formation to a halt within just a few hundred Myr.
Both of these processes are likely to operate in any given star-forming galaxy at high-redshift. We show that, for more intensely star-forming galaxies, timescales would indicate that CR feedback arises preferentially through strangulation. 
In less active systems, quenching arises first (with strangulation operating over a longer period of time).
We note that, at lower redshifts, the indirect X-ray mechanism does not operate as effectively as it does in the early Universe. This is because it becomes dependent on the up-scattering of stellar photons rather than the CMB, and this yields a substantially diminished X-ray glow compared to early-Universe counterparts. 
This promotes the idea that high-redshift galaxies are very different environments to their low-redshift descendents and, on galactic scales, gives some insight into how these systems may have evolved.


\begin{acknowledgements}
      ERO is supported by a UK Science and Technology Facilities Council PhD Studentship and the International Exchange Studentship of National Tsing Hua University (NTHU), where his visit was kindly hosted by Prof. Shih-Ping Lai. ERO also acknowledges support from a Royal Astronomical Society travel grant and the Institute of Physics C R Barber trust fund and Research Student Conference Fund which supported travel to conferences and meetings where discussion helped to inspire and inform this work. NK's visit to the Mullard Space Science Laboratory (MSSL) was supported by a Kyoto Sangyo University visiting summer studentship.
      We thank B. P. Brian Yu (MSSL) and Qin Han (Nanjing University) for critical review of the manuscript, and for constructive comments. We also thank the anonymous referee for their comments and feedback.
\end{acknowledgements}


\bibliographystyle{aa} 
\bibliography{references} 

\begin{thebibliography}{247}
\expandafter\ifx\csname natexlab\endcsname\relax\def\natexlab#1{#1}\fi

\bibitem[{{Abdo} {et~al.}(2010{\natexlab{a}}){Abdo}, {Ackermann}, {Ajello},
  {Allafort}, {Atwood}, {Baldini}, {Ballet}, {Barbiellini}, {Bastieri},
  {Bechtol}, {Bellazzini}, {Berenji}, {Blandford}, {Bloom}, {Bonamente},
  {Borgland}, {Bouvier}, {Brandt}, {Bregeon}, {Brigida}, {Bruel}, {Buehler},
  {Burnett}, {Buson}, {Caliandro}, {Cameron}, {Cannon}, {Caraveo},
  {Casandjian}, {Cecchi}, {{\c C}elik}, {Charles}, {Chekhtman}, {Chiang},
  {Ciprini}, {Claus}, {Cohen-Tanugi}, {Conrad}, {Dermer}, {de Angelis}, {de
  Palma}, {Digel}, {Silva}, {Drell}, {Drlica-Wagner}, {Dubois}, {Favuzzi},
  {Fegan}, {Fortin}, {Frailis}, {Fukazawa}, {Funk}, {Fusco}, {Gargano},
  {Germani}, {Giglietto}, {Giordano}, {Giroletti}, {Glanzman}, {Godfrey},
  {Grenier}, {Grondin}, {Guiriec}, {Gustafsson}, {Hadasch}, {Harding},
  {Hayashi}, {Hayashida}, {Hays}, {Healey}, {Jean}, {J{\'o}hannesson},
  {Johnson}, {Johnson}, {Johnson}, {Kamae}, {Katagiri}, {Kataoka}, {Kerr},
  {Kn{\"o}dlseder}, {Kuss}, {Lande}, {Latronico}, {Lee}, {Lemoine-Goumard},
  {Longo}, {Loparco}, {Lott}, {Lovellette}, {Lubrano}, {Madejski}, {Makeev},
  {Martin}, {Mazziotta}, {Mehault}, {Michelson}, {Mitthumsiri}, {Mizuno},
  {Moiseev}, {Monte}, {Monzani}, {Morselli}, {Moskalenko}, {Murgia},
  {Naumann-Godo}, {Nolan}, {Norris}, {Nuss}, {Ohsugi}, {Okumura}, {Omodei},
  {Orlando}, {Ormes}, {Ozaki}, {Paneque}, {Panetta}, {Parent}, {Pepe},
  {Persic}, {Pesce-Rollins}, {Piron}, {Porter}, {Rain{\`o}}, {Rando},
  {Razzano}, {Reimer}, {Reimer}, {Ritz}, {Romani}, {Sadrozinski}, {Saz
  Parkinson}, {Sgr{\`o}}, {Siskind}, {Smith}, {Smith}, {Spandre}, {Spinelli},
  {Strickman}, {Strigari}, {Strong}, {Suson}, {Takahashi}, {Takahashi},
  {Tanaka}, {Thayer}, {Thompson}, {Tibaldo}, {Torres}, {Tosti}, {Tramacere},
  {Uchiyama}, {Usher}, {Vandenbroucke}, {Vianello}, {Vilchez}, {Vitale},
  {Waite}, {Wang}, {Winer}, {Wood}, {Yang}, \& {Ziegler}}]{Abdo2010A&A-b}
{Abdo}, A.~A., {Ackermann}, M., {Ajello}, M., {et~al.} 2010{\natexlab{a}},
  \aap, 523, L2

\bibitem[{{Abdo} {et~al.}(2010{\natexlab{b}}){Abdo}, {Ackermann}, {Ajello},
  {Atwood}, {Axelsson}, {Baldini}, {Ballet}, {Barbiellini}, {Bastieri},
  {Bechtol}, {Bellazzini}, {Berenji}, {Bloom}, {Bonamente}, {Borgland},
  {Bregeon}, {Brez}, {Brigida}, {Bruel}, {Burnett}, {Caliandro}, {Cameron},
  {Caraveo}, {Casandjian}, {Cavazzuti}, {Cecchi}, {{\c C}elik}, {Charles},
  {Chekhtman}, {Cheung}, {Chiang}, {Ciprini}, {Claus}, {Cohen-Tanugi},
  {Conrad}, {Dermer}, {de Angelis}, {de Palma}, {Digel}, {Silva}, {Drell},
  {Drlica-Wagner}, {Dubois}, {Dumora}, {Farnier}, {Favuzzi}, {Fegan}, {Focke},
  {Foschini}, {Frailis}, {Fukazawa}, {Funk}, {Fusco}, {Gargano}, {Gasparrini},
  {Gehrels}, {Germani}, {Giebels}, {Giglietto}, {Giordano}, {Glanzman},
  {Godfrey}, {Grenier}, {Grondin}, {Grove}, {Guillemot}, {Guiriec}, {Hanabata},
  {Harding}, {Hayashida}, {Hays}, {Hughes}, {J{\'o}hannesson}, {Johnson},
  {Johnson}, {Johnson}, {Kamae}, {Katagiri}, {Kataoka}, {Kawai}, {Kerr},
  {Kn{\"o}dlseder}, {Kocian}, {Kuss}, {Lande}, {Latronico}, {Lemoine-Goumard},
  {Longo}, {Loparco}, {Lott}, {Lovellette}, {Lubrano}, {Madejski}, {Makeev},
  {Mazziotta}, {McConville}, {McEnery}, {Meurer}, {Michelson}, {Mitthumsiri},
  {Mizuno}, {Moiseev}, {Monte}, {Monzani}, {Morselli}, {Moskalenko}, {Murgia},
  {Nakamori}, {Nolan}, {Norris}, {Nuss}, {Ohsugi}, {Omodei}, {Orlando},
  {Ormes}, {Ozaki}, {Paneque}, {Panetta}, {Parent}, {Pelassa}, {Pepe},
  {Pesce-Rollins}, {Piron}, {Porter}, {Rain{\`o}}, {Rando}, {Razzano},
  {Reimer}, {Reimer}, {Reposeur}, {Ritz}, {Rodriguez}, {Romani}, {Roth},
  {Ryde}, {Sadrozinski}, {Sander}, {Saz Parkinson}, {Scargle}, {Sellerholm},
  {Sgr{\`o}}, {Shaw}, {Smith}, {Smith}, {Spandre}, {Spinelli}, {Strickman},
  {Strong}, {Suson}, {Takahashi}, {Tanaka}, {Thayer}, {Thayer}, {Thompson},
  {Tibaldo}, {Tibolla}, {Torres}, {Tosti}, {Tramacere}, {Uchiyama}, {Usher},
  {Vasileiou}, {Vilchez}, {Vitale}, {Waite}, {Wang}, {Winer}, {Wood}, {Ylinen},
  {Ziegler}, \& {Fermi LAT Collaboration}}]{Abdo2010ApJ}
{Abdo}, A.~A., {Ackermann}, M., {Ajello}, M., {et~al.} 2010{\natexlab{b}},
  \apjl, 709, L152

\bibitem[{{Abdo} {et~al.}(2010{\natexlab{c}}){Abdo}, {Ackermann}, {Ajello},
  {Atwood}, {Baldini}, {Ballet}, {Barbiellini}, {Bastieri}, {Baughman},
  {Bechtol}, {Bellazzini}, {Berenji}, {Blandford}, {Bloom}, {Bonamente},
  {Borgland}, {Bregeon}, {Brez}, {Brigida}, {Bruel}, {Burnett}, {Buson},
  {Caliandro}, {Cameron}, {Caraveo}, {Casandjian}, {Cecchi}, {{\c C}elik},
  {Chekhtman}, {Cheung}, {Chiang}, {Ciprini}, {Claus}, {Cohen-Tanugi},
  {Cominsky}, {Conrad}, {Cutini}, {Dermer}, {de Angelis}, {de Palma}, {Digel},
  {Silva}, {Drell}, {Dubois}, {Dumora}, {Farnier}, {Favuzzi}, {Fegan}, {Focke},
  {Fortin}, {Frailis}, {Fukazawa}, {Fusco}, {Gargano}, {Gasparrini}, {Gehrels},
  {Germani}, {Giavitto}, {Giebels}, {Giglietto}, {Giordano}, {Glanzman},
  {Godfrey}, {Gotthelf}, {Grenier}, {Grondin}, {Grove}, {Guillemot}, {Guiriec},
  {Hanabata}, {Harding}, {Hayashida}, {Hays}, {Horan}, {Hughes}, {Jackson},
  {Jean}, {J{\'o}hannesson}, {Johnson}, {Johnson}, {Johnson}, {Johnson},
  {Kamae}, {Katagiri}, {Kataoka}, {Kawai}, {Kerr}, {Kn{\"o}dlseder}, {Kocian},
  {Kuss}, {Lande}, {Latronico}, {Lemoine-Goumard}, {Longo}, {Loparco}, {Lott},
  {Lovellette}, {Lubrano}, {Madejski}, {Makeev}, {Marshall}, {Martin},
  {Mazziotta}, {McConville}, {McEnery}, {Meurer}, {Michelson}, {Mitthumsiri},
  {Mizuno}, {Moiseev}, {Monte}, {Monzani}, {Morselli}, {Moskalenko}, {Murgia},
  {Nolan}, {Norris}, {Nuss}, {Ohsugi}, {Omodei}, {Orlando}, {Ormes}, {Paneque},
  {Parent}, {Pelassa}, {Pepe}, {Pesce-Rollins}, {Piron}, {Porter}, {Rain{\`o}},
  {Rando}, {Razzano}, {Reimer}, {Reimer}, {Reposeur}, {Ritz}, {Rodriguez},
  {Romani}, {Roth}, {Ryde}, {Sadrozinski}, {Sanchez}, {Sander}, {Saz
  Parkinson}, {Scargle}, {Sellerholm}, {Sgr{\`o}}, {Siskind}, {Smith}, {Smith},
  {Spandre}, {Spinelli}, {Starck}, {Strickman}, {Strong}, {Suson}, {Tajima},
  {Takahashi}, {Tanaka}, {Thayer}, {Thayer}, {Thompson}, {Tibaldo}, {Torres},
  {Tosti}, {Tramacere}, {Uchiyama}, {Usher}, {Vasileiou}, {Venter}, {Vilchez},
  {Vitale}, {Waite}, {Wang}, {Weltevrede}, {Winer}, {Wood}, {Ylinen}, \&
  {Ziegler}}]{Abdo2010A&A}
{Abdo}, A.~A., {Ackermann}, M., {Ajello}, M., {et~al.} 2010{\natexlab{c}},
  \aap, 512, A7

\bibitem[{{Abdo} {et~al.}(2010{\natexlab{d}}){Abdo}, {Ackermann}, {Ajello},
  {Baldini}, {Ballet}, {Barbiellini}, {Bastieri}, {Bechtol}, {Bellazzini},
  {Berenji}, {Blandford}, {Bloom}, {Bonamente}, {Borgland}, {Bouvier},
  {Brandt}, {Bregeon}, {Brez}, {Brigida}, {Bruel}, {Buehler}, {Buson},
  {Caliandro}, {Cameron}, {Caraveo}, {Carrigan}, {Casandjian}, {Cecchi}, {{\c
  C}elik}, {Charles}, {Chekhtman}, {Cheung}, {Chiang}, {Ciprini}, {Claus},
  {Cohen-Tanugi}, {Conrad}, {Dermer}, {de Palma}, {Digel}, {Silva}, {Drell},
  {Dubois}, {Dumora}, {Favuzzi}, {Fegan}, {Fukazawa}, {Funk}, {Fusco},
  {Gargano}, {Gasparrini}, {Gehrels}, {Germani}, {Giglietto}, {Giordano},
  {Giroletti}, {Glanzman}, {Godfrey}, {Grenier}, {Grondin}, {Grove}, {Guiriec},
  {Hadasch}, {Harding}, {Hayashida}, {Hays}, {Horan}, {Hughes}, {Jean},
  {J{\'o}hannesson}, {Johnson}, {Johnson}, {Kamae}, {Katagiri}, {Kataoka},
  {Kerr}, {Kn{\"o}dlseder}, {Kuss}, {Lande}, {Latronico}, {Lee},
  {Lemoine-Goumard}, {Llena Garde}, {Longo}, {Loparco}, {Lovellette},
  {Lubrano}, {Makeev}, {Martin}, {Mazziotta}, {McEnery}, {Michelson},
  {Mitthumsiri}, {Mizuno}, {Monte}, {Monzani}, {Morselli}, {Moskalenko},
  {Murgia}, {Nakamori}, {Naumann-Godo}, {Nolan}, {Norris}, {Nuss}, {Ohsugi},
  {Okumura}, {Omodei}, {Orlando}, {Ormes}, {Panetta}, {Parent}, {Pelassa},
  {Pepe}, {Pesce-Rollins}, {Piron}, {Porter}, {Rain{\`o}}, {Rando}, {Razzano},
  {Reimer}, {Reimer}, {Reposeur}, {Ripken}, {Ritz}, {Romani}, {Sadrozinski},
  {Sander}, {Saz Parkinson}, {Scargle}, {Sgr{\`o}}, {Siskind}, {Smith},
  {Smith}, {Spandre}, {Spinelli}, {Strickman}, {Strong}, {Suson}, {Takahashi},
  {Takahashi}, {Tanaka}, {Thayer}, {Thayer}, {Thompson}, {Tibaldo}, {Torres},
  {Tosti}, {Tramacere}, {Uchiyama}, {Usher}, {Vandenbroucke}, {Vasileiou},
  {Vilchez}, {Vitale}, {Waite}, {Wang}, {Winer}, {Wood}, {Yang}, {Ylinen}, \&
  {Ziegler}}]{Abdo2010A&A-a}
{Abdo}, A.~A., {Ackermann}, M., {Ajello}, M., {et~al.} 2010{\natexlab{d}},
  \aap, 523, A46

\bibitem[{{Abe} {et~al.}(1990){Abe}, {Amidei}, {Apollinari}, {Ascoli}, {Atac},
  {Auchincloss}, {Baden}, {Barbaro-Galtieri}, {Barnes}, {Bedeschi}, {Belforte},
  {Bellettini}, {Bellinger}, {Bensinger}, {Beretvas}, {Berge}, {Bertolucci},
  {Bhadra}, {Binkley}, {Blair}, {Blocker}, {Bofill}, {Booth}, {Brandenburg},
  {Brown}, {Byon}, {Byrum}, {Campbell}, {Carey}, {Carithers}, {Carlsmith},
  {Carroll}, {Cashmore}, {Cervelli}, {Chadwick}, {Chapin}, {Chiarelli},
  {Chinowsky}, {Cihangir}, {Cline}, {Connor}, {Contreras}, {Cooper},
  {Cordelli}, {Curatolo}, {Day}, {Delfabbro}, {dell'orso}, {Demortier},
  {Devlin}, {Dibitonto}, {Diebold}, {Dittus}, {Divirgilio}, {Elias}, {Ely},
  {Errede}, {Esposito}, {Flaugher}, {Focardi}, {Foster}, {Franklin}, {Freeman},
  {Frisch}, {Fukui}, {Garfinkel}, {Giannetti}, {Giokaris}, {Giromini},
  {Gladney}, {Gold}, {Goulianos}, {Grosso-Pilcher}, {Haber}, {Hahn}, {Handler},
  {Harris}, {Hauser}, {Hessing}, {Hollebeek}, {Hu}, {Hubbard}, {Hurst}, {Huth},
  {Jensen}, {Johnson}, {Joshi}, {Kadel}, {Kamon}, {Kanda}, {Kardelis},
  {Karliner}, {Kearns}, {Kephart}, {Kesten}, {Keutelian}, {Kim}, {Kirsch},
  {Kondo}, {Kruse}, {Kuhlmann}, {Laasanen}, {Li}, {Liss}, {Lockyer},
  {Marchetto}, {Markeloff}, {Markosky}, {McIntyre}, {Menzione}, {Meyer},
  {Mikamo}, {Miller}, {Mimashi}, {Miscetti}, {Mishina}, {Miyashita}, {Mondal},
  {Mori}, {Morita}, {Mukherjee}, {Newman-Holmes}, {Nodulman}, {Paoletti},
  {Para}, {Patrick}, {Phillips}, {Piekarz}, {Plunkett}, {Pondrom}, {Proudfoot},
  {Punzi}, {Quarrie}, {Ragan}, {Redlinger}, {Rhoades}, {Rimondi}, {Ristori},
  {Rohaly}, {Roodman}, {Sansoni}, {Sard}, {Scarpine}, {Schlabach}, {Schmidt},
  {Schoessow}, {Schub}, {Schwitters}, {Scribano}, {Segler}, {Sekiguchi},
  {Sestini}, {Shapiro}, {Sheaff}, {Shibata}, {Shochet}, {Siegrist}, {Sinervo},
  {Skarha}, {Smith}, {Snider}, {St.~Denis}, {Stefanini}, {Takaiwa}, {Takikawa},
  {Tarem}, {Theriot}, {Tollestrup}, {Tonelli}, {Tsay}, {Ukegawa}, {Underwood},
  {Vidal}, {Wagner}, {Wagner}, {Walsh}, {Watts}, {Webb}, {Westhusing}, {White},
  {Wicklund}, {Williams}, {Yamanouchi}, {Yamashita}, {Yasuoka}, {Yeh}, {Yoh},
  \& {Zetti}}]{Abe1990PhRvD}
{Abe}, F., {Amidei}, D., {Apollinari}, G., {et~al.} 1990, \prd, 41, 2330

\bibitem[{{Acero} {et~al.}(2009){Acero}, {Aharonian}, {Akhperjanian}, {Anton},
  {Barres de Almeida}, {Bazer-Bachi}, {Becherini}, {Behera}, {Bernl{\"o}hr},
  {Bochow}, {Boisson}, {Bolmont}, {Borrel}, {Brucker}, {Brun}, {Brun},
  {B{\"u}hler}, {Bulik}, {B{\"u}sching}, {Boutelier}, {Chadwick},
  {Charbonnier}, {Chaves}, {Cheesebrough}, {Chounet}, {Clapson}, {Coignet},
  {Dalton}, {Daniel}, {Davids}, {Degrange}, {Deil}, {Dickinson},
  {Djannati-Ata{\"i}}, {Domainko}, {Drury}, {Dubois}, {Dubus}, {Dyks}, {Dyrda},
  {Egberts}, {Emmanoulopoulos}, {Espigat}, {Farnier}, {Fegan}, {Feinstein},
  {Fiasson}, {F{\"o}rster}, {Fontaine}, {F{\"u}{\ss}ling}, {Gabici}, {Gallant},
  {G{\'e}rard}, {Gerbig}, {Giebels}, {Glicenstein}, {Gl{\"u}ck}, {Goret},
  {G{\"o}ring}, {Hauser}, {Hauser}, {Heinz}, {Heinzelmann}, {Henri}, {Hermann},
  {Hinton}, {Hoffmann}, {Hofmann}, {Hofverberg}, {Hoppe}, {Horns},
  {Jacholkowska}, {de Jager}, {Jahn}, {Jung}, {Katarzy{\'n}ski}, {Katz},
  {Kaufmann}, {Kerschhaggl}, {Khangulyan}, {Kh{\'e}lifi}, {Keogh}, {Klochkov},
  {Klu{\'z}niak}, {Kneiske}, {Komin}, {Kosack}, {Kossakowski}, {Lamanna},
  {Lenain}, {Lohse}, {Marandon}, {Martineau-Huynh}, {Marcowith}, {Masbou},
  {Maurin}, {McComb}, {Medina}, {M{\'e}hault}, {Moderski}, {Moulin},
  {Naumann-Godo}, {de Naurois}, {Nedbal}, {Nekrassov}, {Nicholas}, {Niemiec},
  {Nolan}, {Ohm}, {Olive}, {Wilhelmi}, {Orford}, {Ostrowski}, {Panter},
  {Arribas}, {Pedaletti}, {Pelletier}, {Petrucci}, {Pita}, {P{\"u}hlhofer},
  {Punch}, {Quirrenbach}, {Raubenheimer}, {Raue}, {Rayner}, {Reimer}, {Renaud},
  {Rieger}, {Ripken}, {Rob}, {Rosier-Lees}, {Rowell}, {Rudak}, {Rulten},
  {Ruppel}, {Sahakian}, {Santangelo}, {Schlickeiser}, {Sch{\"o}ck}, {Schwanke},
  {Schwarzburg}, {Schwemmer}, {Shalchi}, {Sikora}, {Skilton}, {Sol}, {Stawarz},
  {Steenkamp}, {Stegmann}, {Stinzing}, {Superina}, {Szostek}, {Tam},
  {Tavernet}, {Terrier}, {Tibolla}, {Tluczykont}, {van Eldik}, {Vasileiadis},
  {Venter}, {Venter}, {Vialle}, {Vincent}, {Vivier}, {V{\"o}lk}, {Volpe},
  {Wagner}, {Ward}, {Zdziarski}, \& {Zech}}]{Acero2009Sci}
{Acero}, F., {Aharonian}, F., {Akhperjanian}, A.~G., {et~al.} 2009, Science,
  326, 1080

\bibitem[{Ackermann {et~al.}(2013)Ackermann, Ajello, Allafort, Baldini, Ballet,
  Barbiellini, Baring, Bastieri, Bechtol, Bellazzini, Blandford, Bloom,
  Bonamente, Borgland, Bottacini, Brandt, Bregeon, Brigida, Bruel, Buehler,
  Busetto, Buson, Caliandro, Cameron, Caraveo, Casandjian, Cecchi, {\c C}elik,
  Charles, Chaty, Chaves, Chekhtman, Cheung, Chiang, Chiaro, Cillis, Ciprini,
  Claus, Cohen-Tanugi, Cominsky, Conrad, Corbel, Cutini,
  D{\textquoteright}Ammando, de~Angelis, de~Palma, Dermer, do~Couto~e Silva,
  Drell, Drlica-Wagner, Falletti, Favuzzi, Ferrara, Franckowiak, Fukazawa,
  Funk, Fusco, Gargano, Germani, Giglietto, Giommi, Giordano, Giroletti,
  Glanzman, Godfrey, Grenier, Grondin, Grove, Guiriec, Hadasch, Hanabata,
  Harding, Hayashida, Hayashi, Hays, Hewitt, Hill, Hughes, Jackson, Jogler,
  J{\'o}hannesson, Johnson, Kamae, Kataoka, Katsuta, Kn{\"o}dlseder, Kuss,
  Lande, Larsson, Latronico, Lemoine-Goumard, Longo, Loparco, Lovellette,
  Lubrano, Madejski, Massaro, Mayer, Mazziotta, McEnery, Mehault, Michelson,
  Mignani, Mitthumsiri, Mizuno, Moiseev, Monzani, Morselli, Moskalenko, Murgia,
  Nakamori, Nemmen, Nuss, Ohno, Ohsugi, Omodei, Orienti, Orlando, Ormes,
  Paneque, Perkins, Pesce-Rollins, Piron, Pivato, Rain{\`o}, Rando, Razzano,
  Razzaque, Reimer, Reimer, Ritz, Romoli, S{\'a}nchez-Conde, Schulz, Sgr{\`o},
  Simeon, Siskind, Smith, Spandre, Spinelli, Stecker, Strong, Suson, Tajima,
  Takahashi, Takahashi, Tanaka, Thayer, Thayer, Thompson, Thorsett, Tibaldo,
  Tibolla, Tinivella, Troja, Uchiyama, Usher, Vandenbroucke, Vasileiou,
  Vianello, Vitale, Waite, Werner, Winer, Wood, Wood, Yamazaki, Yang, \&
  Zimmer}]{Ackermann2013Sci}
Ackermann, M., Ajello, M., Allafort, A., {et~al.} 2013, Science, 339, 807

\bibitem[{Aharonian {et~al.}(2012)Aharonian, Bykov, Parizot, Ptuskin, \&
  Watson}]{Aharonian2012SSR}
Aharonian, F., Bykov, A., Parizot, E., Ptuskin, V., \& Watson, A. 2012, \ssr,
  166, 97

\bibitem[{{Aharonian} \& {Atoyan}(2000)}]{Aharonian2000A&A}
{Aharonian}, F.~A. \& {Atoyan}, A.~M. 2000, \aap, 362, 937

\bibitem[{{Alatalo} {et~al.}(2016){Alatalo}, {Lisenfeld}, {Lanz}, {Appleton},
  {Ardila}, {Cales}, {Kewley}, {Lacy}, {Medling}, {Nyland}, {Rich}, \&
  {Urry}}]{Alatalo2016ApJ}
{Alatalo}, K., {Lisenfeld}, U., {Lanz}, L., {et~al.} 2016, \apj, 827, 106

\bibitem[{{Albajar} {et~al.}(1990){Albajar}, {Albrow}, {Allkofer}, {Andrieu},
  {Ankoviak}, {Apsimon}, {Astbury}, {Aubert}, {Bacci}, {Bacon}, {Bains},
  {Bauer}, {Beingessner}, {Bettini}, {Bezaguet}, {Biddulph}, {Bohn},
  {B{\"o}hrer}, {Bonino}, {Bos}, {Botlo}, {Buschbeck}, {Busetto}, {Caner},
  {Casoli}, {Cavanna}, {Cennini}, {Centro}, {Ceradini}, {Charlton}, {Ciapetti},
  {Cittolin}, {Clayton}, {Cline}, {Colas}, {Colas}, {Conte}, {Coughlan}, {Cox},
  {Dau}, {Debrion}, {Degiorgi}, {Della Negra}, {Demoulin}, {Denegri}, {Dibon},
  {Diciaccio}, {Diez Hedo}, {Dobrzynski}, {Dorenbosch}, {Dowell}, {Eggert},
  {Eisenhandler}, {Ellis}, {Erhard}, {Faissner}, {Fensome}, {Ferrando},
  {Fincke-Keeler}, {Fortson}, {Fuess}, {Garvey}, {Geer}, {Geiser}, {Ghiglino},
  {Giraud-Heraud}, {Givernaud}, {Gonidec}, {Gregory}, {Haynes}, {Holthuizen},
  {Ikeda}, {Jank}, {Jimack}, {Jorat}, {Joyce}, {Kalmus}, {Karim{\"a}ki},
  {Keeler}, {Kenyon}, {Kernan}, {Khan}, {Kienzle}, {Kinnunen}, {Krammer},
  {Kroll}, {Kryn}, {Lacava}, {Lammel}, {Landon}, {Leuchs}, {Levegr{\"u}n},
  {Lindgren}, {Linglin}, {Lipa}, {Markou}, {Markytan}, {Marquina}, {Maurin},
  {McMahon}, {Mendiburu}, {Meneguzzo}, {Merlo}, {Meyer}, {Moers}, {Mohammadi},
  {Morgan}, {Moser}, {Moulin}, {Mours}, {Muller}, {Naumann}, {Nedelec},
  {Nikitas}, {Nisati}, {Norton}, {O'Dell}, {Pancheri}, {Pauss}, {Petrolo},
  {Piano Mortari}, {Pietarinen}, {Pimi{\"a}}, {Placci}, {Porte}, {Preischl},
  {Prosi}, {Radermacher}, {Redelberger}, {Reithler}, {Revol}, {Robinson},
  {Rodrigo}, {Rohlf}, {Rubbia}, {Sajot}, {Salvini}, {Sass}, {Samyn},
  {Schinzel}, {Schr{\"o}der}, {Schwartz}, {Scott}, {Seez}, {Shah}, {Siotis},
  {Smith}, {Sphicas}, {Stubenrauch}, {Sumorok}, {Szoncso}, {Taurok}, {Taylor},
  {Ten Have}, {Tether}, {Thompson}, {Tscheslog}, {Tuominiemi}, {van de Guchte},
  {van Dijk}, {Veneziano}, {Vialle}, {Virdee}, {von Schlippe}, {Vrana},
  {Vuillemin}, {Wacker}, {Walzel}, {Wingerter}, {Wu}, {Wulz}, {Yvert},
  {Zaccardelli}, {Zacharov}, {Zanello}, {Zotto}, \& {UA1
  Collaboration}}]{Albajar1990NuPhB}
{Albajar}, C., {Albrow}, M.~G., {Allkofer}, O.~C., {et~al.} 1990, Nuclear
  Physics B, 335, 261

\bibitem[{{Albini} {et~al.}(1976){Albini}, {Capiluppi}, {Giacomelli}, \&
  {Rossi}}]{Albini1976NCimA}
{Albini}, E., {Capiluppi}, P., {Giacomelli}, G., \& {Rossi}, A.~M. 1976, Nuovo
  Cimento A Serie, 32, 101

\bibitem[{{Alexopoulos} {et~al.}(1998){Alexopoulos}, {Anderson}, {Biswas},
  {Bujak}, {Carmony}, {Erwin}, {Gutay}, {Hirsch}, {Hojvat}, {Kenney},
  {Lindsey}, {Losecco}, {Matinyan}, {Morgan}, {Oh}, {Porile}, {Scharenberg},
  {Stringfellow}, {Thompson}, {Turkot}, {Walker}, \&
  {Wang}}]{Alexopoulos1998PhLB}
{Alexopoulos}, T., {Anderson}, E.~W., {Biswas}, N.~N., {et~al.} 1998, Physics
  Letters B, 435, 453

\bibitem[{Allard {et~al.}(2007)Allard, Parizot, \& Olinto}]{Allard2007APh}
Allard, D., Parizot, E., \& Olinto, A. 2007, Astroparticle Physics, 27, 61

\bibitem[{Almeida {et~al.}(1968)Almeida, Rushbrooke, Scharenguivel, Behrens,
  Blobel, Borecka, Dehne, Dfaz, Knies, Schmitt, Str\"omer, \&
  Swanson}]{Almeida1968PR}
Almeida, S.~P., Rushbrooke, J.~G., Scharenguivel, J.~H., {et~al.} 1968, Phys.
  Rev., 174, 1638

\bibitem[{{Alner} {et~al.}(1985){Alner}, {Alpg{\aa}rd}, {Anderer}, {Ansorge},
  {{\AA}sman}, {B{\"o}ckmann}, {Booth}, {Burow}, {Carlson}, {Chevalley},
  {Declercq}, {Dewolf}, {Eckart}, {Ekspong}, {Evangelou}, {Eyring}, {Fabre},
  {French}, {Fr{\"o}bel}, {Gauden}, {Geich-Gimbel}, {Gijsen}, {Von Holt},
  {Hospes}, {Johnson}, {Jon-And}, {Kokott}, {Langer}, {Meinke}, {M{\"u}ller},
  {Munday}, {Ovens}, {Rushbrooke}, {Schmickler}, {Triantis}, {Van Hamme},
  {Walck}, {Ward}, {Ward}, {White}, {Wilquet}, {Yamdagni}, \& {UA5
  Collaboration}}]{Alner1985PhLB}
{Alner}, G.~J., {Alpg{\aa}rd}, K., {Anderer}, P., {et~al.} 1985, Physics
  Letters B, 160, 193

\bibitem[{{Alner} {et~al.}(1984){Alner}, {Alpg{\aa}rd}, {Ansorge}, {{\AA}sman},
  {B{\"o}ckmann}, {Booth}, {Burow}, {Carlson}, {Chevalley}, {DeClercq},
  {DeWolf}, {Eckart}, {Ekspong}, {Evangelou}, {Eyring}, {Fabre}, {French},
  {Gaudaen}, {Geich-Gimbel}, {Gijsen}, {von Holt}, {Hospes}, {Johnson},
  {Jon-And}, {Kokott}, {Mackenzi}, {Maggs}, {Meinke}, {M{\"u}ller}, {Munday},
  {Ovens}, {Rushbrooke}, {Saarikko}, {Saarikko}, {Triantis}, {Walck}, {Ward},
  {Ward}, {White}, {Wilquet}, {Yamdagni}, \& {UA5
  Collaboration}}]{Alner1984PhLB}
{Alner}, G.~J., {Alpg{\aa}rd}, K., {Ansorge}, R.~E., {et~al.} 1984, Physics
  Letters B, 138, 304

\bibitem[{Ansorge {et~al.}(1989)}]{Ansorge1988ZPhys}
Ansorge, R.~E. {et~al.} 1989, Z. Phys., C43, 357

\bibitem[{{Aoyama} {et~al.}(2017){Aoyama}, {Hou}, {Shimizu}, {Hirashita},
  {Todoroki}, {Choi}, \& {Nagamine}}]{Aoyama2017MNRAS}
{Aoyama}, S., {Hou}, K.-C., {Shimizu}, I., {et~al.} 2017, \mnras, 466, 105

\bibitem[{{Arnison} {et~al.}(1983){Arnison}, {Astbury}, {Aubert}, {Bacci},
  {Bernabei}, {B{\'e}zaguet}, {B{\"o}ck}, {Bowcock}, {Calvetti}, {Catz},
  {Centro}, {Ceradini}, {Cittolin}, {Cnops}, {Cochet}, {Colas}, {Corden},
  {Dallman}, {D'Angelo}, {DeBeer}, {Della Negra}, {Demoulin}, {Denegri},
  {DiBitonto}, {Dobrzynski}, {Dowell}, {Edwards}, {Eggert}, {Eisenhandler},
  {Ellis}, {Erhard}, {Faissner}, {Fontaine}, {Fournier}, {Frey},
  {Fr{\"u}hwirth}, {Garvey}, {Geer}, {Ghesqui{\`e}re}, {Ghez}, {Giboni},
  {Gibson}, {Giraud-Heraud}, {Givernaud}, {Gonidec}, {Grayer}, {Gutierrez},
  {Haidan}, {Hansl-Kozanecka}, {Haynes}, {Hertzberger}, {Hodges}, {Hoffmann},
  {Hoffmann}, {Holthuizen}, {Homer}, {Honma}, {Jank}, {Kalmus}, {Karim{\"a}ki},
  {Keeler}, {Kenyon}, {Kernan}, {Kinnunen}, {Kowalski}, {Kozanecki}, {Kryn},
  {Lacava}, {Laugier}, {Lees}, {Lehmann}, {Leuchs}, {L{\'e}vq{\circ}ue},
  {Linglin}, {Locci}, {Markiewicz}, {Maurin}, {McMahon}, {Mendiburu}, {Minard},
  {Moricca}, {Muller}, {Nandi}, {Naumann}, {Norton}, {Orkin-Lecourtois},
  {Paoluzi}, {Piano Mortari}, {Pimia}, {Placci}, {Radermacher}, {Ransdell},
  {Reithler}, {Rich}, {Rijssenbeek}, {Roberts}, {Rubbia}, {Sadoulet}, {Sajot},
  {Salvi}, {Salvini}, {Sass}, {Saudraix}, {Savoy-Navarro}, {Schinzel}, {Scott},
  {Shah}, {Skimming}, {Spiro}, {Strauss}, {Sumorok}, {Szoncso}, {Tao},
  {Thompson}, {Timmer}, {Tscheslog}, {Tuominiemi}, {Vialle}, {Vrana},
  {Vuillemin}, {Wahl}, {Watkins}, {Wilson}, {Yvert}, {Zurfluh}, \& {UA1
  Collaboration}}]{Arnison1983PhLB}
{Arnison}, G., {Astbury}, A., {Aubert}, B., {et~al.} 1983, Physics Letters B,
  123, 108

\bibitem[{Atoyan \& Dermer(2003)}]{Dermer2003ApJ}
Atoyan, A.~M. \& Dermer, C.~D. 2003, \apj, 586, 79

\bibitem[{Baldini {et~al.}(1987)Baldini, Flaminio, Moorhead, \&
  Morrison}]{Baldini1987book}
Baldini, A., Flaminio, V., Moorhead, W., \& Morrison, D. 1987, Total
  Cross-Sections for Reactions of High Energy Particles, Landolt-B{\"o}rnstein:
  Numerical Data and Functional Relationships in Science and Technology - New
  Series / Elementary Particles, Nuclei and Atoms (Springer)

\bibitem[{Balsara {et~al.}(2004)Balsara, Kim, Low, \& Mathews}]{Balsara2004ApJ}
Balsara, D.~S., Kim, J., Low, M.-M.~M., \& Mathews, G.~J. 2004, \apj, 617, 339

\bibitem[{Beck {et~al.}(2012)Beck, Lesch, Dolag, Kotarba, Geng, \&
  Stasyszyn}]{Beck2012MNRAS}
Beck, A.~M., Lesch, H., Dolag, K., {et~al.} 2012, \mnras, 422, 2152

\bibitem[{{Begelman}(1995)}]{Begelman1995ASPC}
{Begelman}, M.~C. 1995, in Astronomical Society of the Pacific Conference
  Series, Vol.~80, The Physics of the Interstellar Medium and Intergalactic
  Medium, ed. A.~{Ferrara}, C.~F. {McKee}, C.~{Heiles}, \& P.~R. {Shapiro}, 545

\bibitem[{{Behroozi} \& {Silk}(2015)}]{Behroozi2015ApJ}
{Behroozi}, P.~S. \& {Silk}, J. 2015, \apj, 799, 32

\bibitem[{{Bell}(1978)}]{Bell1978MNRAS}
{Bell}, A.~R. 1978, \mnras, 182, 147

\bibitem[{Bell {et~al.}(2013)Bell, Schure, Reville, \&
  Giacinti}]{Bell2013MNRAS}
Bell, A.~R., Schure, K.~M., Reville, B., \& Giacinti, G. 2013, \mnras, 431, 415

\bibitem[{{Benhabiles-Mezhoud} {et~al.}(2013){Benhabiles-Mezhoud}, {Kiener},
  {Tatischeff}, \& {Strong}}]{Benhanbiles-Mezhoud2013ApJ}
{Benhabiles-Mezhoud}, H., {Kiener}, J., {Tatischeff}, V., \& {Strong}, A.~W.
  2013, \apj, 763, 98

\bibitem[{Berezhko \& Ellison(1999)}]{Berezhko1999ApJ}
Berezhko, E.~G. \& Ellison, D.~C. 1999, \apj, 526, 385

\bibitem[{{Berezinskii} {et~al.}(1990){Berezinskii}, {Bulanov}, {Dogiel}, \&
  {Ptuskin}}]{Berezinskii1990book}
{Berezinskii}, V.~S., {Bulanov}, S.~V., {Dogiel}, V.~A., \& {Ptuskin}, V.~S.
  1990, {Astrophysics of cosmic rays} (Amsterdam: North-Holland)

\bibitem[{Berezinsky {et~al.}(2006)Berezinsky, Gazizov, \&
  Grigorieva}]{Berezinsky2006PRD}
Berezinsky, V., Gazizov, A., \& Grigorieva, S. 2006, \prd, 74, 043005

\bibitem[{Berezinsky \& Gazizov(1993)}]{Berezinsky1993PRD}
Berezinsky, V.~S. \& Gazizov, A.~Z. 1993, \prd, 47, 4206

\bibitem[{{Berezinsky} \& {Grigor'eva}(1988)}]{Berezinsky1988A&A}
{Berezinsky}, V.~S. \& {Grigor'eva}, S.~I. 1988, \aap, 199, 1

\bibitem[{{Bernet} {et~al.}(2008){Bernet}, {Miniati}, {Lilly}, {Kronberg}, \&
  {Dessauges-Zavadsky}}]{Bernet2008Nat}
{Bernet}, M.~L., {Miniati}, F., {Lilly}, S.~J., {Kronberg}, P.~P., \&
  {Dessauges-Zavadsky}, M. 2008, \nat, 454, 302

\bibitem[{Bethe \& Heitler(1934)}]{Bethe1934ProcRSA}
Bethe, H. \& Heitler, W. 1934, Proc. R. Soc. A, 146, 83

\bibitem[{{Bethe} \& {Maximon}(1954)}]{Bethe1954PR}
{Bethe}, H.~A. \& {Maximon}, L.~C. 1954, Physical Review, 93, 768

\bibitem[{Bhattacharjee \& Sigl(2000)}]{Bhattacharjee2000PhysRep}
Bhattacharjee, P. \& Sigl, G. 2000, Physics Reports, 327, 109

\bibitem[{{Bianchi} \& {Schneider}(2007)}]{Bianchi2007MNRAS}
{Bianchi}, S. \& {Schneider}, R. 2007, \mnras, 378, 973

\bibitem[{{Binney} \& {Tremaine}(2008)}]{Binney2008_book}
{Binney}, J. \& {Tremaine}, S. 2008, {Galactic Dynamics: Second Edition}
  (Princeton University Press)

\bibitem[{{Birnboim} \& {Dekel}(2003)}]{Birnboim2003MNRAS}
{Birnboim}, Y. \& {Dekel}, A. 2003, \mnras, 345, 349

\bibitem[{{Blattnig} {et~al.}(2000){Blattnig}, {Swaminathan}, {Kruger}, {Ngom},
  \& {Norbury}}]{BlattnigPRD2000}
{Blattnig}, S.~R., {Swaminathan}, S.~R., {Kruger}, A.~T., {Ngom}, M., \&
  {Norbury}, J.~W. 2000, \prd, 62, 094030

\bibitem[{{Blumenthal}(1970)}]{Blumenthal1970PRD}
{Blumenthal}, G.~R. 1970, \prd, 1, 1596

\bibitem[{Blumenthal \& Gould(1970)}]{Blumenthal1970RMP}
Blumenthal, G.~R. \& Gould, R.~J. 1970, Rev. Mod. Phys., 42, 237

\bibitem[{Borsellino(1947)}]{Borsellino1947INC}
Borsellino, A. 1947, Il Nuovo Cimento (1943-1954), 4, 112

\bibitem[{{Boselli} \& {Gavazzi}(2006)}]{Boselli2006PASP}
{Boselli}, A. \& {Gavazzi}, G. 2006, \pasp, 118, 517

\bibitem[{{Bournaud} \& {Elmegreen}(2009)}]{Bournaud2009ApJ}
{Bournaud}, F. \& {Elmegreen}, B.~G. 2009, \apjl, 694, L158

\bibitem[{{Bouwens} \& {Illingworth}(2006)}]{Bouwens2006Natur}
{Bouwens}, R.~J. \& {Illingworth}, G.~D. 2006, \nat, 443, 189

\bibitem[{{Bouwens} {et~al.}(2004){Bouwens}, {Thompson}, {Illingworth},
  {Franx}, {van Dokkum}, {Fan}, {Dickinson}, {Eisenstein}, \&
  {Rieke}}]{Bouwens2004ApJ}
{Bouwens}, R.~J., {Thompson}, R.~I., {Illingworth}, G.~D., {et~al.} 2004,
  \apjl, 616, L79

\bibitem[{{Bradley} {et~al.}(2008){Bradley}, {Bouwens}, {Ford}, {Illingworth},
  {Jee}, {Ben{\'{\i}}tez}, {Broadhurst}, {Franx}, {Frye}, {Infante}, {Motta},
  {Rosati}, {White}, \& {Zheng}}]{Bradley2008ApJ}
{Bradley}, L.~D., {Bouwens}, R.~J., {Ford}, H.~C., {et~al.} 2008, \apj, 678,
  647

\bibitem[{{Brandenburg} \& {Lazarian}(2013)}]{Brandenburg2013SSRv}
{Brandenburg}, A. \& {Lazarian}, A. 2013, \ssr, 178, 163

\bibitem[{{Breakstone} {et~al.}(1984){Breakstone}, {Campanini}, {Crawley},
  {Dallavalle}, {Deninno}, {Doroba}, {Drijard}, {Fabbri}, {Faessler},
  {Firestone}, {Fischer}, {Frehse}, {Geist}, {Giacomelli}, {Gokieli},
  {Gorbics}, {Gruhn}, {Hanke}, {Heiden}, {Herr}, {Innocenti}, {Kluge}, {Lamsa},
  {Lohse}, {Meyer}, {Mornacchi}, {Nakada}, {Panter}, {Putzer}, {Rauschnabel},
  {Rensch}, {Rimondi}, {Sosnowski}, {Stenlund}, {Symons}, {Szczekowski},
  {Szwed}, {Ullaland}, {Wegener}, \& {Wunsch}}]{Breakstone1984PhRvD}
{Breakstone}, A., {Campanini}, R., {Crawley}, H.~B., {et~al.} 1984, \prd, 30,
  528

\bibitem[{{Caprioli}(2012)}]{Caprioli2012JCAP}
{Caprioli}, D. 2012, \jcap, 7, 038

\bibitem[{{Ceverino} {et~al.}(2010){Ceverino}, {Dekel}, \&
  {Bournaud}}]{Ceverino2010MNRAS}
{Ceverino}, D., {Dekel}, A., \& {Bournaud}, F. 2010, \mnras, 404, 2151

\bibitem[{{Couch} \& {Sharples}(1987)}]{Couch1987MNRAS}
{Couch}, W.~J. \& {Sharples}, R.~M. 1987, \mnras, 229, 423

\bibitem[{{Cowie} \& {Songaila}(1977)}]{Cowie1977Natur}
{Cowie}, L.~L. \& {Songaila}, A. 1977, \nat, 266, 501

\bibitem[{{Cowie} {et~al.}(1996){Cowie}, {Songaila}, {Hu}, \&
  {Cohen}}]{Cowie1996AJ}
{Cowie}, L.~L., {Songaila}, A., {Hu}, E.~M., \& {Cohen}, J.~G. 1996, \aj, 112,
  839

\bibitem[{Crutcher {et~al.}(2010)Crutcher, Wandelt, Heiles, Falgarone, \&
  Troland}]{Crutcher2010ApJ}
Crutcher, R.~M., Wandelt, B., Heiles, C., Falgarone, E., \& Troland, T.~H.
  2010, \apj, 725, 466

\bibitem[{{Dar} \& {de R{\'u}jula}(2008)}]{Dar2008PhR}
{Dar}, A. \& {de R{\'u}jula}, A. 2008, \physrep, 466, 179

\bibitem[{{de Cea del Pozo} {et~al.}(2009){de Cea del Pozo}, {Torres}, \&
  {Rodriguez Marrero}}]{de_Cea_del_Pozo2009ApJ}
{de Cea del Pozo}, E., {Torres}, D.~F., \& {Rodriguez Marrero}, A.~Y. 2009,
  \apj, 698, 1054

\bibitem[{Dehnen(1993)}]{Dehnen1993MNRAS}
Dehnen, W. 1993, \mnras, 265, 250

\bibitem[{{Dekel} \& {Birnboim}(2006)}]{Dekel2006MNRAS}
{Dekel}, A. \& {Birnboim}, Y. 2006, \mnras, 368, 2

\bibitem[{{Dekel} {et~al.}(2009{\natexlab{a}}){Dekel}, {Birnboim}, {Engel},
  {Freundlich}, {Goerdt}, {Mumcuoglu}, {Neistein}, {Pichon}, {Teyssier}, \&
  {Zinger}}]{Dekel2009Natur}
{Dekel}, A., {Birnboim}, Y., {Engel}, G., {et~al.} 2009{\natexlab{a}}, \nat,
  457, 451

\bibitem[{{Dekel} {et~al.}(2009{\natexlab{b}}){Dekel}, {Sari}, \&
  {Ceverino}}]{Dekel2009ApJ}
{Dekel}, A., {Sari}, R., \& {Ceverino}, D. 2009{\natexlab{b}}, \apj, 703, 785

\bibitem[{Dermer \& Menon(2009)}]{Dermer2009book}
Dermer, C.~D. \& Menon, G. 2009, {High energy radiation from black holes: gamma
  rays, cosmic rays, and neutrinos}, Princeton series in astrophysics
  (Princeton Univ. Press: Princeton, NJ)

\bibitem[{{Dermer} \& {Powale}(2013)}]{Dermer2013A&A}
{Dermer}, C.~D. \& {Powale}, G. 2013, \aap, 553, A34

\bibitem[{{Di Fazio} {et~al.}(1979){Di Fazio}, {Occhionero}, \&
  {Vagnetti}}]{DiFazio1979A&A}
{Di Fazio}, A., {Occhionero}, F., \& {Vagnetti}, F. 1979, \aap, 72, 204

\bibitem[{{Dijkstra} {et~al.}(2006){Dijkstra}, {Haiman}, \&
  {Spaans}}]{Dijkstra2006ApJ}
{Dijkstra}, M., {Haiman}, Z., \& {Spaans}, M. 2006, \apj, 649, 37

\bibitem[{{Domingo-Santamar{\'{\i}}a} \&
  {Torres}(2005)}]{Domingo-Santamaria2005A&A}
{Domingo-Santamar{\'{\i}}a}, E. \& {Torres}, D.~F. 2005, \aap, 444, 403

\bibitem[{{Draine}(2011)}]{Draine2011book}
{Draine}, B.~T. 2011, {Physics of the Interstellar and Intergalactic Medium}
  (Princeton University Press)

\bibitem[{{Dressler} \& {Gunn}(1983)}]{Dressler1983ApJ}
{Dressler}, A. \& {Gunn}, J.~E. 1983, \apj, 270, 7

\bibitem[{{Elmegreen} \& {Elmegreen}(2005)}]{Elmegreen2005ApJ}
{Elmegreen}, B.~G. \& {Elmegreen}, D.~M. 2005, \apj, 627, 632

\bibitem[{{Elmegreen} {et~al.}(2004{\natexlab{a}}){Elmegreen}, {Elmegreen}, \&
  {Hirst}}]{Elmegreen2004ApJa}
{Elmegreen}, D.~M., {Elmegreen}, B.~G., \& {Hirst}, A.~C. 2004{\natexlab{a}},
  \apjl, 604, L21

\bibitem[{{Elmegreen} {et~al.}(2009){Elmegreen}, {Elmegreen}, {Marcus},
  {Shahinyan}, {Yau}, \& {Petersen}}]{Elmegreen2009ApJ}
{Elmegreen}, D.~M., {Elmegreen}, B.~G., {Marcus}, M.~T., {et~al.} 2009, \apj,
  701, 306

\bibitem[{{Elmegreen} {et~al.}(2007){Elmegreen}, {Elmegreen}, {Ravindranath},
  \& {Coe}}]{Elmegreen2007ApJ}
{Elmegreen}, D.~M., {Elmegreen}, B.~G., {Ravindranath}, S., \& {Coe}, D.~A.
  2007, \apj, 658, 763

\bibitem[{{Elmegreen} {et~al.}(2004{\natexlab{b}}){Elmegreen}, {Elmegreen}, \&
  {Sheets}}]{Elmegreen2004ApJb}
{Elmegreen}, D.~M., {Elmegreen}, B.~G., \& {Sheets}, C.~M. 2004{\natexlab{b}},
  \apj, 603, 74

\bibitem[{{En{\ss}lin} {et~al.}(2011){En{\ss}lin}, {Pfrommer}, {Miniati}, \&
  {Subramanian}}]{Ensslin2011A&A}
{En{\ss}lin}, T., {Pfrommer}, C., {Miniati}, F., \& {Subramanian}, K. 2011,
  \aap, 527, A99

\bibitem[{{En{\ss}lin} {et~al.}(2007){En{\ss}lin}, {Pfrommer}, {Springel}, \&
  {Jubelgas}}]{Ensslin2007A&A}
{En{\ss}lin}, T.~A., {Pfrommer}, C., {Springel}, V., \& {Jubelgas}, M. 2007,
  \aap, 473, 41

\bibitem[{{Epinat} {et~al.}(2009){Epinat}, {Contini}, {Le F{\`e}vre},
  {Vergani}, {Garilli}, {Amram}, {Queyrel}, {Tasca}, \&
  {Tresse}}]{Epinat2009A&A}
{Epinat}, B., {Contini}, T., {Le F{\`e}vre}, O., {et~al.} 2009, \aap, 504, 789

\bibitem[{{Falgarone} {et~al.}(2017){Falgarone}, {Zwaan}, {Godard}, {Bergin},
  {Ivison}, {Andreani}, {Bournaud}, {Bussmann}, {Elbaz}, {Omont}, {Oteo}, \&
  {Walter}}]{Falgarone2017Natur}
{Falgarone}, E., {Zwaan}, M.~A., {Godard}, B., {et~al.} 2017, \nat, 548, 430

\bibitem[{{Farber} {et~al.}(2018){Farber}, {Ruszkowski}, {Yang}, \&
  {Zweibel}}]{Farber2018ApJ}
{Farber}, R., {Ruszkowski}, M., {Yang}, H.-Y.~K., \& {Zweibel}, E.~G. 2018,
  \apj, 856, 112

\bibitem[{{Faucher-Gigu{\`e}re} \& {Kere{\v
  s}}(2011)}]{FaucherGiguere2011MNRAS}
{Faucher-Gigu{\`e}re}, C.-A. \& {Kere{\v s}}, D. 2011, \mnras, 412, L118

\bibitem[{Federrath {et~al.}(2011)Federrath, Chabrier, Schober, Banerjee,
  Klessen, \& Schleicher}]{Federrath2011PRL}
Federrath, C., Chabrier, G., Schober, J., {et~al.} 2011, \prl, 107, 114504

\bibitem[{{Fermi}(1949)}]{Fermi1949PhRv}
{Fermi}, E. 1949, Phys. Rev., 75, 1169

\bibitem[{{Ferrara} {et~al.}(2016){Ferrara}, {Viti}, \&
  {Ceccarelli}}]{Ferrara2016MNRAS}
{Ferrara}, A., {Viti}, S., \& {Ceccarelli}, C. 2016, \mnras, 463, L112

\bibitem[{{Field} {et~al.}(1969){Field}, {Goldsmith}, \&
  {Habing}}]{Field1969ApJ}
{Field}, G.~B., {Goldsmith}, D.~W., \& {Habing}, H.~J. 1969, \apjl, 155, L149

\bibitem[{{Fields} {et~al.}(2001){Fields}, {Olive}, {Cass{\'e}}, \&
  {Vangioni-Flam}}]{Fields2001A&A}
{Fields}, B.~D., {Olive}, K.~A., {Cass{\'e}}, M., \& {Vangioni-Flam}, E. 2001,
  \aap, 370, 623

\bibitem[{{Fiete Grosse-Oetringhaus} \& {Reygers}(2010)}]{Fiete2010JPhG}
{Fiete Grosse-Oetringhaus}, J. \& {Reygers}, K. 2010, Journal of Physics G
  Nuclear Physics, 37, 083001

\bibitem[{{F{\"o}rster Schreiber} {et~al.}(2009){F{\"o}rster Schreiber},
  {Genzel}, {Bouch{\'e}}, {Cresci}, {Davies}, {Buschkamp}, {Shapiro},
  {Tacconi}, {Hicks}, {Genel}, {Shapley}, {Erb}, {Steidel}, {Lutz},
  {Eisenhauer}, {Gillessen}, {Sternberg}, {Renzini}, {Cimatti}, {Daddi},
  {Kurk}, {Lilly}, {Kong}, {Lehnert}, {Nesvadba}, {Verma}, {McCracken},
  {Arimoto}, {Mignoli}, \& {Onodera}}]{Forster-Schreiber2009ApJ}
{F{\"o}rster Schreiber}, N.~M., {Genzel}, R., {Bouch{\'e}}, N., {et~al.} 2009,
  \apj, 706, 1364

\bibitem[{{F{\"o}rster Schreiber} {et~al.}(2011){F{\"o}rster Schreiber},
  {Shapley}, {Genzel}, {Bouch{\'e}}, {Cresci}, {Davies}, {Erb}, {Genel},
  {Lutz}, {Newman}, {Shapiro}, {Steidel}, {Sternberg}, \&
  {Tacconi}}]{ForsterSchreiber2011ApJ}
{F{\"o}rster Schreiber}, N.~M., {Shapley}, A.~E., {Genzel}, R., {et~al.} 2011,
  \apj, 739, 45

\bibitem[{{French} {et~al.}(2015){French}, {Yang}, {Zabludoff}, {Narayanan},
  {Shirley}, {Walter}, {Smith}, \& {Tremonti}}]{French2015ApJ}
{French}, K.~D., {Yang}, Y., {Zabludoff}, A., {et~al.} 2015, \apj, 801, 1

\bibitem[{{French} {et~al.}(2018){French}, {Zabludoff}, {Yoon}, {Shirley},
  {Yang}, {Smercina}, {Smith}, \& {Narayanan}}]{French2018ApJ}
{French}, K.~D., {Zabludoff}, A.~I., {Yoon}, I., {et~al.} 2018, \apj, 861, 123

\bibitem[{{Fryer}(1999)}]{Fryer1999ApJ}
{Fryer}, C.~L. 1999, \apj, 522, 413

\bibitem[{{Fumagalli} {et~al.}(2011){Fumagalli}, {Prochaska}, {Kasen}, {Dekel},
  {Ceverino}, \& {Primack}}]{Fumagalli2011MNRAS}
{Fumagalli}, M., {Prochaska}, J.~X., {Kasen}, D., {et~al.} 2011, \mnras, 418,
  1796

\bibitem[{Gaggero(2012)}]{Gaggero2012thesis}
Gaggero, D. 2012, Cosmic Ray Diffusion in the Galaxy and Diffuse Gamma
  Emission, Springer Theses (Springer: Berlin)

\bibitem[{{Galli} \& {Padovani}(2015)}]{Galli2015arXiv}
{Galli}, D. \& {Padovani}, M. 2015, ArXiv e-prints [\eprint[arXiv]{1502.03380}]

\bibitem[{{Gnerucci} {et~al.}(2011){Gnerucci}, {Marconi}, {Cresci}, {Maiolino},
  {Mannucci}, {Schreiber}, {Davies}, {Shapiro}, \& {Hicks}}]{Gnerucci2011A&A}
{Gnerucci}, A., {Marconi}, A., {Cresci}, G., {et~al.} 2011, \aap, 533, A124

\bibitem[{{Goerdt} \& {Ceverino}(2015)}]{Goerdt2015MNRAS}
{Goerdt}, T. \& {Ceverino}, D. 2015, \mnras, 450, 3359

\bibitem[{{Goerdt} {et~al.}(2015){Goerdt}, {Ceverino}, {Dekel}, \&
  {Teyssier}}]{Goerdt2015MNRAS_b}
{Goerdt}, T., {Ceverino}, D., {Dekel}, A., \& {Teyssier}, R. 2015, \mnras, 454,
  637

\bibitem[{{Goerdt} {et~al.}(2012){Goerdt}, {Dekel}, {Sternberg}, {Gnat}, \&
  {Ceverino}}]{Goerdt2012MNRAS}
{Goerdt}, T., {Dekel}, A., {Sternberg}, A., {Gnat}, O., \& {Ceverino}, D. 2012,
  \mnras, 424, 2292

\bibitem[{{Goldreich} \& {Sridhar}(1995)}]{Goldreich1995ApJ}
{Goldreich}, P. \& {Sridhar}, S. 1995, \apj, 438, 763

\bibitem[{{Gonz{\'a}lez} {et~al.}(2010){Gonz{\'a}lez}, {Labb{\'e}}, {Bouwens},
  {Illingworth}, {Franx}, {Kriek}, \& {Brammer}}]{Gonzalez2010ApJ}
{Gonz{\'a}lez}, V., {Labb{\'e}}, I., {Bouwens}, R.~J., {et~al.} 2010, \apj,
  713, 115

\bibitem[{{Goto}(2006)}]{Goto2006MNRAS}
{Goto}, T. 2006, \mnras, 369, 1765

\bibitem[{{Gould}(1975)}]{Gould1975ApJ}
{Gould}, R.~J. 1975, \apj, 196, 689

\bibitem[{{Gould} \& {Burbidge}(1965)}]{Gould1965AnAp}
{Gould}, R.~J. \& {Burbidge}, G.~R. 1965, Annales d'Astrophysique, 28, 171

\bibitem[{{Grazian} {et~al.}(2012){Grazian}, {Castellano}, {Fontana},
  {Pentericci}, {Dunlop}, {McLure}, {Koekemoer}, {Dickinson}, {Faber},
  {Ferguson}, {Galametz}, {Giavalisco}, {Grogin}, {Hathi}, {Kocevski}, {Lai},
  {Newman}, \& {Vanzella}}]{Grazian2012A&A}
{Grazian}, A., {Castellano}, M., {Fontana}, A., {et~al.} 2012, \aap, 547, A51

\bibitem[{{Griffin} {et~al.}(2016){Griffin}, {Dai}, \&
  {Thompson}}]{Griffin2016ApJ}
{Griffin}, R.~D., {Dai}, X., \& {Thompson}, T.~A. 2016, \apjl, 823, L17

\bibitem[{{Gullberg} {et~al.}(2018){Gullberg}, {Swinbank}, {Smail}, {Biggs},
  {Bertoldi}, {De Breuck}, {Chapman}, {Chen}, {Cooke}, {Coppin}, {Cox},
  {Dannerbauer}, {Dunlop}, {Edge}, {Farrah}, {Geach}, {Greve}, {Hodge}, {Ibar},
  {Ivison}, {Karim}, {Schinnerer}, {Scott}, {Simpson}, {Stach}, {Thomson}, {van
  der Werf}, {Walter}, {Wardlow}, \& {Weiss}}]{Gullberg2018ApJ}
{Gullberg}, B., {Swinbank}, A.~M., {Smail}, I., {et~al.} 2018, \apj, 859, 12

\bibitem[{{Gunn} \& {Gott}(1972)}]{Gunn1972ApJ}
{Gunn}, J.~E. \& {Gott}, III, J.~R. 1972, \apj, 176, 1

\bibitem[{Guo \& Oh(2008)}]{Guo2008MNRAS}
Guo, F. \& Oh, S.~P. 2008, \mnras, 384, 251

\bibitem[{{Hammond} {et~al.}(2012){Hammond}, {Robishaw}, \&
  {Gaensler}}]{Hammond2012arXiv}
{Hammond}, A.~M., {Robishaw}, T., \& {Gaensler}, B.~M. 2012, ArXiv e-prints
  [\eprint[arXiv]{1209.1438}]

\bibitem[{Hashimoto {et~al.}(2018)Hashimoto, Laporte, Mawatari, Ellis, Inoue,
  Zackrisson, Roberts-Borsani, Zheng, Tamura, Bauer, Fletcher, Harikane,
  Hatsukade, Hayatsu, Matsuda, Matsuo, Okamoto, Ouchi, Pell{\'o}, Rydberg,
  Shimizu, Taniguchi, Umehata, \& Yoshida}]{Hashimoto2018Nat}
Hashimoto, T., Laporte, N., Mawatari, K., {et~al.} 2018, Nature, 557, 392

\bibitem[{Haug(1981)}]{Haug1981Znat}
Haug, E. 1981, Z. Naturforsch, 36, 413

\bibitem[{{Hayashida} {et~al.}(2013){Hayashida}, {Stawarz}, {Cheung},
  {Bechtol}, {Madejski}, {Ajello}, {Massaro}, {Moskalenko}, {Strong}, \&
  {Tibaldo}}]{Hayashida2013ApJ}
{Hayashida}, M., {Stawarz}, {\L}., {Cheung}, C.~C., {et~al.} 2013, \apj, 779,
  131

\bibitem[{{Heger} {et~al.}(2003){Heger}, {Fryer}, {Woosley}, {Langer}, \&
  {Hartmann}}]{Heger2003ApJ}
{Heger}, A., {Fryer}, C.~L., {Woosley}, S.~E., {Langer}, N., \& {Hartmann},
  D.~H. 2003, \apj, 591, 288

\bibitem[{{Hillas}(1984)}]{Hillas1984ARAA}
{Hillas}, A.~M. 1984, \araa, 22, 425

\bibitem[{{Inoue} {et~al.}(2016){Inoue}, {Tamura}, {Matsuo}, {Mawatari},
  {Shimizu}, {Shibuya}, {Ota}, {Yoshida}, {Zackrisson}, {Kashikawa}, {Kohno},
  {Umehata}, {Hatsukade}, {Iye}, {Matsuda}, {Okamoto}, \&
  {Yamaguchi}}]{Inoue2016Sci}
{Inoue}, A.~K., {Tamura}, Y., {Matsuo}, H., {et~al.} 2016, Science, 352, 1559

\bibitem[{Iwamoto \& Kunugise(2006)}]{Iwamoto2006AIPC}
Iwamoto, K. \& Kunugise, T. 2006, AIP Conference Proceedings, 847, 406

\bibitem[{{Janka}(2012)}]{Janka2012ARNPS}
{Janka}, H.-T. 2012, Ann. Rev. Nucl. and Part. Sci., 62, 407

\bibitem[{Joseph \& Rohrlich(1958)}]{Joseph1958RevModPhys}
Joseph, J. \& Rohrlich, F. 1958, Rev. Mod. Phys., 30, 354

\bibitem[{Jost {et~al.}(1950)Jost, Luttinger, \& Slotnick}]{Jost1950PR}
Jost, R., Luttinger, J.~M., \& Slotnick, M. 1950, Phys. Rev., 80, 189

\bibitem[{{Juvela} \& {Ysard}(2011)}]{Juvela2011ApJ}
{Juvela}, M. \& {Ysard}, N. 2011, \apj, 739, 63

\bibitem[{{Kafexhiu} {et~al.}(2014){Kafexhiu}, {Aharonian}, {Taylor}, \&
  {Vila}}]{Kafexhiu2014PRD}
{Kafexhiu}, E., {Aharonian}, F., {Taylor}, A.~M., \& {Vila}, G.~S. 2014, \prd,
  90, 123014

\bibitem[{Kamae {et~al.}(2006)Kamae, Karlsson, Mizuno, Abe, \&
  Koi}]{Kamae2006ApJ}
Kamae, T., Karlsson, N., Mizuno, T., Abe, T., \& Koi, T. 2006, The
  Astrophysical Journal, 647, 692

\bibitem[{Karlsson(2008)}]{Karlsson2008AIPC}
Karlsson, N. 2008, AIP Conference Proceedings, 1085, 561

\bibitem[{{Kaviraj} {et~al.}(2007){Kaviraj}, {Kirkby}, {Silk}, \&
  {Sarzi}}]{Kaviraj2007MNRAS}
{Kaviraj}, S., {Kirkby}, L.~A., {Silk}, J., \& {Sarzi}, M. 2007, \mnras, 382,
  960

\bibitem[{{Kelner} {et~al.}(2006){Kelner}, {Aharonian}, \&
  {Bugayov}}]{Kelner2006PhRvD}
{Kelner}, S.~R., {Aharonian}, F.~A., \& {Bugayov}, V.~V. 2006, \prd, 74, 034018

\bibitem[{{Kere{\v s}} {et~al.}(2009){Kere{\v s}}, {Katz}, {Fardal},
  {Dav{\'e}}, \& {Weinberg}}]{Keres2009MNRAS}
{Kere{\v s}}, D., {Katz}, N., {Fardal}, M., {Dav{\'e}}, R., \& {Weinberg},
  D.~H. 2009, \mnras, 395, 160

\bibitem[{{Kere{\v s}} {et~al.}(2005){Kere{\v s}}, {Katz}, {Weinberg}, \&
  {Dav{\'e}}}]{Keres2005MNRAS}
{Kere{\v s}}, D., {Katz}, N., {Weinberg}, D.~H., \& {Dav{\'e}}, R. 2005,
  \mnras, 363, 2

\bibitem[{{Klein}(2006)}]{Klein2006RadPhyChem}
{Klein}, S.~R. 2006, Radiation Physics and Chemistry, 75, 696

\bibitem[{Knudsen {et~al.}(2016)Knudsen, Richard, Kneib, Jauzac, Claement,
  Drouart, Egami, \& Lindroos}]{Knudsen2016MNRAS}
Knudsen, K.~K., Richard, J., Kneib, J.-P., {et~al.} 2016, \mnras, 462, L6

\bibitem[{{Knudsen} {et~al.}(2017){Knudsen}, {Watson}, {Frayer}, {Christensen},
  {Gallazzi}, {Micha{\l}owski}, {Richard}, \& {Zavala}}]{Knudsen2017MNRAS}
{Knudsen}, K.~K., {Watson}, D., {Frayer}, D., {et~al.} 2017, \mnras, 466, 138

\bibitem[{{Kobayashi} {et~al.}(2016){Kobayashi}, {Murata}, {Koekemoer},
  {Murayama}, {Taniguchi}, {Kajisawa}, {Shioya}, {Scoville}, {Nagao}, \&
  {Capak}}]{Kobayashi2016ApJ}
{Kobayashi}, M.~A.~R., {Murata}, K.~L., {Koekemoer}, A.~M., {et~al.} 2016,
  \apj, 819, 25

\bibitem[{{Kotera} {et~al.}(2010){Kotera}, {Allard}, \&
  {Olinto}}]{Kotera2010JCAP}
{Kotera}, K., {Allard}, D., \& {Olinto}, A.~V. 2010, \jcap, 10, 013

\bibitem[{{Kotera} \& {Olinto}(2011)}]{Kotera2011ARAA}
{Kotera}, K. \& {Olinto}, A.~V. 2011, \araa, 49, 119

\bibitem[{{Krumholz} \& {Dekel}(2010)}]{Krumholz2010MNRAS}
{Krumholz}, M.~R. \& {Dekel}, A. 2010, \mnras, 406, 112

\bibitem[{Lacki \& Beck(2013)}]{Lacki2013MNRAS}
Lacki, B.~C. \& Beck, R. 2013, \mnras, 430, 3171

\bibitem[{Lacki \& Thompson(2012)}]{Lacki2012AIPC}
Lacki, B.~C. \& Thompson, T.~A. 2012, AIP Conference Proceedings, 1505, 56

\bibitem[{{Lacki} {et~al.}(2011){Lacki}, {Thompson}, {Quataert}, {Loeb}, \&
  {Waxman}}]{Lacki2011ApJ}
{Lacki}, B.~C., {Thompson}, T.~A., {Quataert}, E., {Loeb}, A., \& {Waxman}, E.
  2011, \apj, 734, 107

\bibitem[{{Larson} {et~al.}(1980){Larson}, {Tinsley}, \&
  {Caldwell}}]{Larson1980ApJ}
{Larson}, R.~B., {Tinsley}, B.~M., \& {Caldwell}, C.~N. 1980, \apj, 237, 692

\bibitem[{{Latif} {et~al.}(2013){Latif}, {Schleicher}, {Schmidt}, \&
  {Niemeyer}}]{Latif2013MNRAS}
{Latif}, M.~A., {Schleicher}, D.~R.~G., {Schmidt}, W., \& {Niemeyer}, J. 2013,
  \mnras, 432, 668

\bibitem[{{Lazarian} \& {Pogosyan}(2000)}]{Lazarian2000ApJ}
{Lazarian}, A. \& {Pogosyan}, D. 2000, \apj, 537, 720

\bibitem[{Leite {et~al.}(2017)Leite, Evoli, D'Angelo, Ciardi, Sigl, \&
  Ferrara}]{Leite2017MNRAS}
Leite, N., Evoli, C., D'Angelo, M., {et~al.} 2017, \mnras, 469, 416

\bibitem[{{Lemoine-Goumard} {et~al.}(2012){Lemoine-Goumard}, {Renaud}, {Vink},
  {Allen}, {Bamba}, {Giordano}, \& {Uchiyama}}]{Lemoine2012A&A}
{Lemoine-Goumard}, M., {Renaud}, M., {Vink}, J., {et~al.} 2012, \aap, 545, A28

\bibitem[{Lipari(2003)}]{Lipari2003CERN}
Lipari, P. 2003

\bibitem[{Loeb \& Waxman(2006)}]{Loeb2006JCAP}
Loeb, A. \& Waxman, E. 2006, Journal of Cosmology and Astroparticle Physics,
  2006, 003

\bibitem[{{Loewenstein} {et~al.}(1991){Loewenstein}, {Zweibel}, \&
  {Begelman}}]{Loewenstein1991ApJ}
{Loewenstein}, M., {Zweibel}, E.~G., \& {Begelman}, M.~C. 1991, \apj, 377, 392

\bibitem[{{Mainali} {et~al.}(2018){Mainali}, {Zitrin}, {Stark}, {Ellis},
  {Richard}, {Tang}, {Laporte}, {Oesch}, \& {McGreer}}]{Mainali2018MNRAS}
{Mainali}, R., {Zitrin}, A., {Stark}, D.~P., {et~al.} 2018, \mnras, 479, 1180

\bibitem[{{Mawatari} {et~al.}(2016){Mawatari}, {Yamada}, {Fazio}, {Huang}, \&
  {Ashby}}]{Mawatari2016PASJ}
{Mawatari}, K., {Yamada}, T., {Fazio}, G.~G., {Huang}, J.-S., \& {Ashby},
  M.~L.~N. 2016, \pasj, 68, 46

\bibitem[{{Meier} {et~al.}(2002){Meier}, {Turner}, \& {Beck}}]{Meier2002AJ}
{Meier}, D.~S., {Turner}, J.~L., \& {Beck}, S.~C. 2002, \aj, 124, 877

\bibitem[{{Moore} {et~al.}(1999){Moore}, {Lake}, {Quinn}, \&
  {Stadel}}]{Moore1999MNRAS}
{Moore}, B., {Lake}, G., {Quinn}, T., \& {Stadel}, J. 1999, \mnras, 304, 465

\bibitem[{{Morlino}(2017)}]{Morlino2017hsn}
{Morlino}, G. 2017, {High-Energy Cosmic Rays from Supernovae}, ed. A.~W.
  {Alsabti} \& P.~{Murdin}, 1711

\bibitem[{{Morlino} \& {Caprioli}(2012)}]{Morlino2012A&A}
{Morlino}, G. \& {Caprioli}, D. 2012, \aap, 538, A81

\bibitem[{{M\"{u}cke} {et~al.}(1999){M\"{u}cke}, {Rachen}, {Engel},
  {Protheroe}, \& {Stanev}}]{Mucke1999PASA}
{M\"{u}cke}, A., {Rachen}, J.~P., {Engel}, R., {Protheroe}, R.~J., \& {Stanev},
  T. 1999, \pasa, 16, 160

\bibitem[{{Murata} {et~al.}(2014){Murata}, {Kajisawa}, {Taniguchi},
  {Kobayashi}, {Shioya}, {Capak}, {Ilbert}, {Koekemoer}, {Salvato}, \&
  {Scoville}}]{Murata2014ApJ}
{Murata}, K.~L., {Kajisawa}, M., {Taniguchi}, Y., {et~al.} 2014, \apj, 786, 15

\bibitem[{Nakamura(2010)}]{Nakamura2010JPG}
Nakamura, K. 2010, J.~Phys.~G, 37, 075021

\bibitem[{{Narayanan} {et~al.}(2008){Narayanan}, {Cox}, {Kelly}, {Dav{\'e}},
  {Hernquist}, {Di Matteo}, {Hopkins}, {Kulesa}, {Robertson}, \&
  {Walker}}]{Narayanan2008ApJS}
{Narayanan}, D., {Cox}, T.~J., {Kelly}, B., {et~al.} 2008, \apjs, 176, 331

\bibitem[{{Nath} \& {Biermann}(1993)}]{Nath1993MNRAS}
{Nath}, B.~B. \& {Biermann}, P.~L. 1993, \mnras, 265, 421

\bibitem[{{Nath} {et~al.}(2008){Nath}, {Laskar}, \& {Shull}}]{Nath2008ApJ}
{Nath}, B.~B., {Laskar}, T., \& {Shull}, J.~M. 2008, \apj, 682, 1055

\bibitem[{{Nozawa} {et~al.}(2007){Nozawa}, {Kozasa}, {Habe}, {Dwek}, {Umeda},
  {Tominaga}, {Maeda}, \& {Nomoto}}]{Nozawa2007ApJ}
{Nozawa}, T., {Kozasa}, T., {Habe}, A., {et~al.} 2007, \apj, 666, 955

\bibitem[{{Nulsen}(1982)}]{Nulsen1982MNRAS}
{Nulsen}, P.~E.~J. 1982, \mnras, 198, 1007

\bibitem[{{Oesch} {et~al.}(2014){Oesch}, {Bouwens}, {Illingworth}, {Labb{\'e}},
  {Smit}, {Franx}, {van Dokkum}, {Momcheva}, {Ashby}, {Fazio}, {Huang},
  {Willner}, {Gonzalez}, {Magee}, {Trenti}, {Brammer}, {Skelton}, \&
  {Spitler}}]{Oesch2014ApJ}
{Oesch}, P.~A., {Bouwens}, R.~J., {Illingworth}, G.~D., {et~al.} 2014, \apj,
  786, 108

\bibitem[{{Oesch} {et~al.}(2016){Oesch}, {Brammer}, {van Dokkum},
  {Illingworth}, {Bouwens}, {Labb{\'e}}, {Franx}, {Momcheva}, {Ashby}, {Fazio},
  {Gonzalez}, {Holden}, {Magee}, {Skelton}, {Smit}, {Spitler}, {Trenti}, \&
  {Willner}}]{Oesch2016ApJ}
{Oesch}, P.~A., {Brammer}, G., {van Dokkum}, P.~G., {et~al.} 2016, \apj, 819,
  129

\bibitem[{{Oesch} {et~al.}(2009){Oesch}, {Carollo}, {Stiavelli}, {Trenti},
  {Bergeron}, {Koekemoer}, {Lucas}, {Pavlovsky}, {Beckwith}, {Dahlen},
  {Ferguson}, {Gardner}, {Lilly}, {Mobasher}, \& {Panagia}}]{Oesch2009ApJ}
{Oesch}, P.~A., {Carollo}, C.~M., {Stiavelli}, M., {et~al.} 2009, \apj, 690,
  1350

\bibitem[{{Oesch} {et~al.}(2015){Oesch}, {van Dokkum}, {Illingworth},
  {Bouwens}, {Momcheva}, {Holden}, {Roberts-Borsani}, {Smit}, {Franx},
  {Labb{\'e}}, {Gonz{\'a}lez}, \& {Magee}}]{Oesch2015ApJ}
{Oesch}, P.~A., {van Dokkum}, P.~G., {Illingworth}, G.~D., {et~al.} 2015,
  \apjl, 804, L30

\bibitem[{{Ono} {et~al.}(2012){Ono}, {Ouchi}, {Mobasher}, {Dickinson},
  {Penner}, {Shimasaku}, {Weiner}, {Kartaltepe}, {Nakajima}, {Nayyeri},
  {Stern}, {Kashikawa}, \& {Spinrad}}]{Ono2012ApJ}
{Ono}, Y., {Ouchi}, M., {Mobasher}, B., {et~al.} 2012, \apj, 744, 83

\bibitem[{{Ouchi} {et~al.}(2013){Ouchi}, {Ellis}, {Ono}, {Nakanishi}, {Kohno},
  {Momose}, {Kurono}, {Ashby}, {Shimasaku}, {Willner}, {Fazio}, {Tamura}, \&
  {Iono}}]{Ouchi2013ApJ}
{Ouchi}, M., {Ellis}, R., {Ono}, Y., {et~al.} 2013, \apj, 778, 102

\bibitem[{{Owen} {et~al.}(2018){Owen}, {Jacobsen}, {Wu}, \&
  {Surajbali}}]{Owen2018MNRAS}
{Owen}, E.~R., {Jacobsen}, I.~B., {Wu}, K., \& {Surajbali}, P. 2018, \mnras,
  481, 666

\bibitem[{{Owen} {et~al.}(2019){Owen}, {Jin}, {Wu}, \& {Chan}}]{Owen2018sub}
{Owen}, E.~R., {Jin}, X., {Wu}, K., \& {Chan}, S. 2019, \mnras, 484, 1645

\bibitem[{{Paglione} {et~al.}(1996){Paglione}, {Marscher}, {Jackson}, \&
  {Bertsch}}]{Paglione1996ApJ}
{Paglione}, T.~A.~D., {Marscher}, A.~P., {Jackson}, J.~M., \& {Bertsch}, D.~L.
  1996, \apj, 460, 295

\bibitem[{Papadopoulos(2010)}]{Papadopoulos2010ApJ}
Papadopoulos, P.~P. 2010, \apj, 720, 226

\bibitem[{Papadopoulos {et~al.}(2011)Papadopoulos, Thi, Miniati, \&
  Viti}]{Papadopoulos2011MNRAS}
Papadopoulos, P.~P., Thi, W.-F., Miniati, F., \& Viti, S. 2011, \mnras, 414,
  1705

\bibitem[{Patrignani {et~al.}(2016)}]{Patrignani2016xqp}
Patrignani, C. {et~al.} 2016, Chin. Phys., C40, 100001

\bibitem[{{Peng} {et~al.}(2016){Peng}, {Wang}, {Liu}, {Tang}, \&
  {Wang}}]{Peng2016ApJ}
{Peng}, F.-K., {Wang}, X.-Y., {Liu}, R.-Y., {Tang}, Q.-W., \& {Wang}, J.-F.
  2016, \apjl, 821, L20

\bibitem[{{Peng} {et~al.}(2015){Peng}, {Maiolino}, \&
  {Cochrane}}]{Peng2015Natur}
{Peng}, Y., {Maiolino}, R., \& {Cochrane}, R. 2015, \nat, 521, 192

\bibitem[{{Persic} {et~al.}(2008){Persic}, {Rephaeli}, \&
  {Arieli}}]{Persic2008A&A}
{Persic}, M., {Rephaeli}, Y., \& {Arieli}, Y. 2008, \aap, 486, 143

\bibitem[{{Pfrommer}(2013)}]{Pfrommer2013ApJ}
{Pfrommer}, C. 2013, \apj, 779, 10

\bibitem[{{Planck Collaboration}(2016)}]{Planck2015A&A}
{Planck Collaboration}. 2016, \aap, 594, A13

\bibitem[{{Pollack} \& {Fazio}(1963)}]{Pollack1963PhRv}
{Pollack}, J.~B. \& {Fazio}, G.~G. 1963, Phys. Rev., 131, 2684

\bibitem[{{Protheroe} \& {Johnson}(1996)}]{Protheroe1996APh}
{Protheroe}, R.~J. \& {Johnson}, P.~A. 1996, Astroparticle Physics, 4, 253

\bibitem[{Pudritz \& Fich(2012)}]{Pudritz2012book}
Pudritz, R. \& Fich, M. 2012, Galactic and Extragalactic Star Formation, Nato
  Science Series C: (Springer: Netherlands)

\bibitem[{{Rees}(1987)}]{Rees1987QJRAS}
{Rees}, M.~J. 1987, \qjras, 28, 197

\bibitem[{{Rephaeli} {et~al.}(2010){Rephaeli}, {Arieli}, \&
  {Persic}}]{Rephaeli2010MNRAS}
{Rephaeli}, Y., {Arieli}, Y., \& {Persic}, M. 2010, \mnras, 401, 473

\bibitem[{Rephaeli \& Persic(2014)}]{Rephaeli2015NucPhysBProc}
Rephaeli, Y. \& Persic, M. 2014, 256

\bibitem[{{Ribaudo} {et~al.}(2011){Ribaudo}, {Lehner}, {Howk}, {Werk}, {Tripp},
  {Prochaska}, {Meiring}, \& {Tumlinson}}]{Ribaudo2011ApJ}
{Ribaudo}, J., {Lehner}, N., {Howk}, J.~C., {et~al.} 2011, \apj, 743, 207

\bibitem[{{Rieder} \& {Teyssier}(2016)}]{Rieder2016MNRAS}
{Rieder}, M. \& {Teyssier}, R. 2016, \mnras, 457, 1722

\bibitem[{Rimondi(1993)}]{Rimondi1993proc}
Rimondi, F. 1993, in {23rd International Symposium on Ultra-High Energy
  Multiparticle Phenomena Aspen, Colorado, September 12-17, 1993}, 0400--404

\bibitem[{{Robertson} {et~al.}(2010){Robertson}, {Ellis}, {Dunlop}, {McLure},
  \& {Stark}}]{Robertson2010Natur}
{Robertson}, B.~E., {Ellis}, R.~S., {Dunlop}, J.~S., {McLure}, R.~J., \&
  {Stark}, D.~P. 2010, \nat, 468, 49

\bibitem[{{Rojas-Bravo} \& {Araya}(2016)}]{Rojas-Bravo2016MNRAS}
{Rojas-Bravo}, C. \& {Araya}, M. 2016, \mnras, 463, 1068

\bibitem[{{Roseboom} {et~al.}(2009){Roseboom}, {Oliver}, \&
  {Farrah}}]{Roseboom2009ApJ}
{Roseboom}, I.~G., {Oliver}, S., \& {Farrah}, D. 2009, \apjl, 699, L1

\bibitem[{{Rowlands} {et~al.}(2015){Rowlands}, {Wild}, {Nesvadba}, {Sibthorpe},
  {Mortier}, {Lehnert}, \& {da Cunha}}]{Rowlands2015MNRAS}
{Rowlands}, K., {Wild}, V., {Nesvadba}, N., {et~al.} 2015, \mnras, 448, 258

\bibitem[{{Ruszkowski} {et~al.}(2017){Ruszkowski}, {Yang}, \&
  {Reynolds}}]{Ruszkowski2017ApJ}
{Ruszkowski}, M., {Yang}, H.-Y.~K., \& {Reynolds}, C.~S. 2017, \apj, 844, 13

\bibitem[{{Ruszkowski} {et~al.}(2018){Ruszkowski}, {Yang}, \&
  {Reynolds}}]{Ruszkowski2018ApJ}
{Ruszkowski}, M., {Yang}, H.-Y.~K., \& {Reynolds}, C.~S. 2018, \apj, 858, 64

\bibitem[{{Rybicki} \& {Lightman}(1979)}]{Rybicki1979book}
{Rybicki}, G.~B. \& {Lightman}, A.~P. 1979, {Radiative processes in
  astrophysics} (Wiley)

\bibitem[{{Salem} \& {Bryan}(2014)}]{Salem2014MNRAS}
{Salem}, M. \& {Bryan}, G.~L. 2014, \mnras, 437, 3312

\bibitem[{{Samui} {et~al.}(2005){Samui}, {Subramanian}, \&
  {Srianand}}]{Samui2005ICRC}
{Samui}, S., {Subramanian}, K., \& {Srianand}, R. 2005, International Cosmic
  Ray Conference, 9, 215

\bibitem[{{S{\'a}nchez Almeida} {et~al.}(2014){S{\'a}nchez Almeida},
  {Elmegreen}, {Mu{\~n}oz-Tu{\~n}{\'o}n}, \& {Elmegreen}}]{Sanchez2014A&ARv}
{S{\'a}nchez Almeida}, J., {Elmegreen}, B.~G., {Mu{\~n}oz-Tu{\~n}{\'o}n}, C.,
  \& {Elmegreen}, D.~M. 2014, \aapr, 22, 71

\bibitem[{{Sancisi} {et~al.}(2008){Sancisi}, {Fraternali}, {Oosterloo}, \& {van
  der Hulst}}]{Sancisi2008A&ARv}
{Sancisi}, R., {Fraternali}, F., {Oosterloo}, T., \& {van der Hulst}, T. 2008,
  \aapr, 15, 189

\bibitem[{{Sazonov} \& {Sunyaev}(2015)}]{Sazonov2015MNRAS}
{Sazonov}, S. \& {Sunyaev}, R. 2015, \mnras, 454, 3464

\bibitem[{{Schleicher} \& {Beck}(2013)}]{Schleicher2013A&A}
{Schleicher}, D. R.~G. \& {Beck}, R. 2013, \aap, 556, A142

\bibitem[{{Schlickeiser}(2002)}]{Schlickeiser2002book}
{Schlickeiser}, R. 2002, {Cosmic Ray Astrophysics} (Berlin: Springer)

\bibitem[{{Schober} {et~al.}(2013){Schober}, {Schleicher}, \&
  {Klessen}}]{Schober2013A&A}
{Schober}, J., {Schleicher}, D.~R.~G., \& {Klessen}, R.~S. 2013, \aap, 560, A87

\bibitem[{Schober {et~al.}(2015)Schober, Schleicher, \&
  Klessen}]{Schober2015MNRAS}
Schober, J., Schleicher, D. R.~G., \& Klessen, R.~S. 2015, \mnras, 446, 2

\bibitem[{{Schure} \& {Bell}(2013)}]{Schure2013MNRAS}
{Schure}, K.~M. \& {Bell}, A.~R. 2013, \mnras, 435, 1174

\bibitem[{{Silvia} {et~al.}(2010){Silvia}, {Smith}, \& {Shull}}]{Silvia2010ApJ}
{Silvia}, D.~W., {Smith}, B.~D., \& {Shull}, J.~M. 2010, \apj, 715, 1575

\bibitem[{{Skorodko} {et~al.}(2008){Skorodko}, {Bashkanov}, {Bogoslawsky},
  {Calen}, {Cappellaro}, {Clement}, {Demiroers}, {Doroshkevich}, {Duniec},
  {Ekstr{\"o}m}, {Franssen}, {Gustafsson}, {H{\"o}istad}, {Ivanov}, {Jacewicz},
  {Jiganov}, {Johansson}, {Khakimova}, {Kaskulov}, {Keleta}, {Koch}, {Kren},
  {Kullander}, {Kup{\'s}{\'c}}, {Kuznetsov}, {Marciniewski}, {Meier},
  {Morosov}, {Pauly}, {Petterson}, {Petukhov}, {Povtorejko}, {Sch{\"o}nning},
  {Scobel}, {Shwartz}, {Sopov}, {Stepeniak}, {Th{\"o}rngren-Engblom},
  {Tikhomirov}, {Wagner}, {Wolke}, {Yamamoto}, {Zabierowski}, \&
  {Z{\l}omanczuk}}]{Skorodko2008EPJA}
{Skorodko}, T., {Bashkanov}, M., {Bogoslawsky}, D., {et~al.} 2008, Eur. Phys.
  J. A, 35, 317

\bibitem[{Slattery(1972)}]{Slattery1972PRL}
Slattery, P. 1972, Phys. Rev. Lett., 29, 1624

\bibitem[{{Smartt}(2009)}]{Smartt2009ARAA}
{Smartt}, S.~J. 2009, \araa, 47, 63

\bibitem[{{Smercina} {et~al.}(2018){Smercina}, {Smith}, {Dale}, {French},
  {Croxall}, {Zhukovska}, {Togi}, {Bell}, {Crocker}, {Draine}, {Jarrett},
  {Tremonti}, {Yang}, \& {Zabludoff}}]{Smercina2018ApJ}
{Smercina}, A., {Smith}, J.~D.~T., {Dale}, D.~A., {et~al.} 2018, \apj, 855, 51

\bibitem[{{Solomon} \& {Vanden Bout}(2005)}]{Solomon2005ARAA}
{Solomon}, P.~M. \& {Vanden Bout}, P.~A. 2005, \araa, 43, 677

\bibitem[{{Stanimirovi{\'c}} \& {Lazarian}(2001)}]{Stanimirovic2001ApJ}
{Stanimirovi{\'c}}, S. \& {Lazarian}, A. 2001, \apjl, 551, L53

\bibitem[{{Stark} {et~al.}(2013){Stark}, {Schenker}, {Ellis}, {Robertson},
  {McLure}, \& {Dunlop}}]{Stark2013ApJ}
{Stark}, D.~P., {Schenker}, M.~A., {Ellis}, R., {et~al.} 2013, \apj, 763, 129

\bibitem[{{Stecker} {et~al.}(1968){Stecker}, {Tsuruta}, \&
  {Fazio}}]{Stecker1968ApJ}
{Stecker}, F.~W., {Tsuruta}, S., \& {Fazio}, G.~G. 1968, \apj, 151, 881

\bibitem[{{Stepney} \& {Guilbert}(1983)}]{Stepney1983MNRAS}
{Stepney}, S. \& {Guilbert}, P.~W. 1983, \mnras, 204, 1269

\bibitem[{{Stewart} {et~al.}(2013){Stewart}, {Brooks}, {Bullock}, {Maller},
  {Diemand}, {Wadsley}, \& {Moustakas}}]{Stewart2013ApJ}
{Stewart}, K.~R., {Brooks}, A.~M., {Bullock}, J.~S., {et~al.} 2013, \apj, 769,
  74

\bibitem[{Strong {et~al.}(2007)Strong, Moskalenko, \&
  Ptuskin}]{Strong2007ARNPS}
Strong, A.~W., Moskalenko, I.~V., \& Ptuskin, V.~S. 2007, Ann. Rev. Nucl. and
  Part. Sci., 57, 285

\bibitem[{{Strong} {et~al.}(2010){Strong}, {Porter}, {Digel},
  {J{\'o}hannesson}, {Martin}, {Moskalenko}, {Murphy}, \&
  {Orlando}}]{Strong2010ApJ}
{Strong}, A.~W., {Porter}, T.~A., {Digel}, S.~W., {et~al.} 2010, \apjl, 722,
  L58

\bibitem[{{Sun} \& {Furlanetto}(2016)}]{Sun2016MNRAS}
{Sun}, G. \& {Furlanetto}, S.~R. 2016, \mnras, 460, 417

\bibitem[{{Sur} {et~al.}(2018){Sur}, {Bhat}, \& {Subramanian}}]{2018MNRASSur}
{Sur}, S., {Bhat}, P., \& {Subramanian}, K. 2018, \mnras, 475, L72

\bibitem[{{Tadaki} {et~al.}(2014){Tadaki}, {Kodama}, {Tanaka}, {Hayashi},
  {Koyama}, \& {Shimakawa}}]{Tadaki2014ApJ}
{Tadaki}, K.-i., {Kodama}, T., {Tanaka}, I., {et~al.} 2014, \apj, 780, 77

\bibitem[{{Tang} {et~al.}(2014){Tang}, {Wang}, \& {Tam}}]{Tang2014ApJ}
{Tang}, Q.-W., {Wang}, X.-Y., \& {Tam}, P.-H.~T. 2014, \apj, 794, 26

\bibitem[{{Temim} {et~al.}(2015){Temim}, {Dwek}, {Tchernyshyov}, {Boyer},
  {Meixner}, {Gall}, \& {Roman-Duval}}]{Temim2015ApJ}
{Temim}, T., {Dwek}, E., {Tchernyshyov}, K., {et~al.} 2015, \apj, 799, 158

\bibitem[{{Thomas} {et~al.}(2017){Thomas}, {Le F{\`e}vre}, {Scodeggio},
  {Cassata}, {Garilli}, {Le Brun}, {Lemaux}, {Maccagni}, {Pforr}, {Tasca},
  {Zamorani}, {Bardelli}, {Hathi}, {Tresse}, {Zucca}, \&
  {Koekemoer}}]{Thomas2017A&A}
{Thomas}, R., {Le F{\`e}vre}, O., {Scodeggio}, M., {et~al.} 2017, \aap, 602,
  A35

\bibitem[{Thome {et~al.}(1977)}]{Thome1977NuclPhys}
Thome, W. {et~al.} 1977, Nucl. Phys., B129, 365

\bibitem[{{Torres}(2004)}]{Torres2004ApJ}
{Torres}, D.~F. 2004, \apj, 617, 966

\bibitem[{{Tremaine} {et~al.}(1994){Tremaine}, {Richstone}, {Byun}, {Dressler},
  {Faber}, {Grillmair}, {Kormendy}, \& {Lauer}}]{Tremaine1994ApJ}
{Tremaine}, S., {Richstone}, D.~O., {Byun}, Y.-I., {et~al.} 1994, \aj, 107, 634

\bibitem[{{Tremonti} {et~al.}(2007){Tremonti}, {Moustakas}, \&
  {Diamond-Stanic}}]{Tremonti2007ApJ}
{Tremonti}, C.~A., {Moustakas}, J., \& {Diamond-Stanic}, A.~M. 2007, \apjl,
  663, L77

\bibitem[{{Turner} {et~al.}(2015){Turner}, {Beck}, {Benford}, {Consiglio},
  {Ho}, {Kov{\'a}cs}, {Meier}, \& {Zhao}}]{Turner2015Natur}
{Turner}, J.~L., {Beck}, S.~C., {Benford}, D.~J., {et~al.} 2015, \nat, 519, 331

\bibitem[{{van den Bergh} {et~al.}(1996){van den Bergh}, {Abraham}, {Ellis},
  {Tanvir}, {Santiago}, \& {Glazebrook}}]{van_den_Bergh1996AJ}
{van den Bergh}, S., {Abraham}, R.~G., {Ellis}, R.~S., {et~al.} 1996, \aj, 112,
  359

\bibitem[{{VERITAS Collaboration} {et~al.}(2009){VERITAS Collaboration},
  {Acciari}, {Aliu}, {Arlen}, {Aune}, {Bautista}, {Beilicke}, {Benbow},
  {Boltuch}, {Bradbury}, {Buckley}, {Bugaev}, {Byrum}, {Cannon}, {Celik},
  {Cesarini}, {Chow}, {Ciupik}, {Cogan}, {Colin}, {Cui}, {Dickherber}, {Duke},
  {Fegan}, {Finley}, {Finnegan}, {Fortin}, {Fortson}, {Furniss}, {Galante},
  {Gall}, {Gibbs}, {Gillanders}, {Godambe}, {Grube}, {Guenette}, {Gyuk},
  {Hanna}, {Holder}, {Horan}, {Hui}, {Humensky}, {Imran}, {Kaaret}, {Karlsson},
  {Kertzman}, {Kieda}, {Kildea}, {Konopelko}, {Krawczynski}, {Krennrich},
  {Lang}, {Lebohec}, {Maier}, {McArthur}, {McCann}, {McCutcheon}, {Millis},
  {Moriarty}, {Mukherjee}, {Nagai}, {Ong}, {Otte}, {Pandel}, {Perkins},
  {Pizlo}, {Pohl}, {Quinn}, {Ragan}, {Reyes}, {Reynolds}, {Roache}, {Rose},
  {Schroedter}, {Sembroski}, {Smith}, {Steele}, {Swordy}, {Theiling},
  {Thibadeau}, {Varlotta}, {Vassiliev}, {Vincent}, {Wagner}, {Wakely}, {Ward},
  {Weekes}, {Weinstein}, {Weisgarber}, {Williams}, {Wissel}, {Wood}, \&
  {Zitzer}}]{VERITAS2009Natur}
{VERITAS Collaboration}, {Acciari}, V.~A., {Aliu}, E., {et~al.} 2009, \nat,
  462, 770

\bibitem[{{Walker}(2016)}]{Walker2016ApJ}
{Walker}, M.~A. 2016, \apj, 818, 23

\bibitem[{Wang(1991)}]{Wang1991thesis}
Wang, C.-H. 1991, PhD thesis, Iowa State University

\bibitem[{Wang \& Fields(2014)}]{Wang2014AIPC}
Wang, X. \& Fields, B.~D. 2014, AIP Conference Proceedings, 1595, 231

\bibitem[{{Wang} \& {Fields}(2018)}]{Wang2018MNRAS}
{Wang}, X. \& {Fields}, B.~D. 2018, \mnras, 474, 4073

\bibitem[{{Wardlow} {et~al.}(2017){Wardlow}, {Cooray}, {Osage}, {Bourne},
  {Clements}, {Dannerbauer}, {Dunne}, {Dye}, {Eales}, {Farrah}, {Furlanetto},
  {Ibar}, {Ivison}, {Maddox}, {Micha{\l}owski}, {Riechers}, {Rigopoulou},
  {Scott}, {Smith}, {Wang}, {van der Werf}, {Valiante}, {Valtchanov}, \&
  {Verma}}]{Wardlow2017ApJ}
{Wardlow}, J.~L., {Cooray}, A., {Osage}, W., {et~al.} 2017, \apj, 837, 12

\bibitem[{{Watson} {et~al.}(2015){Watson}, {Christensen}, {Knudsen}, {Richard},
  {Gallazzi}, \& {Micha{\l}owski}}]{Watson2015Natur}
{Watson}, D., {Christensen}, L., {Knudsen}, K.~K., {et~al.} 2015, \nat, 519,
  327

\bibitem[{{Wentzel}(1971)}]{Wentzel1971ApJ}
{Wentzel}, D.~G. 1971, \apj, 163, 503

\bibitem[{{Whitmore}(1974)}]{Whitmore1974PhR}
{Whitmore}, J. 1974, \physrep, 10, 273

\bibitem[{{Wiener} {et~al.}(2013){Wiener}, {Zweibel}, \& {Oh}}]{Wiener2013ApJ}
{Wiener}, J., {Zweibel}, E.~G., \& {Oh}, S.~P. 2013, \apj, 767, 87

\bibitem[{{Wright}(2006)}]{Wright2006PASP}
{Wright}, E.~L. 2006, \pasp, 118, 1711

\bibitem[{{Yamasawa} {et~al.}(2011){Yamasawa}, {Habe}, {Kozasa}, {Nozawa},
  {Hirashita}, {Umeda}, \& {Nomoto}}]{Yamasawa2011ApJ}
{Yamasawa}, D., {Habe}, A., {Kozasa}, T., {et~al.} 2011, \apj, 735, 44

\bibitem[{{Yan} \& {Lazarian}(2004)}]{Yan2004ApJ}
{Yan}, H. \& {Lazarian}, A. 2004, \apj, 614, 757

\bibitem[{{Yan} {et~al.}(2006){Yan}, {Newman}, {Faber}, {Konidaris}, {Koo}, \&
  {Davis}}]{Yan2006ApJ}
{Yan}, R., {Newman}, J.~A., {Faber}, S.~M., {et~al.} 2006, \apj, 648, 281

\bibitem[{{Yang} {et~al.}(2006){Yang}, {Tremonti}, {Zabludoff}, \&
  {Zaritsky}}]{Yang2006ApJ}
{Yang}, Y., {Tremonti}, C.~A., {Zabludoff}, A.~I., \& {Zaritsky}, D. 2006,
  \apjl, 646, L33

\bibitem[{{Yoast-Hull} {et~al.}(2017){Yoast-Hull}, {Gallagher}, {Aalto}, \&
  {Varenius}}]{Yoast-Hull2017MNRAS}
{Yoast-Hull}, T.~M., {Gallagher}, III, J.~S., {Aalto}, S., \& {Varenius}, E.
  2017, \mnras, 469, L89

\bibitem[{{Zweibel}(2016)}]{Zweibel2016APS}
{Zweibel}, E. 2016, in APS Meeting Abstracts, SR1.001

\bibitem[{{Zweibel}(2013)}]{Zweibel2013PhPl}
{Zweibel}, E.~G. 2013, Physics of Plasmas, 20, 055501

\end{thebibliography}

\begin{appendix}

 \section{Cosmic ray interactions}
\label{sec:interaction_mechanisms}

Hadronic and leptonic CRs interact through different mechanisms. 
For instance, their comparatively low mass makes CR electrons considerably more susceptible to scattering and cooling processes compared to their heavier proton counterparts. 
For protons, absorption processes are more important. 
We define a `cooling' process as one in which the original CR survives the interaction but loses some fraction of its energy and thus can continue to scatter many times. 
Conversely, we specify an `absorption' process to be one where either a substantial fraction of the original CR energy is lost in an interaction (such that the particle can no longer be considered high energy afterwards, and is not realistically able to undergo another interaction of the same type), or where the particle is truly destroyed in the process.

\subsection{High energy proton interactions}
\label{sec:cosmic_ray_proton_interactions}

The relevant processes governing hadronic interactions of CRs with radiation and matter fields in astrophysical environments are Photo-pion (${\rm p}\gamma$) production, Bethe-Heitler pair production and ${\rm pp}$ pion-production interactions. The pion-producing processes may be considered as absorption mechanisms for hadronic CRs, as they result in substantial energy loss of the particle in a single interaction. 
The Bethe-Heitler pair production process results in a lower energy loss fraction per interaction and so is better regarded as a continuous cooling mechanism rather than a stochastic absorption process. Hence, we separate our discussion of these mechanisms accordingly. 

Each of these interactions produce neutrons, protons, leptons and other hadrons including charged and neutral pions
    \citep[see][]{Pollack1963PhRv, Gould1965AnAp, Stecker1968ApJ, 
     Mucke1999PASA, Berezinsky2006PRD, Dermer2009book}. The protons can continue to propagate and undergo further interactions like that of the primary, until the particle energy is below the respective interaction threshold. The neutrons have a half life of 880 s~\citep{Nakamura2010JPG}, and so are most likely to undergo a $\beta^{-}$ decay ${\rm n}\rightarrow {\rm p}{\rm e}^{-}\bar{\nu}_{\rm e}$.  The charged and neutral pions only have a fleeting lifetime before they undergo decays, so these cannot realistically undergo further interactions. The decay of neutral pions is dominated by the electromagnetic process $\pi^0 \rightarrow 2\gamma$ with branching ratio (BR) of 98.8\%~\citep{Patrignani2016xqp}, arising on a timescale of $8.5 \times 10^{-17}\;\!{\rm s}$. Charged pions instead decay to produce leptons (predominantly electrons and positrons) and neutrinos by the weak interaction $\pi^+ \rightarrow \rm{e}^+ \nu_{\rm e} \bar{\nu}_{\rm \mu} \nu_{\rm \mu}$ for positively charged pions, or $\pi^- \rightarrow \rm{e}^- \bar{\nu}_{\rm e} \nu_{\rm \mu} \bar{\nu}_{\rm \mu}$\footnote{These processes arise via the formation and decay of a $\mu^{\pm}$.} for negatively charged pions, both with BR = 99.9\%~\citep{Patrignani2016xqp} and arising over a longer timescale of $2.6\times 10^{-8}\;\!{\rm s}$. This is the main mechanism by which secondary CR electrons are injected into the environment.

In the following, we denote the energy of the primary CR proton as $E_{\rm p} = \gamma_{\rm p} m_{\rm p} \rm{c}^2$, or $\epsilon_{\rm p} = \gamma_{\rm p} (m_{\rm p}/m_{\rm e})$ if normalised to electron rest mass, where $\gamma_{\rm p}$ is the Lorentz factor. Where interactions involve a target field of photons, the normalised photon energy of frequency $\nu$ is written as $\epsilon = {\rm h}\nu/m_{\rm e}{\rm c}^2$. Alternatively, for CR-baryon interactions, the target baryon (presumed to be a hydrogen nucleus) has a negligible energy compared to the CR primary, and so may be regarded as `at rest', with $\gamma_{\rm p} = 1$. In this case, the normalised energy of the target baryon is $\epsilon \approx m_{\rm p}/m_{\rm e}$. It is also useful to define the invariant normalised interaction energy as $\epsilon_{\rm r} = \gamma_{\rm p}  (1-\beta_{\rm p}\cos\theta) \epsilon$, for $\theta$ as the angle between the momentum vectors of the incident proton and target field particle (photon or proton). $\beta_{\rm p}$ is the CR primary velocity in units of ${\rm c}$.
  
\subsubsection{Absorption processes}
\label{sec:proton_absorption}

Photo-pion production and ${\rm pp}$-pion production are both inelastic processes which ultimately lead to the absorption of the original CR particle and a reprocessing of its energy. While fundamentally similar, the first of these involves the interaction between a CR proton and a low-energy photon, while the second involves a low energy baryon as the target.
\bigskip

  {\sffamily Photo-pion production} \newline
 
The dominant interactions in this process are single-pion at lower energies, and multiple-pion production above $\epsilon_{\rm r} \approx 3500$. This occurs both via resonant production and direct production
  \citep[see][]{Mucke1999PASA}. The resonant production channel (which occurs at around three times the rate of direct production at $\epsilon_{\rm r} \approx 500$) progresses with the formation of the $\Delta^+$ resonance which decays by one of two channels, either $\Delta^{+} \rightarrow \rm{p} \pi^0$ (BR = 2/3), or $\Delta^{+} \rightarrow\rm{n} \pi^+$ (BR = 1/3)~\citep[see][]{Berezinsky1993PRD}, where the pion decay then proceeds as described before. Residual interactions (e.g. ${\rm p}\gamma \rightarrow \Delta^{++} \pi^{-}, \Delta^{0}\pi^{+}$) provide additional charged pions and, when taking all channels into account, the production of each type of pion arises at a similar rate~\citep[see][]{Dermer2009book}. Based on experimental data  \citep{Baldini1987book}, the inelastic cross section for the photo-pion interaction can be reasonably approximated as
  $\hat{\sigma}_{\rm \gamma \pi} \approx  \hat{\sigma}_{\rm \gamma \pi}^*\;\! {\cal H}(\epsilon_{\rm r} - \epsilon_{\rm th})$, 
  where ${\cal H}(...)$ is the Heaviside step function, with $ \hat{\sigma}_{\rm \gamma \pi}^* = 70~\rm{\mu b}$ and 
 $\epsilon_{\rm th} = 390$. These are the same values as suggested in~\citet{Dermer2009book}, which are also adopted in~\citet[][]{Dermer2003ApJ}. 
    
   The collision rate is defined by the collision timescale as $\dot{N}(\gamma_{\rm p})_{\gamma \pi, {\rm coll}} = [\tau_{\gamma \pi, {\rm coll}}(\gamma_{\rm p})]^{-1}$ for a CR proton with Lorentz factor $\gamma_{\rm p}$. Some fraction of these collisions will be elastic, while others will contribute to the pion-producing inelastic pathway. The total collision rate accounts for both of these but, for our purposes, we require the CR proton absorption rate - this is the fraction of collisions that successfully yield pion-production and subsequent CR absorption. The inelastic collision rate follows as:
  \begin{align}
  \dot{N}_{\gamma \pi, {\rm abs}}(\gamma_{\rm p}) \approx \frac{{\rm c}}{2 \gamma_{\rm p}} \int_0^{\infty} {\rm d}\epsilon \frac{n_{\rm ph}(\epsilon)}{\epsilon} \hat{\sigma}_{\gamma \pi}(\gamma_{\rm p}) \ ,
  \end{align}
  which is the inverse timescale for the photo-pion interaction. Substituting the dominant CMB radiation field for the photon number density,
  \begin{equation}
  n_{\rm ph}(\epsilon) = \frac{8 \pi}{\lambda_{\rm C}^3}\frac{\epsilon^2}{\exp\left[{\epsilon}/{\Theta(z)}\right] - 1} \ ,
  \label{eq:blackbody}
  \end{equation}
    where symbols retain their earlier definitions, gives
        \begin{align}
    \dot{N}_{\gamma \pi, {\rm abs}}(\gamma_{\rm p}) &\approx \frac{{\rm c}}{2 \gamma_{\rm p}} \int_0^{\infty} {\rm d}\epsilon \frac{8 \pi}{\lambda_{\rm C}^3}\frac{\epsilon}{\exp\left[{\epsilon}/{\Theta(z)}\right] - 1} \hat{\sigma}_{\gamma \pi}(\gamma_{\rm p}) \nonumber \\
   & \approx  \frac{4 \pi \hat{\sigma}_{\gamma \pi}(\gamma_{\rm p}) \Theta^2(z)}{\gamma_{\rm p} \lambda_{\rm C}^3}\int_0^{\infty} \frac{x {\rm d}x}{\exp(x)-1} \ ,
    \end{align}
    where the substitution $x = \epsilon/\Theta$ has been applied. The integral has a standard result of $\pi^2/6$, yielding the final absorption rate as
    \begin{equation}
     \dot{N}_{\gamma \pi, {\rm abs}}(\gamma_{\rm p}) = \frac{2 \pi^3 \hat{\sigma}_{\gamma \pi}(\gamma_{\rm p}) \Theta^2(z)}{3 \gamma_{\rm p} \lambda_{\rm C}^3} \ .
    \end{equation}
\bigskip

  {\sffamily ${\rm pp}$-pion production} \newline
 
The ${\rm pp}$ mechanism is similar to the photo-pion process, but with the interaction target as the low energy baryon field of the ISM. The dominant channels proceed via $\Delta^+$ and $\Delta^{++}$ resonance production (${\rm pp} \rightarrow {\rm p}\Delta^+$ and ${\rm pp} \rightarrow {\rm n} \Delta^{++}$) which then decay according to
\begin{align}\label{eq:pp_interaction1}%
		\rm{p}  \Delta^{+~\;} \rightarrow\begin{cases}%
				\rm{p} \rm{p} \pi^{0}  \xi_{0}(\pi^{0}) \xi_{\pm}(\pi^{+} \pi^{-}) \\[0.5ex]%
				\rm{p} \rm{p}  \pi^{+}  \pi^{-}  \xi_{0}(\pi^{0}) \xi_{\pm}(\pi^{+} \pi^{-}) \\[0.5ex]%
				\rm{p} \rm{n}  \pi^{+}  \xi_{0}(\pi^{0}) \xi_{\pm}(\pi^{+} \pi^{-})\\[0.5ex]%
			\end{cases}
\end{align}%
and
\begin{align}\label{eq:pp_interaction2}%
		\rm{n} \Delta^{++} \rightarrow\begin{cases}%
				\rm{n} \rm{p} \pi^{+} \xi_{0}(\pi^{0}) \xi_{\pm}(\pi^{+} \pi^{-}) \\[0.5ex]%
				\rm{n} \rm{n} 2\pi^{+} \xi_{0}(\pi^{0}) \xi_{\pm}(\pi^{+} \pi^{-})\\[0.5ex]%
			\end{cases}
\end{align}%
with $\xi_{0}$ and $\xi_{\pm}$ as the  neutral and charged pion multiplicities respectively which themselves are a function of the CR primary energy 
  \citep[see][]{Almeida1968PR, Skorodko2008EPJA}. Using the empirical parameterisations given by~\citet{BlattnigPRD2000}, the inelastic cross section for pion production indicates branching ratios for each of the pion species $\{\pi^{+}, \pi^{-}, \pi^{0}\}$ as $\{0.6, 0.1, 0.3\}$ at 1 GeV, levelling out to $\{0.3, 0.4, 0.3\}$ at higher energies, above 50 GeV. These suggest that the production rate of each type of pion is similar (as with the photo-pion process). The total inelastic cross section, which is required for the ${\rm pp}$ absorption rate of CR primaries, is well described by the \citet{Kafexhiu2014PRD} parameterisation: 
\begin{equation}%
\label{eq:pp_cs}%
\hat{\sigma}_{\rm p\pi} = \left( 30.7 - 0.96\ln(\chi) + 0.18(\ln\chi)^{2} \right)\left( 1 - \chi^{-1.9}   \right)^{3}~\rm{mb} \ ,
\end{equation}%
which we adopt for subsequent calculations in this work. Here, $\chi =  E_{\rm p}/E^{\rm{th}}_{\rm p}$, for $E^{\rm{th}}$ as the threshold energy for the process (0.28 GeV\footnote{$E^{\rm{th}}_{\rm p} = (2m_{\rm \pi^{0}}+m^{2}_{\rm \pi^{0}}/2m_{\rm p}) {\rm c}^2 \approx0.28~\rm{GeV}$ - i.e the energy required for the production of the lowest energy pion (the $\pi^0$).}). The rate of CR absorption then follows as:
\begin{equation}
\label{eq:cr_absorption}
\dot{N}_{\rm p\pi, abs}(\gamma_{\rm p}) = {\rm c} ~\hat{\sigma}_{\rm p\pi}(\gamma_{\rm p})~n_{\rm b} \ .
\end{equation}

\subsubsection{Cooling processes}

Compared to the pion production mechanisms described in the previous section, photo-pair production is better described as a cooling process as each interaction only results in a small change in the CR primary energy. Many cooling processes usually attributed to charged particles (e.g. radiative cooling via inverse-Compton and Synchrotron emission) are not important for protons due to their greater mass. Instead, these are more relevant to the electronic secondaries -- see section~\ref{sec:cosmic_ray_electron_interactions}. Thus, unabsorbed free-streaming protons would more likely to be cooled by the adiabatic cosmological expansion, particularly if they are able to escape from their source galaxy to traverse intergalactic space.
\bigskip

 {\sffamily Photo-pair production} \newline

The Photo-pair Bethe-Heitler process \citep{Bethe1934ProcRSA} produces predominantly electron-positron pairs over the energies of interest in this study~\citep{Blumenthal1970PRD, Klein2006RadPhyChem} and, for a CR proton, proceeds as
${\rm p}\gamma \rightarrow {\rm p} {e}^+ {e}^-$. When the interaction energy $\epsilon_{\rm r} \gtrsim 60$ (as may be assumed in this work), a reasonable approximation to the fitted inelastic cross section~\citep[see][]{Jost1950PR, Bethe1954PR, Blumenthal1970PRD, Stepney1983MNRAS} was found to be:
\begin{equation}
\label{approx-sigma}
\hat{\sigma}_{\rm \gamma e}(\epsilon_{\rm r})\approx %
\left\{	\frac{7}{6\pi}\alpha_{\rm f} \ln\left[\frac{\epsilon_{\rm r}}{k_{\rm \gamma e}}\right] \right\} \sigma_{\rm T}
\end{equation}%
where $\alpha_{\rm f}$ is the fine structure constant, 
$\sigma_{\rm T}$ is the Thomson cross-section, 
 and the fitting constant $k_{\rm \gamma e}$ is set to 6.7~\citep[see][]{Owen2018MNRAS}\footnote{
  \citet{Dermer2009book} suggests a lower value of $k_{\rm \gamma e} = 2$, but we note that this was in the case of $\epsilon_{\rm r} \gtrsim 40$.}. The timescale for this interaction is
  \begin{equation}
\label{eq:photopair_inv_t}%
t_{\rm \gamma e} \approx 
 \frac{m_{\rm p}}{m_{\rm e}} 
  \frac{ \gamma_{\rm p}^2}{ \rm{c}} 
\left\{ \int_{\gamma_{\rm p}^{-1}}^{\infty} {\rm d}\epsilon\;\! 
   \frac{n_{\rm ph}(\epsilon)}{\epsilon^2}
    \int_{2}^{2\gamma_{\rm p}\epsilon} {\rm d} \epsilon_{\rm r}\;\! 
        \frac{\epsilon_{\rm r}\;\! \hat{\sigma}_{\rm \gamma e}(\epsilon_{\rm r})}{\sqrt{1+2\epsilon_{\rm r}}}  \right\}^{-1}
\end{equation}%
\citep[see][]{Protheroe1996APh, Dermer2009book}, if assuming that the electron-positron pairs are produced in the zero-momentum frame of the interaction, and the interaction energy is completely dominated by the contribution from the CR primary. The energy loss rate, $b_{\gamma e}(\gamma_{\rm p}) = m_{\rm p} {\rm c}^2 \gamma_{\rm p} t_{\rm \gamma e}^{-1}$, is then
\begin{equation}%
\label{eq:photopair_losses1}%
b_{\rm \gamma e}(\gamma_{\rm p}) \approx %
\frac{112}{9}
	\frac{{\rm c}^3 m_{\rm e} \alpha_{\rm f} \sigma_{\rm T}\mathcal{F}_{\rm \gamma e}(u)}{\lambda_{\rm C}^3 \gamma_{\rm p}^2 u^{5/3}} \ ,%
\end{equation}%
where $u =  m_{\rm e} {\rm c}^2/\gamma_{\rm p} {\rm k}_{\rm B} T$ and the function
\begin{equation}%
\label{eq:f_func}
\mathcal{F}_{\rm \gamma e}(u) = %
	\mathcal{C}(u) + \mathcal{D}(u)\ln\left[\frac{1}{0.974 u k_{\rm \gamma e}}\right]+ \left(0.382 u k_{\rm \gamma e}\right)^{3/2} \mathcal{E}(u) %
\end{equation}%
results from evaluating the integral in equation~\ref{eq:photopair_inv_t}\footnote{We note that equation~\ref{eq:photopair_losses1} requires a multiplicative factor if the radiation field is diluted, for example as would be the case with light emitted from a distribution of stars. For the CMB, the field is not diluted and the equation holds without modification.}, where
   $\mathcal{C}(u) = 0.74$ and $\mathcal{D}(u) = \Gamma(5/2)\zeta(5/2)$ 
   when $b\ll1$, $\mathcal{C}(u) = u^{3/2}\ln(u)e^{-u}$ 
   and $\mathcal{D}(u) = u^{3/2}e^{-u}$ 
   when $u\gg 1$, and $\mathcal{E}(u) = -\ln[1-e^{-u}]$ for all values of $u$ \citep{Dermer2009book}. 
\bigskip

 {\sffamily Adiabatic cosmological expansion} \newline

If propagating over cosmological scales, the expansion of the Universe will cool a particle. The rate at which this arises is
\begin{align}%
\label{eq:cosmological_losses}%
b_{\rm exp}(\gamma_{\rm p})  = \gamma_{\rm p} m_{\rm p}{\rm c}^2 {\rm{H_0}}\sqrt{\Omega_{\rm m}(1+z)^3 + \Omega_{\rm \Lambda}} 
\end{align}%
\citep[see][]{Gould1975ApJ, Berezinsky1988A&A, Berezinsky2006PRD}, 
where $\rm{H_0} = (67.8\pm 0.9)$ km s$^{-1}$ Mpc$^{-1}$, $\Omega_{\rm m} = 0.308 \pm 0.012$ 
   and $\Omega_{\rm \Lambda} = 0.691 \pm 0.0062$, 
   and negligible curvature and radiation energy densities~\citep{Planck2015A&A}. This is only important for particles which are not liable to more rapid cooling by other processes (see, e.g. section~\ref{sec:cosmic_ray_electron_interactions}).

\subsection{Electron injection}
\label{sec:particle_injection}

Secondary high energy ($>\text{GeV}$) CR electrons are injected 
by the pion-yielding interactions of CR protons with the host environment (e.g. via ${\rm pp}$- and photo-pion processes). 
Such injection of electrons is thought to be even more important than their direct acceleration, particularly in starburst environments where observations indicate that 60-80\% of high energy electrons are secondaries~\citep[e.g.][]{Loeb2006JCAP, Lacki2011ApJ, Lacki2013MNRAS, Schober2015MNRAS}. This high secondary fraction is also consistent with actively star-forming regions in the more local Universe~\citep[see][]{Paglione1996ApJ, Torres2004ApJ, Domingo-Santamaria2005A&A, Persic2008A&A, de_Cea_del_Pozo2009ApJ, Rephaeli2010MNRAS} where conditions would be ideal for the hadronic CR interactions thought to produce these electron CR secondaries. 

The energy spectrum of the CR secondary particles is governed by the interaction cross sections of the relevant processes driving their formation. These are photo-pion production in radiation dominated systems (i.e. where $n_{\rm ph} \gg n_{\rm b}$), or ${\rm pp}$-pion production in matter dominated systems ($n_{\rm b} \gg n_{\rm ph}$). In general, we can consider both of these together with their combined effective absorption coefficient as the sum of the contributions from these two processes:
\begin{equation}
\label{eq:alpha_star}
\alpha^*(E_{\rm p}, r) = \left(n_{\rm b}\hat{\sigma}_{\rm p\pi} + n_{\rm ph}\hat{\sigma}_{\rm p\gamma}\right)\vert_{E_{\rm p}, r} \ .
\end{equation}
This, coupled with the (energy dependent) multiplicity of the products and the energy spectrum of the primary CRs allows a reasonable model for the differential CR secondary electron production rate~\citep{Kelner2006PhRvD, Kamae2006ApJ} to be calculated:
\begin{equation}
\label{eq:injection1}
Q_{\rm e}(E_{\rm e}, r) = {\rm c}~\int_{E_{\rm e}}^{\infty}~\alpha^*(E_{\rm p}, r) \; F_{\rm e} \left(\frac{E_{\rm e}}{E_{\rm p}}, E_{\rm p} \right) \; \frac{\Phi_{\rm p}(E_{\rm p}, r)}{E_{\rm p}} \; {\rm d}E_{\rm p} \ .
\end{equation}
This is the injection rate of electrons per unit volume per energy interval (between $E_{\rm e}$ and $E_{\rm e} + {\rm d}E_{\rm e}$), and introduces the differential CR proton energy density as
\begin{equation}
\Phi_{\rm p}(E_{\rm p}, r) = \frac{E_{\rm p} \; {\rm d} N_{\rm p}}{{\rm d}V \; {\rm d}E_{\rm p}}\biggr\vert_{E_{\rm p}, r} \ .
\end{equation}
This is related to the differential number density of protons (number density per energy interval) by the relation $\Phi_{\rm p} = E_{\rm p} \cdot n_{\rm p}$.
In equation~\ref{eq:injection1}, we introduced $F_{\rm e}$ as the number of electrons produced per interaction per energy interval -- effectively the differential multiplicity of electron production. At high energies, above 100 GeV, a parametrisation of the function $F_{\rm e}$ following~\citet{Kelner2006PhRvD} may be used:
\begin{equation}
F_{\rm e} \left(x, \frac{E_{\rm e}}{x} \right) = -\frac{\mathcal{B}_{\rm e} \left[1+\mathcal{K}_{\rm e}\left(\ln x \right)^2 \right]^3~\ln^5(x)}{x \left[1+0.3 x^{-\beta_e}\right]}
\end{equation}
where we have used the substitution $x = E_{\rm e}/E_{\rm p}$. Here,
\begin{equation}
\mathcal{B}_{\rm e} =  \left[ 68.5 - 2.26 \ln\chi + 0.3 \left(\ln \chi\right)^2 \right]^{-1} \ ,
\end{equation}
where, for convenience, we retain the earlier definition for $\chi$ as the CR proton energy normalised to the ${\rm pp}$ interaction threshold energy, $E_{\rm p}/E_{\rm p}^{\rm th}$. Also
\begin{equation}
\mathcal{K}_{\rm e} = \frac{0.0172 \left(\ln\chi\right)^2  - 0.141 \ln\chi - 1.02}{0.3 + \left( \ln\chi - 5.88 \right)^2} \ ,
\end{equation}
and $\beta_e = \left[  0.00042 \left(\ln\chi\right)^2 + 0.063 \ln\chi - 0.28 \right]^{-1/4}$.
While valid for high energies, \citet{Kelner2006PhRvD} shows that this form of $F_{\rm e}$ to be unsuitable at the lower energies in our range of interest, below 100 GeV. Instead, for this regime they suggest the use of a delta-function~\citep[also see][]{Aharonian2000A&A}. In this case, the we break down the process into its composite mechanisms (charged pion production, and their subsequent decay), and model the pion production number by 1-100 GeV CRs per interaction per energy interval (this time between $E_{\rm \pi^{\pm}}$ and $E_{\rm \pi^{\pm}} + {\rm d} E_{\rm \pi^{\pm}}$) as:
\begin{equation}
\label{eq:f_function}
F_{\pi^{\pm}}\left(\frac{E_{\pi^{\pm}}}{E_{\rm p}}, E_{\rm p} \right) \approx \delta \left( E_{\pi^{\pm}}-K_{\pi^{\pm}}(E_{\rm kin})\right) \ ,
\end{equation}
where we have introduced the contraction $E_{\rm kin} =  E_{\rm p}-m_{\rm p} {\rm c}^2$. The efficiency term $K_{\pi^{\pm}}$ encodes the fraction of energy passed to pion production from the CR proton. The branching ratios for the formation of each type of pion are all approximately 1/3 in both the ${\rm pp}$-pion and photo-pion interaction channels (see section~\ref{sec:cosmic_ray_proton_interactions}), and this is not strongly dependent on energy~\citep{Aharonian2000A&A, BlattnigPRD2000}. Accounting also for neutrino production, the total fraction of the CR primary energy passed to charged pions is around 0.6-0.7~\citep[see][]{Dermer2009book}. Here we conservatively adopt the lowest value in this range and set $K_{\pi^{\pm}} = 0.6$. Using the delta function approximation~\ref{eq:f_function} in equation~\ref{eq:injection1}, we arrive at the injection rate of charged pions per unit volume per energy interval as:
\begin{equation}
\label{eq:injection_pion}
Q_{\rm \pi^{\pm}}(E_{\rm \pi^{\pm}}, r) = {\rm c} \; \int_{E_{\rm \pi^{\pm}}}^{\infty}~\alpha^* ~\delta \left( E_{\pi^{\pm}}-K_{\pi^{\pm}} E_{\rm kin}\right) \Phi_{\rm p} \; E_{\rm p}^{-1} \; {\rm d} E_{\rm p}
\end{equation}
where we have used the contractions $\alpha^* = \alpha^*(E_{\rm p}, r)$ and $\Phi_{\rm p} = \Phi_{\rm p}(E_{\rm p}, r)$ and other symbols retain their earlier definitions. Evaluating the integral gives:
\begin{equation}
\label{eq:injection_pion}
Q_{\rm \pi^{\pm}}(E_{\rm \pi^{\pm}}, r) = \frac{{\rm c}}{K_{\pi^{\pm}}} \; \alpha^*\left(L, r\right) \; \Phi_{\rm p}\left(L, r\right)
\end{equation}
\citep{Aharonian2000A&A}, where the energy parameter $L$ is defined as $L = m_{\rm p} {\rm c}^2+\frac{E_{\pi^{\pm}}}{K_{\pi^{\pm}}}$.
The charged pions formed in this process rapidly decay into leptons and neutrinos (see section~\ref{sec:cosmic_ray_proton_interactions}), with the pion energy roughly split equally between the neutrino and lepton products -- these form in the ratio of three neutrinos to each lepton. Detailed studies find that the leptons inherit around 26.5\% of the pion energy~\citep[e.g.][]{Lipari2003CERN, Lacki2013MNRAS}, with the remainder passed to neutrinos and effectively lost from the system. 

We may write the rate of energy injection by pions of a specified energy as $E_{\rm \pi^{\pm}}^2 Q_{\rm \pi^{\pm}}(E_{\rm \pi^{\pm}})$. By conservation of energy, it follows that this energy is provided by the corresponding decays of electrons such that $\mathcal{M}\vert_{E_{\rm p}} E_{\rm e}^2 Q_{\rm e}(E_{\rm e}) \approx K_{\rm e} E_{\rm \pi^{\pm}}^2 Q_{\rm \pi^{\pm}}(E_{\rm \pi^{\pm}})$, where $\mathcal{M}\vert_{E_{\rm p}}$ is introduced as the electron multiplicity, and $K_{\rm e} = 0.265$ is the efficiency of energy transfer from pions to electrons. The multiplicity is determined empirically using a fitting function of the form $c_1+c_2 s^{c_3}$ as specified in ~\citet{Fiete2010JPhG} and~\citet{Albini1976NCimA} with $c_1 = 0.0$, $c_2=3.102$ and $c_3=0.178$ ($s$ is the GeV centre-of-mass interaction energy)\footnote{This fitting function is based on experimental pion-production data from~\citet{Slattery1972PRL, Whitmore1974PhR, Thome1977NuclPhys, Arnison1983PhLB, Breakstone1984PhRvD, Alner1984PhLB, Alner1985PhLB, Ansorge1988ZPhys, Albajar1990NuPhB, Abe1990PhRvD, Wang1991thesis, Rimondi1993proc, Alexopoulos1998PhLB}.}. At 1 GeV, the electron production multiplicity is four, with the centre-of-mass energy governed by $E_{\rm p}$. This means each injected CR electron inherits around 3-5\% of the energy of the CR proton primary initiating the interaction. The electron injection rate (per unit volume) thus follows as:
\begin{align}
Q_{\rm e}(E_{\rm e}, r) &= \frac{K_{\rm e} }{\mathcal{M}\vert_{E_{\rm p}}} \frac{E_{\rm \pi^{\pm}}^2}{E_{\rm e}^2}\; Q_{\rm \pi^{\pm}}(E_{\rm \pi^{\pm}}, r) \nonumber \\
&= \frac{K_{\rm e} }{\mathcal{M}\vert_{E_{\rm p}}} \frac{E_{\rm \pi^{\pm}}^2}{E_{\rm e}^2} \; \frac{{\rm c}}{K_{\pi^{\pm}}}~\alpha^*\left(L, r\right) ~\Phi_{\rm p}\left(L, r \right) \ .
\end{align}
The pion energy is equal to the fraction of energy passed from the primary proton, as specified by the efficiency parameter $K_{\pi^{\pm}}$. Therefore, $E_{\rm \pi^{\pm}} = K_{\pi^{\pm}} \; E_{\rm p}$ and so we may write
\begin{equation}
\label{eq:cre_injection_spec1}
Q_{\rm e}(E_{\rm e}, r) = \frac{\bar{K}}{\mathcal{M}\vert_{E_{\rm p}}} \frac{E_{\rm p}^2}{E_{\rm e}^2} \; {\rm c} \; \alpha^*\left(L, r\right) \; \Phi_{\rm p}\left(L, r\right)  \ ,
\end{equation}
where $\bar{K} = K_{\rm e} K_{\pi^{\pm}}$ is the effective overall energy efficiency parameter for the production of electrons from CR protons, accounting for both the efficiencies in the pion production and pion decay steps. In terms of the electron and proton Lorentz factors, equation~\ref{eq:cre_injection_spec1} may be written as:
\begin{equation}
\label{eq:cre_injection_spec2}
Q_{\rm e}(\gamma_{\rm e}, r) = \frac{\bar{K}}{\mathcal{M}\vert_{\gamma_{\rm p}}} \left(\frac{\gamma_{\rm p}}{\gamma_{\rm e}}\right)^2 \; \left(\frac{m_{\rm p}}{m_{\rm e}}\right)^2 \; {\rm c} \; \alpha^*\left(L, r\right) \; \Phi_{\rm p}\left(L, r\right)  \ .
\end{equation}

\subsection{High energy electron interactions}
\label{sec:cosmic_ray_electron_interactions}

The injected electron secondaries will also interact with their environment. 
While the hadronic primaries are found to be predominantly absorbed, 
CR electrons are more susceptible to continuous cooling processes. 
The main means by which the electrons can lose their energy are: radiatively (by inverse-Compton or synchrotron emission), electro-statically (by Coulomb interactions and free-free processes) and via triplet pair-production (a Bethe-Heitler process). 
\bigskip

  {\sffamily Coulomb and free-free interactions} \newline
  
The cooling rate of electrons due to Coulomb interactions with a low-density fully-ionised plasma is given by:
\begin{equation}
b_{\rm C} \approx m_{\rm e} {\rm c}^2 \; n_{\rm b} \; {\rm c} \; \sigma_{\rm T} \ln \Lambda \ ,
\end{equation}
\citep[see, e.g.][]{Dermer2009book, Schleicher2013A&A} which we note is independent of the electron energy. Here, $\ln \Lambda\simeq 30$ is the Coulomb logarithm (which accounts for the ratio between the maximum and minimum impact parameters in a particle-particle interaction). The electron cooling rate due to bremsstrahlung (free-free interactions) is
\begin{equation}
b_{\rm brem}(\gamma_{\rm e}) \approx \alpha_{\rm f} \; {\rm c} \; \sigma_{\rm T} \; n_{\rm b} \; \gamma_{\rm e} m_{\rm e} {\rm c}^2 \ ,
\end{equation}
\citep[e.g.][]{Dermer2009book, Schleicher2013A&A}.
\bigskip

 {\sffamily Inverse-Compton and synchrotron} \newline
 
In the Thomson limit, cooling rates for inverse-Compton and synchrotron both take the form 
\begin{equation}
\label{eq:synch_ic_cooling}
b_{\rm rad}(\gamma_{\rm e}) = \frac{4}{3} \; \sigma_{\rm T} {\rm c} \; \gamma_{\rm e}^2 \; U_{i}
\end{equation}
\citep[e.g.][]{Blumenthal1970PRD, Rybicki1979book} where $U_{i}$ is the energy density of the relevant field. For inverse-Compton scattering, the high energy electrons interact with lower energy photons, hence the energy density can be taken as that of the dominant radiation field -- in the case of the CMB, $U_{\rm CMB}$, this can be found by integrating over the black-body spectrum given by equation~\ref{eq:blackbody}:
\begin{align}
U_{\rm CMB}(z) &= \int_0^{\infty} n_{\rm ph}(\epsilon)\cdot m_{\rm e} {\rm c}^2 {\rm d}\epsilon \nonumber \\
& = \frac{8 \pi^5 \Theta^4(z) \; m_{\rm e} {\rm c}^2}{15 \; \lambda_{\rm C}^3} \ ,
\end{align}
which is a function of redshift (and where symbols retain their earlier meanings).  The inverse-Compton cooling function has a more complicated form in the high energy Klein-Nishina limit, as shown in Fig.~\ref{fig:total_losses} -- see also~\citet{Blumenthal1970RMP, Dermer2009book}.

In synchrotron emission, electrons cool in a magnetic field. The energy density in this case is given by
$U_{\rm B} = \frac{B^2}{8\pi}$
for a (local) magnetic field strength of $B$.
\bigskip

 {\sffamily Triplet pair-production} \newline
 
At energies above a GeV, triplet pair-production can provide some contribution to the cooling energy losses of electrons. This pair-production process operates in a broadly similar manner to the Bethe-Heitler photo-pair production arising with protons. For $\epsilon_{\rm r} \geq 30$, an approximation correct to within 0.1\% for the cross section can be found:
\begin{equation}%
\begin{split}
\hat{\sigma}_{\rm TPP}(\epsilon_{\rm r}) \approx  & %
	\bigg\{ \frac{7}{6\pi} \alpha_{\rm f}  
	\left( \psi - \frac{109}{42}  \right)  +\big[ 24.3 -9.57 \psi   \\ 
	&  +3.36 \psi^2 -1.16 \psi^3 
	  \big] \left( \frac{10^{-3}}{\epsilon_{\rm r}} \right)     \bigg\}  \sigma_{\rm T} 
\end{split} 
\end{equation}
\citep{Borsellino1947INC, Joseph1958RevModPhys, Haug1981Znat}, where $\psi = \ln(2 \epsilon_{\rm r})$. We note that for the energies of interest here ($\epsilon_{\rm r} \geq 60$), we may adopt the same form of approximation as we did with the photo-pair interaction (for the CR protons):
\begin{equation}%
\label{approx-sigma}
\hat{\sigma}_{\rm TPP}(\epsilon_{\rm r}) \approx %
\left\{	\frac{7}{6\pi}\alpha_{\rm f} \ln\left[\frac{\epsilon_{\rm r}}{k_{\rm TPP}}\right] \right\} \sigma_{\rm T} \ .
\end{equation}%
  The principal difference is that the fitting constant is set to a lower value of $k_{\rm TPP} = 0.05$ here. This gives a path length consistent with~\citep{Bhattacharjee2000PhysRep} at $10^{17}$ eV and $z=0$. The cooling rate then follows as:
 \begin{equation}%
\label{eq:photopair_losses}%
b_{\rm TPP}(\gamma_{\rm e}) \approx %
\frac{112}{9} \frac{{\rm c} \; \alpha_{\rm f} \; \sigma_{\rm T}  u^{5/2} \; \mathcal{F}_{\rm TPP}}{\lambda_{\rm C}^3 } \ ,%
\end{equation}%
where $\mathcal{F}_{\rm TPP}$ takes the same form as the function $\mathcal{F}_{\rm \gamma e}$ in eq. ~\ref{eq:f_func}, but with $k_{\rm \gamma e} = 6.7$ replaced with $k_{\rm TPP} = 0.05$ in line with the corresponding cross sections, and $u =  m_{\rm e} {\rm c}^2/\gamma_{\rm e} {\rm k}_{\rm B} T$.


\section{Inverse-Compton X-ray luminosity}
\label{sec:xray_lum}

\begin{figure}
  \vspace*{0.1cm}
	\includegraphics[width=\columnwidth]{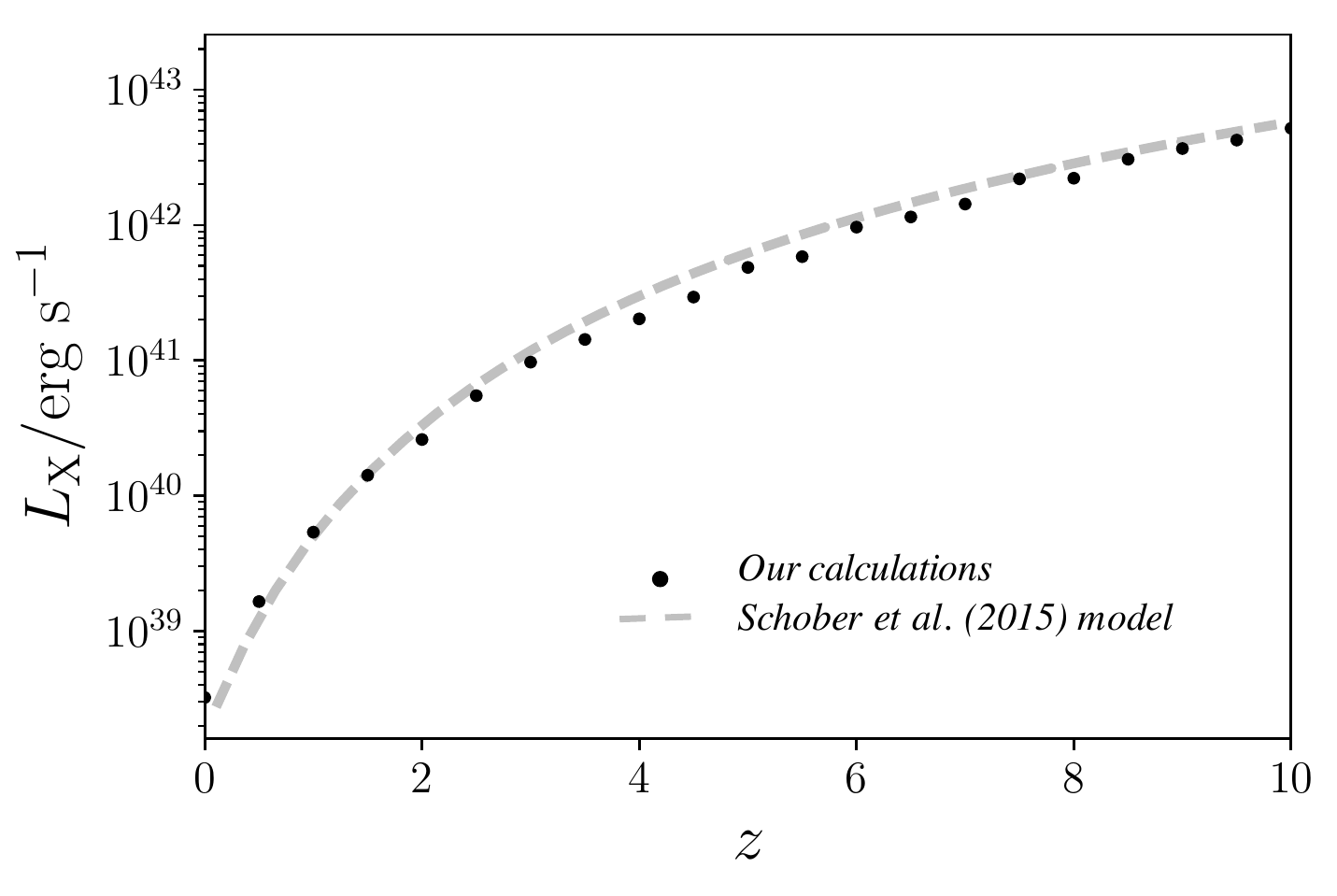}
    \caption{\small Inverse-Compton protogalaxy X-ray luminosity evolution over redshift, as calculated using the approach detailed in this paper with $\mathcal{R}_{\rm SN} \approx 1\;\!\text{yr}^{-1}$ (semi-analytical Monte-Carlo simulation) compared to the earlier study,~\cite{Schober2015MNRAS} (dashed grey line), which adopts an analytical model. Our results are consistent with those obtained by~\cite{Schober2015MNRAS}.}
    \label{fig:cross_check}
\end{figure}

The inverse-Compton X-ray luminosity of starburst galaxies at high-redshift was calculated analytically by~\cite{Schober2015MNRAS} in the 0.2 to 10 keV band, although this previous study invoked a redshift evolution in the volume of the host galaxy following $V\propto \;\! (1+z)^{-3}$. 
The model used in this paper is otherwise comparable and our benchmark model is of volume $1~\text{kpc}^3$, being similar to the volume of the~\cite{Schober2015MNRAS} model at $z=7$.
We run our inverse-Compton X-ray luminosity simulation with $\mathcal{R}_{\rm SN} \approx 1\;\!\text{yr}^{-1}$ and sum over the galaxy to find the total luminosity of the starburst X-ray glow up to $z=10$ to compare with the previous model as a cross-check. We plot our results (black dots) and the previous model when adjusted to account for its additional redshift evolution (grey dashed line) in Fig.~\ref{fig:cross_check}, which shows good consistency between the two approaches.

\section{Heating channel parametrisation}
\label{sec:param_xcheck}

Here we show a cross-check for the two CR heating rate parameterisations introduced in section~\ref{sec:parametrisation}. We may check these parameterisations by taking two extreme cases which have been calculated in full, in section~\ref{sec:cr_heating_power_calc}:
Consider the $\mathcal{R}_{\rm SN} = 0.1~\text{yr}^{-1} \equiv \mathcal{R}_{\rm SF} = 16~\text{M}_{\odot}~\text{yr}^{-1}$ systems at $z=0$ and $z=12$ in Fig.~\ref{fig:z_evolution}. 
The central DC and IX heating levels are shown to be $1.7\times 10^{-24}~\text{erg}\;\text{cm}^{-3}\;\text{s}^{-1}$ and $1.4\times 10^{-29}~\text{erg}\;\text{cm}^{-3}\;\text{s}^{-1}$ respectively at $z=0$, with the IX heating power rising to $4.2\times 10^{-25}~\text{erg}\;\text{cm}^{-3}\;\text{s}^{-1}$ by $z=12$ (DC being unchanged). 
Using the parameterisations~\ref{eq:central_coulomb} and~\ref{eq:central_xray}, our scalings give 
a central DC and IX heating level of $1.7\times 10^{-24}~\text{erg}\;\text{cm}^{-3}\;\text{s}^{-1}$ and $1.6\times 10^{-29}~\text{erg}\;\text{cm}^{-3}\;\text{s}^{-1}$ at $z=0$ respectively, with the X-ray heating power raising to $4.5\times 10^{-25}~\text{erg}\;\text{cm}^{-3}\;\text{s}^{-1}$ by $z=12$, which reconcile 
to within a sufficient degree of accuracy (around 30\%).
At large distances, the X-ray heating rates at 20 kpc are found to be $6.2\times 10^{-38}~\text{erg}\;\text{cm}^{-3}\;\text{s}^{-1}$ and $1.8\times 10^{-33}~\text{erg}\;\text{cm}^{-3}\;\text{s}^{-1}$ for $z=0$ and $z=12$ respectively from our earlier section~\ref{sec:cr_heating_power_calc} calculation. 
Instead, from the scaling parametrisation~\ref{eq:outer_xray}, these levels can be estimated as $4.8\times 10^{-38}~\text{erg}\;\text{cm}^{-3}\;\text{s}^{-1}$ and $1.4\times 10^{-33}~\text{erg}\;\text{cm}^{-3}\;\text{s}^{-1}$ -- again, consistent to within similar error bounds. 
We may thus adopt the parameterisations above to apply our model to the observed galaxies introduced in section~\ref{sec:star_forming_protogalaxies}, avoiding the need to run multiple computationally expensive calculations.

\section{Star-formation rates and feedback}
\label{sec:sf_data_app}

 \begin{figure}
 \centering
	\includegraphics[width=0.9\columnwidth]{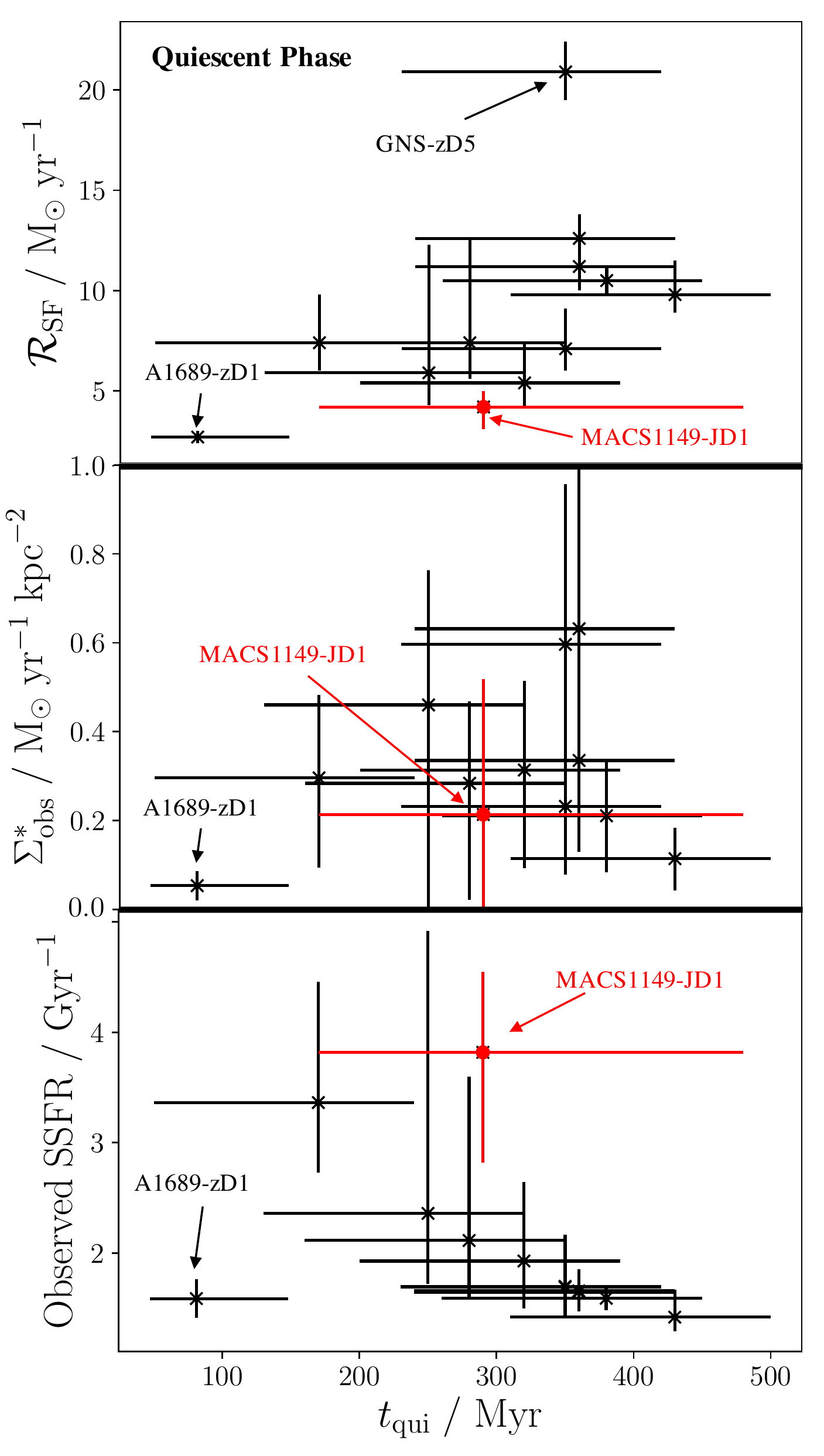}
    \caption{\small Star-formation rates in the sample of 12 post-starburst galaxies. {MACS1149-JD1} is highlighted in red in all panels as this galaxy is at substantially higher redshift ($z=9.11$) than the others in the sample. \textbf{Upper Panel}: Observed quiescent star-formation rate $\mathcal{R}_{\rm SF}$ plotted against the length of quiescence period. This period is measured up to the point of observation, so $t_{\rm qui}$ effectively indicates the time elapsed since the end of starburst activity. This shows a strong positive correlation, with a Pearson coefficient (weighted by the error bounds) of 0.93. 
    \textbf{Middle Panel}: As above, except the $y$-axis is star-formation rate surface density ($\Sigma_{\rm obs}^{*} = \mathcal{R}_{\rm SF}/4\pi r_{\rm gal}^2$). This removes the correlation seen in the upper panel, with a Pearson coefficient of -0.02 (when removing the clear outlier A1689-zD1, as labelled). 
    \textbf{Lower Panel}: $y$-axis as specific star-formation rate, which is adjusted for galaxy mass where $\text{SSFR} = \mathcal{R}_{\rm SF}/M_{*}$. If removing the labelled outlier A1689-zD1, this now gives a non-negligible negative correlation of weighted Pearson coefficient -0.64.}
    \label{fig:sfr_data}
\end{figure}
Residual star-formation appears to remain in our sample of post-starburst systems, or at least has re-emerged by the time of observation. 
This may be indicative of a smoother star-formation history, arising as inflows begin to re-emerge and strengthen over time.
In such a scenario, we would expect to see higher rates of star-formation for systems observed at a greater time after the onset of quenching. 
Indeed, a first view of the data would appear to support this, as seen in the top panel of Fig.~\ref{fig:sfr_data} where the star-formation rate $\mathcal{R}_{\rm SF}$ at the point of observation for each of the high-redshift galaxies is plotted against the time elapsed since the estimated end of their starburst episode, $t_{\rm qui}$.
However, some degeneracies with other effects may be responsible for this trend: in the middle panel we show the star-formation rate surface density, $\Sigma^{*}_{\rm obs} = \mathcal{R}_{\rm SF}/4\pi r_{\rm gal}^2$ and specific star-formation rate, $\text{SSFR} = \mathcal{R}_{\rm SF}/M_{*}$ in the lower panel.
The observed post-starburst $\Sigma^{*}_{\rm obs}$ shows a more complicated behaviour:
while, up to around $t_{\rm qui} \approx 350 \;\! \text{M}_{\odot}\;\text{yr}^{-1}\;\text{kpc}^{-2}$, $\Sigma^{*}_{\rm obs}$ increases with $t_{\rm qui}$. However, for larger $t_{\rm qui}$ values, the star-formation surface density reduces again. 
This could result from the compactness of some of the galaxies: the behaviour of the SSFR in the lower panel (which does not depend on the physical extent of the system) is different and shows that the longer the time elapsed since the end of star-formation, the lower the observed SSFR. 
This is intriguing behaviour, and in tension with our first view of the data, and may suggest $t_{\rm qui}$ is better considered as an indicator for the level of feedback experienced by a system -- see section~\ref{sec:timescales_discussion} for further discussion.

One notable outlier in the top panel of Fig.~\ref{fig:sfr_data}, GNS-zD5, exhibits a remarkably high star-formation rate. This is predominantly due to its large mass -- indeed, it falls in line with the rest of the sample in the lower panels when normalised by size or mass. A1689-zD1 is arguably less conformant, with a particularly low star-formation rate for its size and mass. It may be that this system does not truly fit into the burst-mode SFH considered for these galaxies, instead experiencing a more transient and gradual SFH. In this figure (and subsequently), we have highlighted the high-redshift system MACS1149-JD1. Despite having been observed at a considerably higher redshift than the rest of the sample, its star-forming behaviour does not appear to be unusual.

\begin{figure}
 \centering
	\includegraphics[width=0.92\columnwidth]{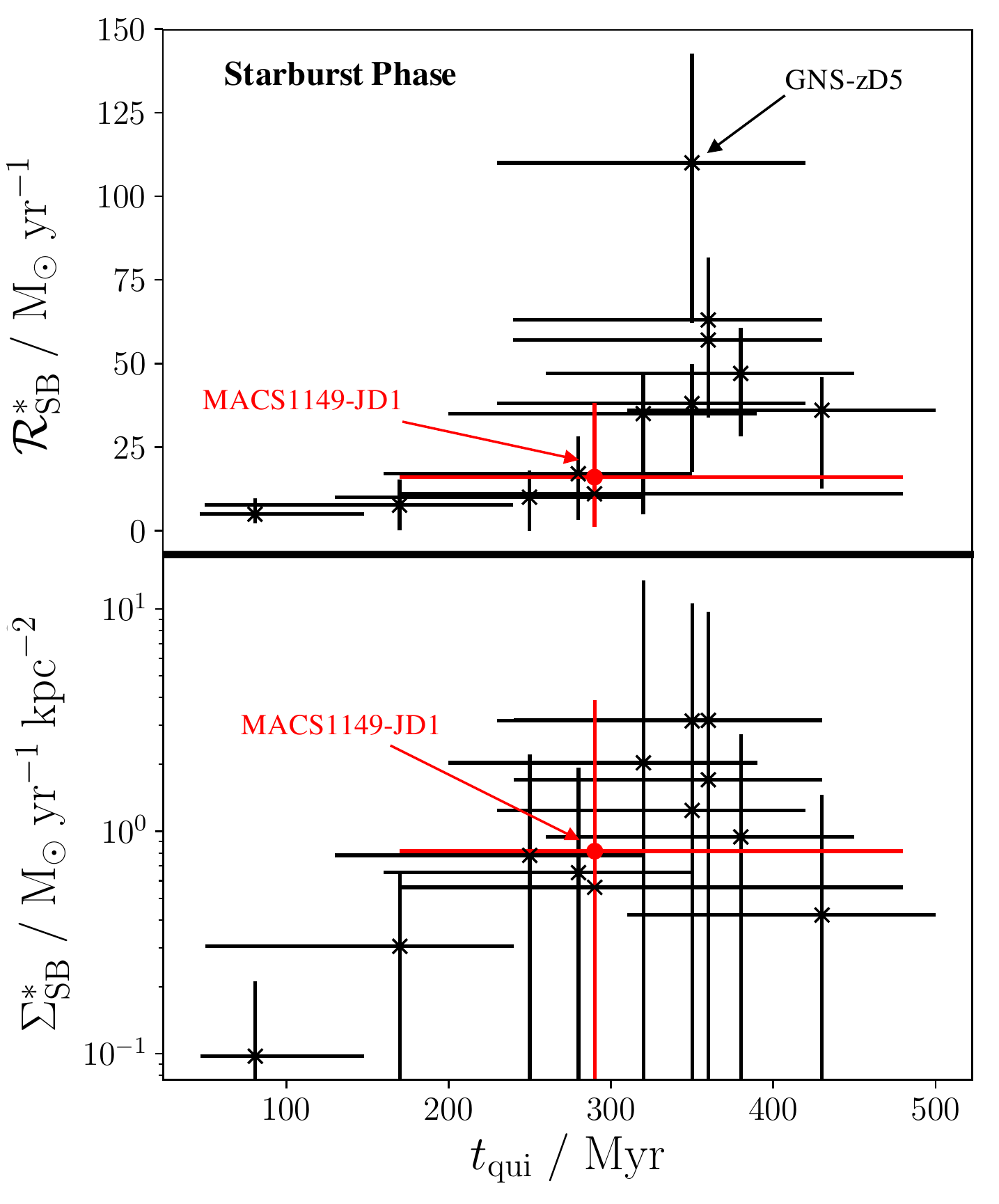}
    \caption{\small As Fig.~\ref{fig:sfr_data}, but for the starburst phase of each system. The weighted Pearson correlation value for the top panel is found to be 0.71, while the lower panel indicates two behaviours with a less clear trend (overall correlation of 0.39) -- see text for details. As SSFR is estimated in the burst phase using $t_{\rm SB}$ and $M_{*}$ rather than being determined independently, it is not physically meaningful to include it here.}
    \label{fig:sfr_data_burst}
\end{figure}
We can also consider the star-formation rate surface density during the starburst episode, $\mathcal{R}_{\rm SB}^{*}$ -- see Fig.~\ref{fig:sfr_data_burst}.
As before, this also correlates with the quiescence timescale, as does the star-formation surface density during the burst. However, the behaviour is again complicated (similar to the middle panel of Fig.~\ref{fig:sfr_data}), increasing to a peak before falling away again. It is unclear whether this pattern is physical in origin, due to random fluctuations with our limited data points, or whether it is a mathematical feature resulting from the way $\mathcal{R}_{\rm B}^{*}$ and associated quantities were calculated:
The star-formation rate during the burst is inversely related to $t_{\rm SB}$ which itself is estimated by $[t(z_{\rm obs}) -t(z_{\rm f})] - t_{\rm qui}$ (see equation~\ref{eq:sb_timescale}). This implies an inverse proportionality between $\mathcal{R}_{\rm SB}^{*}$ (as well as quantities derived from this) and $t_{\rm SB}$ when $t_{\rm qui} > t(z_{\rm obs}) -t(z_{\rm f})$, which occurs after around 350 Myr (which is indeed the location of the peaks in Fig.~\ref{fig:sfr_data_burst})\footnote{This would not account for the similar behaviour seen in Fig.~\ref{fig:sfr_data} where star-formation rates are determined independently. However it is unclear whether this is an indication of a physical process or simply due to low statistics, with some data points aligning to give a false correlation.}. 
We do not plot the SSFR in Fig.~\ref{fig:sfr_data_burst}, given that $\mathcal{R}_{\rm SB}^{*}$ is derived from $M_{*}$ rather than measured independently. 

\end{appendix}

\end{document}